\DeclareUnicodeCharacter{0301}{\'{e}}

\PassOptionsToPackage{dvipsnames}{xcolor}
\PassOptionsToPackage{sort,comma,round,authoryear}{natbib}
\PassOptionsToPackage{colorlinks=True,allcolors=RoyalBlue}{hyperlink}
\documentclass{article}  

\usepackage{aux/preprint}
\usepackage{authblk}

\let\address\affil
\usepackage[
natbib=true,
  isbn=false,
  giveninits=true,
  url=false,
  doi=false,
  eprint=false,
  uniquename=false,  
  uniquelist=false,
  style=authoryear-comp,
  backend=biber,
  maxbibnames=5,
  maxcitenames=2,
  dashed=false,
  hyperref=true,
  sortcites=true,
  sorting=nyt
]{biblatex}

\bibliography{aux/bib}

\AtEveryBibitem{%
  \clearname{translator}%
  \clearlist{publisher}%
  \clearfield{pagetotal}%
}

\AtEveryBibitem{\clearfield{month}}
\AtEveryCitekey{\clearfield{month}}
\renewbibmacro{in:}{%
  \ifentrytype{article}{}{\printtext{\bibstring{in}\intitlepunct}}}
\DefineBibliographyStrings{english}{
  phdthesis = {PhD\addabbrvspace Thesis}
}
\DeclareFieldFormat*{title}{#1}
\DeclareSourcemap{
  \maps[datatype=bibtex]{
    \map{
      \pertype{article}
       \step[fieldset=title, null]
    }
 }
}
\AtEveryBibitem{%
  \clearlist{language}%
}
\DeclareNameAlias{sortname}{last-first}
\newbibmacro{string+doiurlisbn}[1]{%
  \iffieldundef{doi}{%
    \iffieldundef{url}{%
      \iffieldundef{isbn}{%
        \iffieldundef{issn}{%
          #1%
        }{%
          \href{http://books.google.com/books?vid=ISSN\thefield{issn}}{#1}%
        }%
      }{%
        \href{http://books.google.com/books?vid=ISBN\thefield{isbn}}{#1}%
      }%
    }{%
      \href{\thefield{url}}{#1}%
    }%
  }{%
    \href{http://dx.doi.org/\thefield{doi}}{#1}%
  }%
}

\DeclareFieldFormat{journaltitle}{\usebibmacro{string+doiurlisbn}{\mkbibemph{#1}}}
\DeclareFieldFormat[article,incollection]{journaltitle}%
    {\usebibmacro{string+doiurlisbn}{#1}}
\DeclareFieldFormat{volume}{\usebibmacro{string+doiurlisbn}{#1}}
\DeclareFieldFormat[article,incollection]{volume}%
    {\usebibmacro{string+doiurlisbn}{#1}}
\DeclareFieldFormat{pages}{\usebibmacro{string+doiurlisbn}{#1}}
\DeclareFieldFormat[article,incollection]{pages}%
    {\usebibmacro{string+doiurlisbn}{#1}}
\DeclareFieldFormat{number}{\usebibmacro{string+doiurlisbn}{#1}}
\DeclareFieldFormat[article,incollection]{number}%
    {\usebibmacro{string+doiurlisbn}{#1}}
\DeclareFieldFormat{issue}{\usebibmacro{string+doiurlisbn}{#1}}
\DeclareFieldFormat[article,incollection]{issue}%
    {\usebibmacro{string+doiurlisbn}{#1}}

\usepackage{lmodern}

\usepackage{txfonts}
\usepackage{amssymb}
\usepackage{amsmath}
\usepackage{bm}
\usepackage{graphicx}
\usepackage{graphbox}
\usepackage{rotating}
\usepackage{environ}
\usepackage{subcaption}
\usepackage{microtype}
\usepackage[title]{appendix}
\usepackage{booktabs}
\usepackage{here}
\usepackage[utf8]{inputenc}
\usepackage{numprint}
\usepackage{makecell}

\usepackage{lipsum}
\usepackage{pgf}
\usepackage{siunitx}
\usepackage{textcomp}
\usepackage{tikz}
\usepackage{url}
\usepackage{xspace}

\usepackage{tikz}
\usetikzlibrary{external}
\usetikzlibrary{backgrounds,calc,chains,external,shadows,shapes,positioning,fit,arrows}

\usepackage[colorlinks=True,allcolors=RoyalBlue]{hyperref}
\usepackage[nameinlink,capitalize]{cleveref}
\crefname{section}{Sect.}{Sects.} 

\usepackage{ulem}  
\newcommand{\nuna}[2]{(#1)\,\textit{#2}}  

\newcommand{\class}[1]{\texttt{#1}}

\newcommand{\abbr}[2]{#1 (#2)}  

\newcommand{\ie}{i.e.\,}
\newcommand{\eg}{e.g.\,}

\newcommand\inputpgf[2]{{
\let\includegraphicsWithoutPath\includegraphics
\renewcommand{\includegraphics}[2][]{\includegraphicsWithoutPath[##1]{#1/##2}}
\input{#1/#2.pgf}
}}

\def\NSamples{2983}
\def\NAllSamples{5906}
\def\NSamplesPrev{2676}
\def\NAsteroidPrev{1852}
\def\NExcludedSamples{2923}
\def\NSamplesVisOnly{2923}
\def\NAlbedoVisOnly{81.423195}
\def\NSamplesBothSamples{328}
\def\NAsteroidsBothSamples{267}
\def\NZVisOnly{2}
\def\NEVisOnly{2}
\def\NSVisOnly{140}
\def\NAsteroids{2125}
\def\NAsteroidsMoreThanOne{549}

\def\NWavelengths{53}
\def\NSpectra{6038}
\def\NAstSpectra{4526}

\def\NObsAlbedos{4704}
\def\NAstAlbedos{3543}
\def\NFactors{4}
\def\NCluster{50}
\def\NCoreCluster{33}

\def\RatioAlbedoVariance{6.829528}

\def\NMultipleSamples{549}

\def\NClasses{17}

\def\NASamples{57}

\def\NAFinal{32}
\def\NAFinalFraction{1.505882}
\def\NBSamples{68}

\def\NBFinal{45}
\def\NBFinalFraction{2.117647}
\def\NCSamples{299}

\def\NCFinal{221}
\def\NCFinalFraction{10.400000}
\def\NChSamples{144}

\def\NChFinal{107}
\def\NChFinalFraction{5.035294}
\def\NDSamples{119}

\def\NDFinal{82}
\def\NDFinalFraction{3.858824}
\def\NESamples{65}

\def\NEFinal{46}
\def\NEFinalFraction{2.164706}
\def\NKSamples{59}

\def\NKFinal{42}
\def\NKFinalFraction{1.976471}
\def\NLSamples{76}

\def\NLFinal{58}
\def\NLFinalFraction{2.729412}
\def\NMSamples{252}

\def\NMFinal{142}
\def\NMFinalFraction{6.682353}
\def\NOSamples{4}

\def\NOFinal{2}
\def\NOFinalFraction{0.094118}
\def\NPSamples{195}

\def\NPFinal{135}
\def\NPFinalFraction{6.352941}
\def\NQSamples{158}

\def\NQFinal{107}
\def\NQFinalFraction{5.035294}
\def\NRSamples{15}

\def\NRFinal{10}
\def\NRFinalFraction{0.470588}
\def\NSSamples{1188}

\def\NSFinal{898}
\def\NSFinalFraction{42.258824}
\def\NVSamples{206}

\def\NVFinal{142}
\def\NVFinalFraction{6.682353}
\def\NXSamples{50}

\def\NXFinal{33}
\def\NXFinalFraction{1.552941}
\def\NZSamples{28}

\def\NZFinal{23}
\def\NZFinalFraction{1.082353}
\def\BCe{0.497250}
\def\BCh{0.693350}
\def\BCk{0.905960}
\def\NFracChC{20.396601}

\def\NFracPh{19.170984}

\def\NATypesFractionDeMeo{1.612903}

\def\NBTypesFractionDeMeo{1.075269}

\def\NCTypesFractionDeMeo{7.258065}

\def\NChTypesFractionDeMeo{4.838710}

\def\NDTypesFractionDeMeo{4.301075}

\def\NKTypesFractionDeMeo{4.301075}

\def\NLTypesFractionDeMeo{5.913978}

\def\NOTypesFractionDeMeo{0.268817}

\def\NQTypesFractionDeMeo{2.150538}

\def\NRTypesFractionDeMeo{0.268817}

\def\NSTypesFractionDeMeo{53.763441}

\def\NVTypesFractionDeMeo{4.569892}

\def\NXTypesFractionDeMeo{8.602151}

\def\pVCCMean{0.052262}
\def\pVCCStdUpper{0.025039}
\def\pVCCStdLower{0.016929}
\def\pVCMMean{0.147091}
\def\pVCMStdUpper{0.056144}
\def\pVCMStdLower{0.040634}
\def\pVCSMean{0.241933}
\def\pVCSStdUpper{0.102781}
\def\pVCSStdLower{0.072135}

\def\NumberFeaturee{13}

\def\NumberUnknownFeaturee{392}

\def\NumberFeatureh{144}

\def\NumberUnknownFeatureh{361}
\def\RatioUnknownFeatureh{12.101911}
\def\NumberFeaturek{135}

\def\NumberUnknownFeaturek{360}

\def\RatioQNea{83.177570}
\def\RatioNea{34.447059}

\def\RatioMk{40.873016}
\def\RatioMknan{30.158730}

\def\RatioEk{30.769231}
\def\RatioEknan{36.923077}
\def\NMCe{13}
\def\NEe{4}

\def\RatioMCenan{65.442404}
\def\NBarbarians{16}
\def\NLBarbarians{7}
\def\NMBarbarians{5}

\begin{document}

  \title{Asteroid Taxonomy from Cluster Analysis of Spectrometry and
  Albedo}

  \author[1]{Max Mahlke}
  \author[1]{Benoit Carry}
  \author[2]{Pierre-Alexandre Mattei}

  \address[1]{Universit{\'e}
    C{\^o}te d'Azur, Observatoire de la C{\^o}te d'Azur, CNRS, Laboratoire Lagrange, France}
  \affil[2]{Université Côte d'Azur, Inria, Maasai project-team, Laboratoire J.A.
    Dieudonné, UMR CNRS 7351, France}
  \twocolumn[
    \begin{@twocolumnfalse}
      \maketitle
      \begin{abstract}
The
classification of the minor bodies of the Solar System based on observables
has been continuously developed and iterated over the past 40
years. While prior iterations followed either the availability of large
observational campaigns or new instrumental capabilities opening
new observational dimensions, we see the opportunity to improve primarily upon the established
methodology.
We developed an iteration of the asteroid taxonomy which allows the
classification of partial and complete observations (i.e. visible,
near-infrared, and visible-near-infrared spectrometry) and which  reintroduces
the visual albedo into the classification observables. The resulting class
assignments are given probabilistically, enabling  the uncertainty of
a classification to be quantified.
We built the taxonomy based on \num{\NSamples} observations of
\num{\NAsteroids} individual asteroids, representing an almost tenfold increase of
sample size compared with the previous taxonomy. The asteroid classes are
identified in a lower-dimensional representation of the observations using a
mixture of common factor analysers model.
We identify \num{\NClasses} classes split into the three complexes \class{C},
\class{M}, and \class{S}, including the new \class{Z}-class for extremely-red
objects in the main belt. The visual albedo information resolves the spectral
degeneracy of the \class{X}-complex and establishes the \class{P}-class as part
of the \class{C}-complex. We present a classification tool which computes
probabilistic class assignments within this taxonomic scheme from asteroid
observations, intrinsically accounting for degeneracies between classes based on
the observed wavelength region. The taxonomic classifications of
  \num{\NSpectra} observations of \num{\NAstSpectra} individual asteroids are
published.
The ability to classify partial observations and the reintroduction of the
visual albedo into the classification provide a taxonomy which is well suited
for the current and future datasets of asteroid observations, in particular
provided by the Gaia,   MITHNEOS,   NEO Surveyor, and   SPHEREx surveys.
      \end{abstract}
    \end{@twocolumnfalse}
  ]





\section{Introduction}
\label{sec:introduction}

The minor planets of the Solar System exhibit a wide range of surface
compositions as outcomes of their diverse formation histories. Mineralogical
insights into the main asteroid belt gained from observing the bodies' exteriors
serve to constrain the dynamic evolution scenarios of our planetary environment
\citep{TheDynamicalEMorbid2015}, to establish relationships in
asteroid families \citep{AsteroidFamilyMasier2015}, and to identify the parent
bodies of the members of the meteorite collection
\citep{2002aste.book..653B,IdentificationGranvi2018}. The conclusion of a static
Solar System formation history \citep{1982Sci...216.1405G} has since been
discarded in favour of a dynamical version
\citep{OriginOfTheCGomes2005,ChaoticCaptureMorbid2005,OriginOfTheOTsigan2005}
following the increasing resolution of the compositional distribution of
asteroids in the  main belt and in near-Earth orbits thanks to a growing number
of minor bodies characterised by dedicated observational efforts
\citep[\eg][]{SmallMainBeltXuSh1995,PhaseIiOfTheBusS2002,VisibleSpectroDevoge2019}. Today,
the majority of the mass in the main belt is thought to have been dynamically
implanted during a later evolutionary stage of the Solar System
\citep{1982Sci...216.1405G,2014Natur.505..629D}, including some of the largest
members of the main belt
\citep{IronMeteoritesBottke2006,AmmoniatedPhylDeSan2015,Vokrouhlick__2016,VltSphereImagVernaz2021}. Evidence
of a dichotomous meteorite population further strengthens this interpretation of
a large compositional variability among minor bodies as result of early-stage
formation processes in the Solar System \citep{StableIsotopicWarren2011}.

To  describe the compositional distribution, a classification scheme based on the
observable features of asteroids is required. A common device used in the
interpretation of observations is asteroid taxonomy. Taxonomic classification
refers to the grouping of objects with shared characteristics
\citep{TheorieElementCandol1813}. For asteroids, these characteristics are the
observable surface properties, such as the absorption bands imprinted into their
reflectance spectra or the surface albedos. The implicit assumption is that the
observables are related to the minor planets' surface mineralogy
\citep{1979aste.book..688G}, though this is not a prerequisite for a practical
taxonomy.

Schemes for the compositional classification of minor planets have been devised
and iterated  regularly since the 1970s
\citep[\eg][]{1971NASSP.267...51C,AsteroidsSpeMccord1975,1978Icar...35..313B}. The
initial division into carbonaceous \class{C}-types and silicaceous
\class{S}-types was readily apparent in different observables, even for a small
number of observed objects and limited observational detail. However, with  an
increasing number of smaller objects   observed, the underlying continuum
distribution between these complexes has been revealed \citep{2002Icar..158..146B}.

The most commonly used taxonomies for minor bodies are the Tholen system
\citep{1984PhDT.........3T} and the Bus-DeMeo system
\citep{2002Icar..158..146B,AnExtensionOfDemeo2009}. While the latter offers a
feature-based classification which encompasses a wide range of the variability
observed in spectral observations and has been adapted  to visible and
\abbr{near-infrared}{NIR} photometric observations
\citep{SdssBasedTaxoCarvan2010,TheTaxonomicDDemeo2013,TaxonomicClassPopesc2018},
the former has not been fully replaced, in part due to two advantages of the
used asteroid observables: the visual albedo $p_V$ and
spectrophotometric observations down to \abbr{ultraviolet}{UV}
wavelengths. Both features increase in particular the resolution of classes
which only show faint features in the visible and NIR wavelength regimes.

In this work we aim to methodologically improve upon the existing taxonomic schemes for minor
bodies  with regard to three aspects. First, we introduce a method
which enables the  classification of  complete and partial observations. This offers
consistent class definitions
across the \abbr{visible-near-infrared}{VisNIR} region. Second, the visual
albedo $p_V$ is reintroduced into the taxonomy observables, solving the
degeneracy of the \class{X}-complex as a primary consequence. Third, asteroids are
classified in a probabilistic model, yielding a vector of class probabilities
rather than a definite class assignment, which enables  taxonomic
outliers and transitional populations to be identified.

In addition to the methodological advancement, we further aim to align
the scheme of taxonomic classes with advancements in the understanding of
asteroid surface compositions acquired over the last decade. Studies such as
\citet{TheFractionOfRivkin2012}, \citet{MultipleAndFaVernaz2014,Vernazza_2015},
and \citet{ARadarSurveyShepar2015} have combined observational evidence for
several asteroid and meteorite connections which show that the classes in the
current schemes do not reflect mineralogical groups. While this is acceptable a priori
 as taxonomies are built on spectroscopic data alone,  by taking into account the multi-observable studies we believe that a
correction is acceptable and
necessary.

In \cref{sec:method} we outline the collection of the observational data used
in this study, as well as the methodological advancement of the clustering
strategy with respect to previous taxonomies. In \Cref{sec:results} we outline the
clustering results and the strategy of identifying compositional classes. These
classes are then discussed in detail in \cref{sec:discussion}. In
\cref{sec:classification} we investigate degeneracies between the classes in
this taxonomy and compare the classifications of asteroids in this study to those in the literature. The \texttt{classy} tool to classify asteroid observations in the
framework of this taxonomy is presented. Finally, we draw conclusions and give
an outlook in \cref{sec:conclusion}.

\section{Method}%
\label{sec:method}

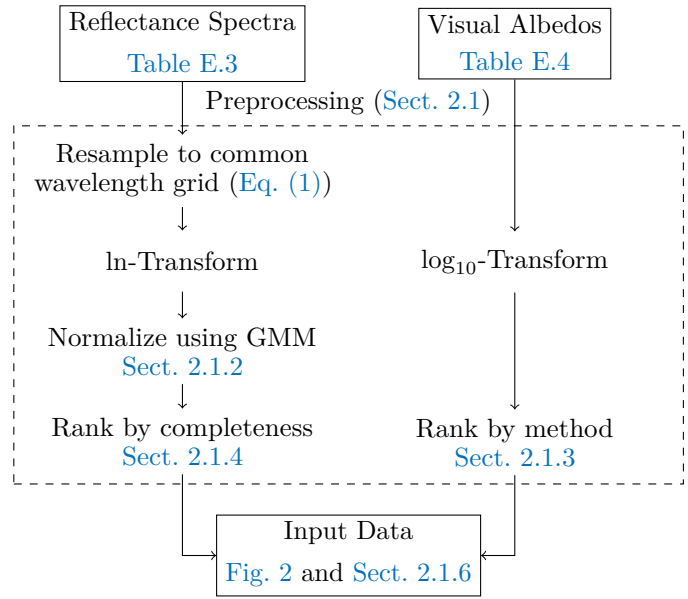
\begin{figure}[t]
  \centering
  \tikzstyle{input} = [rectangle split, rectangle split parts=2, rectangle split draw splits=false, text centered, text height=0.2cm, draw=black]

\tikzstyle{preprocessing} = [rectangle, text centered, text width=4cm, minimum height=0.8cm]

\tikzstyle{process} = [rectangle split, rectangle split parts=2, text centered, draw=black]

\begin{tikzpicture}[node distance=1.2cm]


  \node (input_spectra) [input]
  {Reflectance Spectra \nodepart{second}{\cref{tab:ref_spectra}}};

  \node (input_albedos) [input, right=1.5cm of input_spectra]
  {Visual Albedos \nodepart{second}{\cref{tab:ref_albedos}}};

  \node (prepro_spec_resample) [preprocessing, below=0.7cm of input_spectra]
  {Resample to common wavelength grid (\cref{equ:sampling_pattern})};
  \node (prepro_spec_log) [preprocessing, below of=prepro_spec_resample]
    {$\ln$-Transform};
  \node (prepro_spec_normalize) [preprocessing, below of=prepro_spec_log]
  {Normalize using GMM\\\cref{subsub:spectra}};
  \node (prepro_spec_rank) [preprocessing, below of=prepro_spec_normalize]
  {Rank by completeness\\\cref{sub:merging}};

  \node (prepro_alb_resample) [preprocessing, below=0.82cm of input_albedos]
    {};
  \node (prepro_alb_log) [preprocessing, below of=prepro_alb_resample]
  {$\log_{10}$-Transform};
  \node (prepro_alb_normalize) [preprocessing, below of=prepro_alb_log]
    {};
  \node (prepro_alb_rank) [preprocessing, below of=prepro_alb_normalize]
  {Rank by method\\\cref{subsub:albedo}};

  \begin{scope}[on background layer]
    \node[label=above:Preprocessing (\cref{sub:input_data}), fit = (prepro_spec_resample)(prepro_alb_rank)(prepro_alb_rank)(prepro_alb_log), rectangle, draw=black, dashed] (preprocessing) {};
  \end{scope}

  \node (input_merging) [input, below=0.4cm of preprocessing]
  {Input Data \nodepart{second}{\cref{fig:input_data} and \cref{subsub:data_availability}}};

  \draw[->] (input_spectra) -- (prepro_spec_resample);
  \draw[->] (prepro_spec_resample) -- (prepro_spec_log);
  \draw[->] (prepro_spec_log) -- (prepro_spec_normalize);
  \draw[->] (prepro_spec_normalize) -- (prepro_spec_rank);
  \draw[->] (prepro_spec_rank) |- (input_merging);

  \draw[->] (input_albedos) -- (prepro_alb_log);
  \draw[->] (prepro_alb_log) -- (prepro_alb_rank);
  \draw[->] (prepro_alb_rank) |- (input_merging);

\end{tikzpicture}
  \caption{Overview of  preprocessing  the input observations. The
    preprocessing steps encompassed in the dashed rectangle can be performed using the
  \texttt{classy} \texttt{python} package described in \cref{sec:classification}.}
  \label{fig:preprocessing}
\end{figure}

In this section we describe the compilation and preprocessing of the
asteroid spectra and albedos for the cluster analysis,  an
overview of which  is provided in \cref{fig:preprocessing}. It is followed by a
description of the issues that arise when working with partial observations (\ie missing data). After motivating the split of the dataset into clustering
and classification data, the section concludes with a description of our
approach to the dimensionality reduction and clustering problem at hand.

\subsection{Input data}
\label{sub:input_data}

\subsubsection{Selecting the observables}%
\label{subsub:selecting_the_feature_space}

The selection of asteroid observables to be  included in a taxonomical system
is a crucial decision in its design. A broad set of observables ensures its
applicability to a large number of asteroids and high compositional resolution;
however, it complicates the derivation of the classification scheme and limits
the number of available observations as only the intersection in terms of
observed asteroids can be considered when combining different
datasets.\footnote{In machine learning literature, the
observables used to identify groups in the input data are referred to as
features, while the observations are referred to as samples. The
input data is a matrix spanned by the features as columns and the samples
as rows.}
This first led \citet{1984PhDT.........3T} to apply the albedo only in a
secondary classification step before the observable was completely
dropped by \citet{2002Icar..158..146B}.

One of our main goals for this iteration of the taxonomy is the possibility
to classify partial observations; we are a priori accepting gaps in the input data, and are thus
not limiting the sample size when
combining datasets and can use the union rather than the intersection of
observations. Nevertheless, while we first considered a classification
system based on spectrometric and photometric observations, and on
visual albedos and phase curve coefficients, we found that including
photometric observations and phase curve coefficients did not add to the
compositional resolution of the resulting scheme as they are effectively
low-resolution versions of the former
\citep{TheTaxonomicDDemeo2013,10.1016/j.pss.2015.11.007,AsteroidPhaseMahlke2021}. Therefore,
we chose to build the taxonomy from VisNIR
spectra and visual albedos.

\subsubsection{Spectra} \label{subsub:spectra} 
\begin{figure}[t]
  \centering
  \input{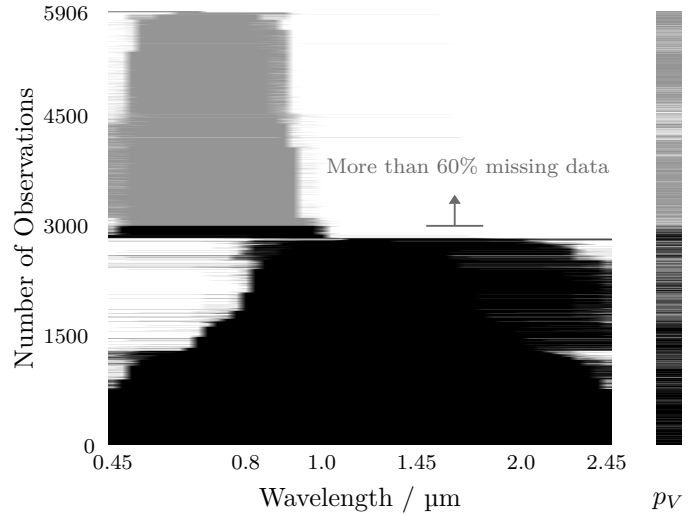}
  \vspace{-0.5em}
  \caption{Input data  shown as a matrix.  The columns represent the
    asteroid observables (\ie the spectral wavelength bins and the
    visual albedo $p_V$) and each row represents one
    observation. The density of sampled wavelength bins is doubled in
    the visible compared to the near-infrared region. The cells are white if
    the corresponding value was not observed. The black cells indicate the samples used in the clustering analysis; the  grey cells are samples that  are
    classified but not used to build the taxonomy itself, due to the
    large degree of missing information in these spectra. \num{2983}
    observations are at least \SI{40}{\percent} complete and were used to train the
  clustering model. The matrix is sorted
    by increasing completeness of the asteroid spectra from top to bottom.}
  \label{fig:input_data}
\end{figure}

Spectrometric observations are
the most compositionally informative asteroid features accessible via remote
sensing. In preparing this work we focused both on building a large repository
of asteroid spectra and on curating the data. In total, we acquired over
\num{7500} spectra from online repositories, archived publications, and directly
from the observers. The majority of spectra are unpublished spectra from the
\abbr{Small Main-belt Asteroid Spectroscopic Survey}{SMASS}
\citep{SmallMainBeltXuSh1995} and \abbr{MIT-Hawaii Near-Earth Object
Spectroscopic Survey}{MITHNEOS}
\citep{CompositionalDBinzel2019,TheDebiasedCoMarsse2022} surveys available
online.\footnote{\url{https://smass.mit.edu}} Literature sources of the spectra
are given in \cref{tab:ref_spectra}. After several iterations of visual
inspection and rejection of low-quality data and duplicated observations,
\num{\NSpectra} spectra of \num{\NAstSpectra} individual asteroids
remained. About \SI{50}{\percent} of spectra cover the visible wavelength range
only, while the sample of VisNIR spectra is about three times as large as in
\citet{AnExtensionOfDemeo2009} (see \cref{fig:input_data} in this paper).

The collection of spectra is heterogeneous in numerous aspects, including but
not limited to their wavelength coverage, resolution, and sampling
patterns. However, a consistent sampling pattern between all spectra is
required for the following numerical analyses. We define this pattern in close
resemblance to the one used by \citet{AnExtensionOfDemeo2009}, though we
 halve the sampling step size in the visible wavelength range as we find
that possible superpositions of absorption features due to mafic minerals
around \SI{1}{\micro\meter} are better described by the finer sampling. The
chosen sampling pattern is

\begin{align}
        \label{equ:sampling_pattern}
        \begin{split}
                \lambda_{S} \in \{&0.45, 0.475, 0.50, \dots, 1.0, 1.025, \\
                &1.05, 1.10, 1.15 \dots, 2.40, 2.45\}\ \mu\textrm{m},
        \end{split}
\end{align}

totaling \NWavelengths\ wavelengths. In the following cluster analysis, each of
these wavelengths represents one data dimension.

Before resampling the spectra, we apply a filter
\citep{1964AnaCh..36.1627S} to smoothen features
present in the spectra (\eg telluric absorption features).
The filter consists of applying least-squares fits of
polynomials to a window of adjacent data points. The window size in
units of data points and the degree of the polynomial dictate the amount of
smoothing that is applied. We set these two parameters for each spectrum
separately by visual inspection of the results.
The smoothened spectra are then linearly interpolated and resampled to the
pattern in \cref{equ:sampling_pattern}.
We then transform the spectra using the natural logarithm,
which serves to approximate a zero mean and
uniform standard deviation of the input spectra as they are
generally normalised to unity at either \SI{0.55}{\micro\meter} or
\SI{1.25}{\micro\meter}. This standardisation transform is generally
beneficial to clustering and dimensionality reduction methods
\citep{bouveyron_celeux_murphy_raftery_2019}.

The inclusion of missing data in the analysis poses a new challenge when it
comes to normalising the spectral data.  The common approach of multiplicatively
setting the reflectance to unity at a shared wavelength is not possible as no
single wavelength is shared among all spectra, as can be seen in
\cref{fig:input_data}. Furthermore, this approach would artificially decrease
the variance in the wavelength chosen for normalisation and the
neighbouring wavelength bins, causing the subsequent clustering analysis to
effectively ignore the normalisation region.

Instead, we prepare the spectra in a way which benefits the following analysis
most by employing a \abbr{Gaussian mixture model}{GMM}. We assume that each
spectrum can be written $\alpha y$, where $\alpha \in \mathbb{R}$ is a
normalisation constant that depends on the considered spectrum, and $y \in
\mathbb{R}^{53}$ is a normalised spectrum. Further assuming that $y$ follows a
mixture of $k$ log-normal distributions with diagonal covariances, all
parameters of the models can be estimated from an incomplete data set via an
expectation-maximisation algorithm \citep{10.2307/2984875}. This allows in
particular to estimate the normalisation constants of all spectra, and to
finally normalise them. By trial and error, we find that $k=30$ mixture
components result in a satisfying normalisation. As outlined in
\cref{sub:dimred_and_clustering}, the assumption of a normal distribution of the
input samples in data space is also made in the clustering analysis.

Finally, we note that \citet{AnExtensionOfDemeo2009} removed the slope component
of the spectra to decrease the influence of space weathering on the taxonomy and to increase the depth of features present in the data. We
cannot do this due to the missing data; however, we consider the presence of
spectral-weathering effects in the taxonomy a beneficial rather than
unfavourable aspect, as we further outline in \cref{sec:discussion}.

\subsubsection{Albedo}
\label{subsub:albedo}

The visual albedos used in this study were compiled for the IMCCE's Solar
system Open Database Network
(SsODNet,\footnote{\url{https://ssp.imcce.fr/webservices/ssodnet/}} Berthier et
al., in prep.). The main contributors in this compilation are the \abbr{Infrared
Astronomical Satellite}{IRAS} \citep{TheSupplementaTedesc2002}, the
\abbr{Wide-field Infrared Survey Explorer}{WISE} \citep{Masiero_2011}, AKARI
\citep{AsteroidCataloUsui2011}, and Spitzer \citep{Neosurvey1InTrilli2016}.  We
use the SsODNet service to collect \num{\NObsAlbedos} albedo
measurements for  the \num{\NAstAlbedos} asteroids of which we have
spectral observations using the \texttt{rocks}\footnote{\url{https://rocks.readthedocs.io}}
\texttt{python}-interface (Berthier et al., in prep.).

When possible, we make use of several albedo measurements per
asteroid when combining the input features (\cref{sub:merging}). To get
the most accurate available value for each asteroid, we first compute the albedo
based on the weighted averages of the asteroid's diameter and absolute magnitude
provided by SsODNet following \citet{2002aste.book..205H}. These averages
consist of the subjectively best available observations (Berthier et
al.{}, in prep{}.). In a second step we compute the weighted mean of any albedo
measurement available in the literature for the given asteroid and use this
value as the second available albedo observation in the input data. Finally, the
non-aggregated albedo observations are appended as additional available
measurements. The literature sources we used to compile available albedo values and recompute
updated ones from absolute magnitude and diameter are given in \cref{tab:ref_albedos}

As for the spectra, we aim to have Gaussian distributions in the albedo
data. \citet{TheAlbedoDistWright2016} shows that the distribution of albedos
follows a double-Rayleigh distribution with a dark peak and a bright peak. To
get a Gaussian distribution, we pass the $\log_{10}$-transform of the albedos to
the clustering algorithm after limiting all albedo values to the interval [0.01,
1). We note that the albedo represents a single data dimension in the
following analysis, compared to the 53 spectral data dimensions.

\subsubsection{Merging of data samples}%
\label{sub:merging}

Previous taxonomies were derived based on photometry or spectrometry from a
single dataset, for example the \abbr{Eight Color Asteroid Survey}{ECAS}
\citep{TheEightColorZellne1985} for \citet{1984PhDT.........3T} or the
SMASS survey for \citet{2002Icar..158..146B}. Individual
observations of a single asteroid were combined in these datasets to give the
single best-possible observation. In this work we do not combine the
observations as they come from numerous different sources; for example,
\num{\NAsteroidsMoreThanOne} of the  \num{\NAsteroids} asteroids have more than one
observation in the input data.

\begin{figure}[t]
  \centering
  \input{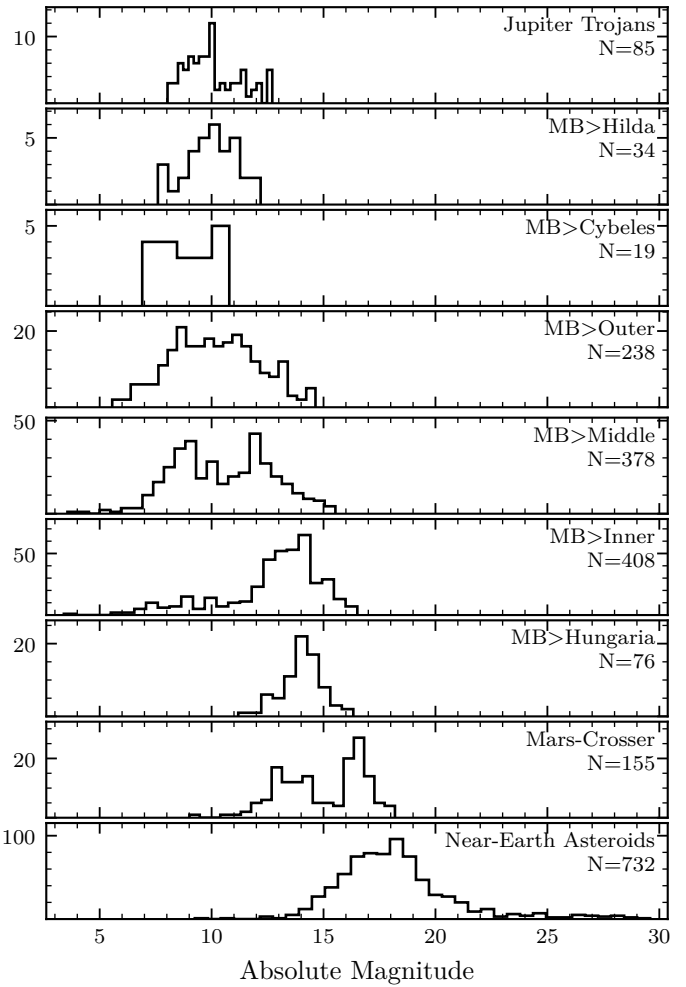}
  \caption{Distribution of the \num{\NAsteroids} individual asteroids used to build the taxonomy over orbital class and absolute magnitude. MB refers to the main
  belt. The number $N$ of asteroids in the orbital population is given below
each orbital class. The bin size of the histograms varies with $N$.}
\label{fig:absolute_magnitude}
\end{figure}

When merging the asteroid spectra and albedo observations for each asteroid, we
aim to create as many complete rows as possible. The array of albedo values is
merged with the asteroid's spectral values in order of most complete
observations. If there are more albedo observations than spectra, we discard the
remaining values. We further set an upper limit of five spectra
for each asteroid, removing any excess ones in order of the fewest
observed wavelength bins. Many spectra of a single asteroid may cause artificial
clusters or trends in the resulting taxonomy. This reduces the number  of
available spectra from \num{\NSpectra} to
\num{\NAllSamples}. \Cref{fig:input_data} shows the final matrix of
observations, colour-coded to differentiate partial and complete observations.

\subsubsection{Clustering versus classification}
\label{sub:clustering_versus_classification}

When clustering the entire input dataset described above with the method
outlined below, we note a population of clusters which contains mostly
visible-only spectra. The intuitive explanation of these clusters is that when
computing dimensionality reduction and projecting the spectra into a
lower-dimensional space, they will be distributed over a smaller volume than
the VisNIR spectra due to their large degree of missing
information. This artificial accumulation of input data in the latent space
disturbs the identification of real clusters in the data. We therefore set an
upper limit of \SI{60}{\percent} of missing values per spectrum to
be included into the clustering input data, which includes \num{\NSamples}
of the \num{\NAllSamples} samples in the input data (see
\cref{fig:input_data}). The remaining \num{\NExcludedSamples} are not used to
derive the taxonomy; instead, they are classified in the resulting scheme
and used to cross-validate the classification, as outlined in
\cref{sec:classification}.
 In \cref{fig:absolute_magnitude}
 we display the distribution of the \num{\NAsteroids} individual asteroids in the
data with which we derive the taxonomy  in the sample over orbital classes and absolute magnitude.

\subsubsection{Data availability}%
\label{subsub:data_availability}

The dataset containing the input data samples, asteroid metadata, and the
resulting classifications as outlined in the next sections is available at the
\abbr{Centre de Données astronomiques de
Strasbourg}{CDS}.\footnote{The table of asteroid classifications and the templates of the defined
taxonomic classes are available in electronic form at the CDS via anonymous ftp
to \url{cdsarc.u-strasbg.fr} (\url{130.79.128.5}) or via
\url{http://cdsweb.u-strasbg.fr/cgi-bin/qcat?J/A+A/665/A26}}

\subsection{Dimensionality reduction and clustering}
\label{sub:dimred_and_clustering}

The derivation of a taxonomy falls into the realm of unsupervised machine
learning. In the context of minor bodies, the approach consists of two steps:
dimensionality reduction followed by clustering. Previous taxonomies have
predominantly chosen \abbr{principal component analysis}{PCA} for the former and
visual clustering for the latter. Given our goals for this new iteration of
the current taxonomy as stated in \cref{sec:introduction}, we need to evolve the
established method, in particular to allow for the classification of partial
observations. In the following we outline this method evolution, which arises
naturally when challenging the PCA method with the requirements of our input
data and the prior knowledge from previous taxonomies. The description of the
resulting model is kept concise; the reader is referred to
\citet{tipping1999probabilistic}, \citet{MixturesOfFacBaek2010}, and
\citet{HeteroscedasticMontan2010} for detailed explanations, and to
\citet{ADataDrivenMCasey2019} for an example application of the same model
but with a different treatment of missing data in the field of stellar
physics.

\subsubsection{Dimensionality reduction}
\label{subsub:dimensionality_reduction}

The necessity of dimensionality reduction derives mainly from the spectrometric
observations, where each bin of the sampling pattern represents one of the  54 data
dimensions. Clustering in such high-dimensional space is not feasible as any
model would be overparametrised given the limited sample size of the input
data. Reducing the dimensionality of the observed data space is achieved by
building linear combinations of the observed variables, referred to as
latent variables,\footnote{\textit{Latent} can here be
understood as a synonym for hidden or underlying as these
variables are not directly observable.} and projecting the input data into the
space spanned by the latent variables, referred to as latent space.

We assume that we have $N$ observations of a $p$-dimensional observable. The
input data $\mathbf{Y}$ is thus of shape $N\times p$, denoted
$\mathbf{Y}_{N\times p}$.\footnote{In the following we state the shape of the
  tensors in this manner when we first introduce them, and drop the notation
afterwards.} PCA can be described as eigendecomposition of the covariance matrix
$\Sigma_{p\times p}$ of $\mathbf{Y}$,

\begin{equation}
  \label{equ:pca}
  \bm{W}^\intercal\Sigma\bm{W} = \mathbf{\Lambda},
\end{equation}where $\mathbf{W}$ and $\mathbf{\Lambda}$ are the eigenvectors and eigenvalues of
$\Sigma$ \citep{pearson1901liii}. Expressing the $p$-dimensional $\Sigma$ by the
$q$-eigenvectors corresponding to the largest eigenvalues of $\Sigma$, where
$q<p$, leads to dimensionality reduction while retaining the largest possible
amount of variance within the data. The lower-dimensional representation
$\mathbf{Z}_{N\times q}$ of the input data $\mathbf{Y}$ is given by the matrix product of
$\mathbf{Y}$ with the matrix of the subset of eigenvectors $\mathbf{W}_{p\times q}$. In the
following   latent components are denoted  $\mathbf{W}$  and
latent scores    $\mathbf{Z}$. The $p$ elements of each latent component are
referred to as latent loadings. They are the coefficients of the linear
combination of dimensions of the input data. Latent components are constrained
to unit L2-norm (\ie the square root of the sum of the squared latent loadings
is one). We note  that the latent components and their loadings are
  determined from the data alone; no a priori information
is used to influence the matrix.

PCA does not allow for missing data as it relies on the eigendecomposition of
$\Sigma$. This limitation is overcome by reformulating PCA as a latent
generative variable model. In essence, while computing the latent components and
scores from the input data, we are making the assumption that there exists a
Gaussian-distributed variable $z$ in the latent space which causes the variance
observed in the  higher-dimensional data (\ie that the resulting latent scores
are normal distributed). A general model of this approach can be expressed as
\citep{tipping1999probabilistic}

 \begin{equation}
   \label{} \bm{Y} = \bm{f}(\bm{Z}, \bm{W}) + \bm{\epsilon},
 \end{equation}

where $f$ is a function of the latent scores $ \bm{Z}$ and the latent
components $\bm{W}$ and $\bm{\epsilon}_{p\times p}$ is a noise matrix independent of $\bm{Z}$.
The reformulation of PCA in this model framework is referred to as \abbr{probabilistic
PCA}{PPCA} \citep{tipping1999probabilistic}. The model parameters are
fit using an expectation-maximisation algorithm \citep{10.2307/2984875} and, when the input
data is complete, gives the same solution as the conventional PCA.

\begin{figure*}[t]
  \centering
  \tikzstyle{single} = [rectangle, text centered, text height=0.2cm, draw=black,
minimum height=0.4cm]
\tikzstyle{double} = [rectangle split, rectangle split parts=2, rectangle split draw splits=false, text centered, text height=0.2cm, draw=black]

\tikzstyle{preprocessing} = [rectangle, text centered, text width=5cm, minimum height=0.8cm]

\tikzstyle{process} = [rectangle split, rectangle split parts=2, text centered, draw=black]

\begin{tikzpicture}[node distance=0.8cm]

  \node (param_init) [double, draw=none] {Parameter Initialization\nodepart{second}{\cref{sub:initialization_and_training}}};
  \node (training) [preprocessing, below=0.2cm of param_init] {Gradient Descent Training};

  \node (latent_factors) [double, below=of training, align=left] {Latent
  Factors (\cref{fig:latent_loadings})\nodepart{second}{Latent Components}};
  \node (latent_scores) [double, right=of latent_factors, align=left] {Latent
  Scores (\cref{fig:latent_scores,fig:latent_scores2})\nodepart{second}{Cluster Probabilities}};
  \begin{scope}[on background layer]
    \node[label={[xshift=-3.3cm, yshift=0cm]Clustering (\cref{sub:dimred_and_clustering})}, fit =
    (param_init)(training)(latent_factors)(latent_scores), rectangle, draw=black, dashed] (mcfa) {};
  \end{scope}

  \node (input_params) [double, right=of mcfa, align=left] {Hyperparameters
  (\cref{subsub:parameters})\nodepart{second}{
\shortstack{4 Latent Factors \\50 Latent Components} }};

  \node (input_data) [double, above=1.4cm of input_params] {Input Data \nodepart{second}{\cref{fig:input_data} and \cref{subsub:data_availability}}};

  \node (new_data) [double, below=1.1cm of input_params] {Input Data or\nodepart{second}{New Observations (\cref{sec:classification})}};

  \node (decision_tree) [rectangle split, rectangle split parts=2, below=0.7cm of mcfa] {Decision Tree \nodepart{second}{\Cref{tab:decision_tree}}};

  \begin{scope}[on background layer]
    \node[label={[xshift=1.9cm, yshift=0cm]Classification (\cref{sec:discussion,sub:classification_tool_classy})}, fit =
    (latent_factors)(latent_scores)(decision_tree), rectangle, draw=black, dashed] (classy) {};
  \end{scope}

  \draw[->] (input_data) -| (mcfa);
  \draw[->] (input_params) -- (mcfa);
  \draw[->] (param_init) -- (training);
  \draw[->] (training) -- (latent_factors);
  \draw[->] (latent_factors) -- (latent_scores);
  \draw[->] (new_data) |- (latent_scores);
  \draw[->] (new_data) |- (decision_tree);
  \draw[->] (mcfa) -- (decision_tree);
\end{tikzpicture}
  \caption{Overview of the clustering and classification of the input observations. The MCFA model encompassed
  in the upper dashed rectangle can be computed using the \texttt{mcfa} \texttt{python} package. The classification of the input
data or new observations in the lower dashed rectangle can be done using the \texttt{classy} \texttt{python} package described in \cref{sec:classification}.}
  \label{fig:clustering}
\end{figure*}
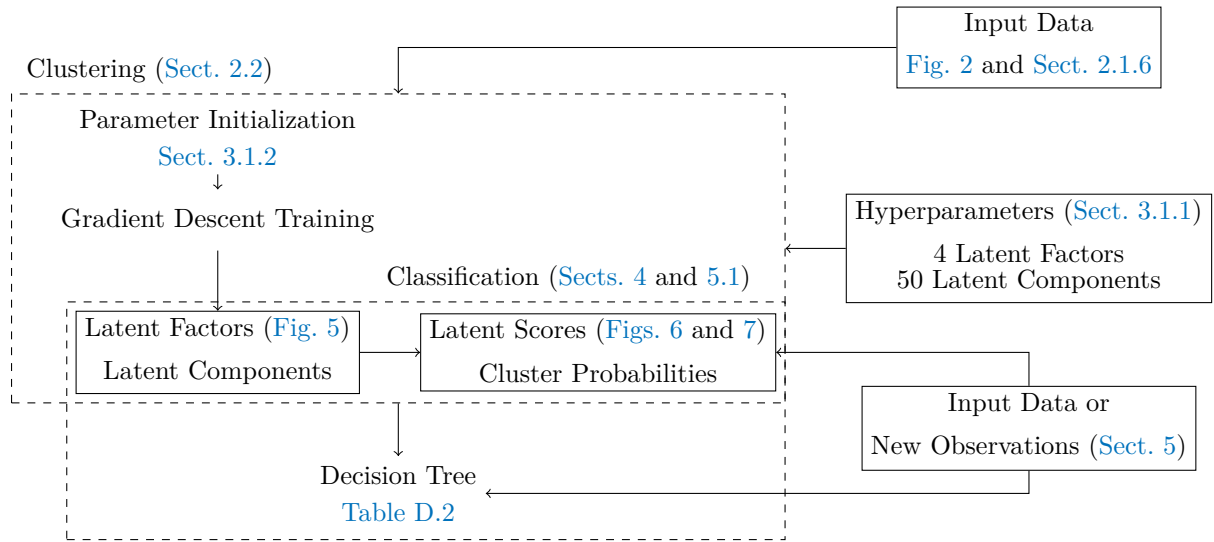

PPCA assumes that the noise matrix $\bm{\epsilon}$ is isotropic (\ie all data
dimensions carry the same noise). This is not necessarily the case for our
observations; the uncertainties between visible and NIR spectra may differ from
one another and  from that of the visual albedo. \abbr{Factor
analysis}{FA} is another latent generative variable model analogous to PPCA
except that the noise matrix $\bm{\epsilon}$ is assumed to be diagonal rather
than isotropic \citep{EmAlgorithmsFRubin1982}. The noise matrix is referred to
as uniqueness as it captures the variance that is unique to each data
dimension, effectively decoupling the measurement uncertainties from the data
covariance.

In FA, the observations $ \bm{Y}$ are modeled as

 \begin{equation}
   \label{eq:factor_analysis} \bm{Y} = \mu + \bm{W}
   \bm{Z} + \bm{\epsilon},
 \end{equation}where $\mu_{p\times 1}$ is a $p$-dimensional vector containing the
 mean values of $ \bm{Y}$ along the feature dimensions; $ \bm{W}$ is the matrix
 of the latent components, as above; and $\bm{\epsilon}$ is a diagonal Gaussian
 noise matrix, $\bm{\epsilon} \sim \mathcal{N}(\bm{0}, {\Psi})$, where $\bm{\Psi}_{p\times p}$ is diagonal.
The latent variables $\bm{Z}$ follow a normal distribution with zero mean and unit
 covariance,  $\bm{Z} \sim \mathcal{N}(\bm{0}, \bm{I})$.
The model parameters can be determined by a maximum-likelihood approach, even in the
case of missing data, under the assumption that the data is missing at random (\ie its probability of  being missing is independent of its value) \citep{StatisticalAnaLittle2019}.

\subsubsection{Clustering}
\label{subsub:clustering}

Using the FA model given in \cref{eq:factor_analysis}, we assume that the
distribution of the latent scores (\ie the asteroid observations mapped into the
latent space) follows a single Gaussian distribution. However, we know a priori from
the previous taxonomic efforts that this is not the case; the \class{C}- and
\class{S}-complexes form separate distributions, and endmember classes such as
\class{A}, \class{K}, and \class{V} follow separate trends in the latent
scores (see   Figure 2 in \citealt{AnExtensionOfDemeo2009}).
Instead of a single Gaussian, we therefore model the data as a mixture of $g$
Gaussian distributions, an approach   referred to as mixture of common factor analysers
\citep[MCFA, ][]{MixturesOfFacBaek2010} in the case where the model components
are fit in the same latent space as is the case here. MCFA can be expressed as
specialisation of the FA model in \cref{eq:factor_analysis} using  \citep{MixturesOfFacBaek2010}

\begin{align}
  \begin{split}
    \mathbf{\mu}_i &= \mathbf{A}\mathbf{ \xi}_i, \\
    \mathbf{\Sigma_i} &= \mathbf{A}\mathbf{\Omega}_i\mathbf{A}^\intercal +
    \mathbf{\epsilon},
  \end{split}
\end{align}

where $i \in (1, \dots, g)$, $\mathbf{A}$ is the common subspace of the mixture
components (\ie the matrix of latent components), $\mathbf{\xi}_i$ is the mean
value of the $i$th mixture components in latent space, and $\mathbf{\Omega}_i$
is its variance. The noise matrix $\mathbf{\epsilon}$ retains its definition as
above, meaning that all mixture components share the same noise.

In MCFA, dimensionality reduction and clustering are achieved concurrently
during the model training. Starting from an initial set of model parameters as
outlined in \cref{sub:initialization_and_training}, at each training epoch
(\ie the optimisation of the log-likelihood of the model against the entire
input dataset), this model searches for the $q$-dimensional latent space and
divides the input
samples into $g$ components, which gives the most likely projection of the input
data assuming that it follows the mixture of $g$ Gaussian distributions in the
reduced space. The hyperparameters in the model are the number $g$ of clusters
and the number $q$ of latent components.

\subsection{Model implementation and availability}
\label{sub:mcfa_model}

Implementations of the MCFA mixture-model approach are available in the
\texttt{R} programming
language\,\footnote{\url{https://github.com/suren-rathnayake/EMMIXmfa}} by
\citet{MixturesOfFacBaek2010} and in the \texttt{python}
language\,\footnote{\url{https://github.com/andycasey/mcfa}} by
\citet{ADataDrivenMCasey2019}. Nevertheless, we chose to write an alternative
implementation in \texttt{python} as the implementation by the latter imputes
the missing data via mean imputation before training the model using an
expectation-maximisation algorithm. Mean imputation is not appropriate for our
dataset as we know that the spectra of different asteroid classes may appear
entirely different in terms of absorption features and slope. Inserting the mean
column value in each empty cell  thus does not represent the missing data
well. Instead, we use the \texttt{tensorflow} library
\citep{tensorflow2015-whitepaper} to implement a stochastic gradient descent
learning strategy which maximises the log-likelihood of the model given the
observed data only, which is statistically sound under the
missing-at-random assumption \citep{StatisticalAnaLittle2019}, contrarily to
using mean imputation. The stochastic gradient descent is of particular interest
here as it estimates the model parameters based on batches of the input data,
meaning that it can scale easily with an increasing number of observations. This
MCFA implementation is independent of the taxonomy itself and may be applied
in different studies. The implementation and documentation are available
online.\footnote{\url{https://github.com/maxmahlke/mcfa}}

\section{Results}
\label{sec:results}

In this section we present the results of fitting the MCFA model outlined in
\cref{sub:mcfa_model} to the input dataset described in
\cref{sub:input_data}. After depicting the latent space and the structure of
the latent dimensions, we explain how the asteroid classes building
this taxonomy are derived from the modelled Gaussian clusters.
An overview of the clustering steps is given in \cref{fig:clustering}.

\subsection{Model fit}
\label{sec:model_fit}

\subsubsection{Parameters}
\label{subsub:parameters}

We choose to cluster the asteroid observations in $q=\num{\NFactors}$ latent dimensions
using $g=\num{\NCluster}$ Gaussian clusters. Both numbers are selected from a wide range
of values after assessing the resulting model fits. Larger values retain
and describe more variability in the data, and  at the same time  increase the
number of free parameters in the model, hence a trade-off is made in both cases. The
model fits obtained with four or five latent factors were comparable in terms
of captured variability in the cluster, thus we opted for the smaller number of
model parameters.

The large initial number of \num{\NCluster} clusters accounts for the model
assumption of Gaussianity in the latent space. We have no reason to expect a
Gaussian distribution of the asteroid classes; therefore, we model them as
superpositions of one or more Gaussian clusters. The modelled clusters are later
joined and mapped to build the asteroid classes using a many-to-many
relationship and following a decision tree.

\subsubsection{Initialisation and training}
\label{sub:initialization_and_training}

The latent loadings and cluster assignments of each observation have to be
initialised at the start of the gradient descent algorithm to train the MCFA
model. The initialisation dictates the global position in the Hamiltonian which
is sampled by the training and thus has a significant impact on the final
result.

A practical issue when reducing the dimensionality of asteroid data made up by
different observables is the feature weighting. In our case the spectra
contribute 53 data dimensions compared to the single dimension of the
albedo. The summed variance in the former is much larger than the variance in
the latter, resulting in a negligible contribution of the albedo to the latent
space computation which does not reflect its actual information
content. \citet{1984PhDT.........3T} therefore chose not to include the albedo
in the dimensionality reduction, using it in a subsequent manual clustering
step instead. We employ an alternative strategy outlined below which allows us to
account for the albedo while building the latent space.

We initialise the latent loadings using PPCA. This approach has
two advantages. First,  the latent loadings are set to the axes of largest variance
in the data, ensuring a high resolution in the latent space, and second,  PPCA is
variant to feature scaling (\ie data dimensions are weighted with respect to their
variance when computing the dimensionality reduction). An effective way to
increase the importance of the albedo information is hence to increase the
variance of albedo values by some transformation prior to model training. We
achieve this by means of the $\log_{10}$ transformation described in
\cref{subsub:albedo}, which increases the variance in the albedo dimension by a
factor of \num[round-mode=places,round-precision=1]{\RatioAlbedoVariance}.
During the gradient-descent model training, we monitor the log-likelihood
of the model given the data. As opposed to PPCA, MCFA is invariant to factor
scaling, which leads to a decrease in the albedo loadings with each training
step. Therefore, we do not train until the model has fully converged, instead
stopping the training when a good balance between the weight of
the albedo and of the spectra has been achieved. This subjective choice of
training epochs is a concession we make to the challenge of combining different
observables in the same model.

The latent cluster memberships are initialised by fitting a Gaussian mixture
model with \num{\NCluster} components to the principal scores of the PPCA and
assigning each sample to its most probable cluster. We train the
MCFA model on the \num{\NSamples} observations of \num{\NAsteroids} individual
asteroids as outlined in \cref{sub:merging}.

\subsection{Latent space}
\label{sub:latent_space}

During the model training the latent components matrix $\bm{W}$ is derived based on the covariance of the input observations.
Each latent component contains one linear coefficient for each input data dimension (\ie the latent loading).
The absolute value of a loading indicates the degree to which the latent
component responds to variance in the corresponding data dimension. Positive
loadings lead to an increase in the latent scores $z$ with increasing value in
the data dimension, negative values to a decrease in $z$. The latent scores $Z$
are essentially a vector product of the input data with the latent components.
As such, both the spectra and the visual albedo of the observations influence the latent scores $Z$
simultaneously.

\begin{figure}[t]
  \centering
  \input{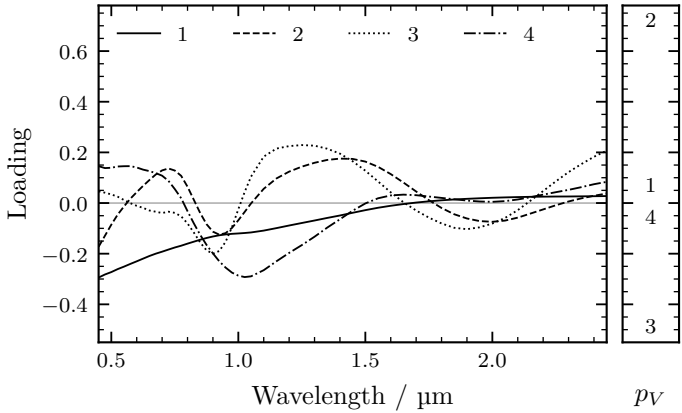}
  \caption{Four latent components of the mixture of common factor
  analysers model trained on the input data. The left   side gives the
  loading of the spectral data dimensions for each latent component, while the right
    side shows the loading corresponding to the albedo.}
  \label{fig:latent_loadings}
\end{figure}

The latent components resulting from the model training are depicted in
\cref{fig:latent_loadings}, with the spectral loadings given on the left
side and the latent loadings corresponding to the albedo dimension on the right  side.
We note that they are displayed separately only for visualisation purposes; for the clustering model itself,
there is no principal distinction between the latent loadings corresponding to the spectra and that corresponding to the albedo.

The spectral loadings in \cref{fig:latent_loadings} resemble different
mineralogical features commonly present in asteroid spectra. The first component
approximates a positive slope,\footnote{The latent loadings represent
the variance of the $\ln$-transformed spectra.} with an inflection point around
\SI{1}{\micro\meter}. Components two and three resemble the spectra
of pyroxene minerals due to their bands at \SI{1}{\micro\meter} and
\SI{2}{\micro\meter}, though the band minima and depths differ between the
components. The strongest distinction between these two components is the
visible slope, which is positive for component two and negative for component
three. Component three has a slight absorption feature at \SI{0.7}{\micro\meter}. The
fourth component depicts an olivine-like \SI{1}{\micro\meter} band
structure. The albedo contributes marginally to the first and fourth latent
component, while component two has a large positive loading and component three
a large negative loading to it.

\begin{figure}[t]
  \centering
  \input{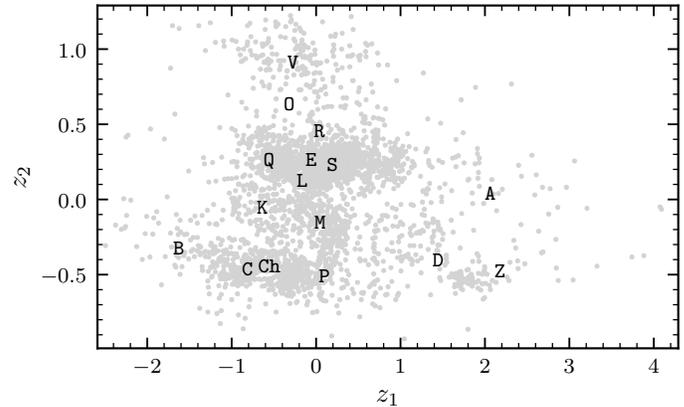}
  \caption{Latent scores of the input data projected along the first two
    latent components (grey circles). The mean score of all asteroids assigned
    to a given class is indicated by the class letter. For better readability, the mean score of class \class{C} has been
  shifted by -0.1 in $z_1$.}
  \label{fig:latent_scores}
\end{figure}

\begin{figure}[t]
  \centering
  \input{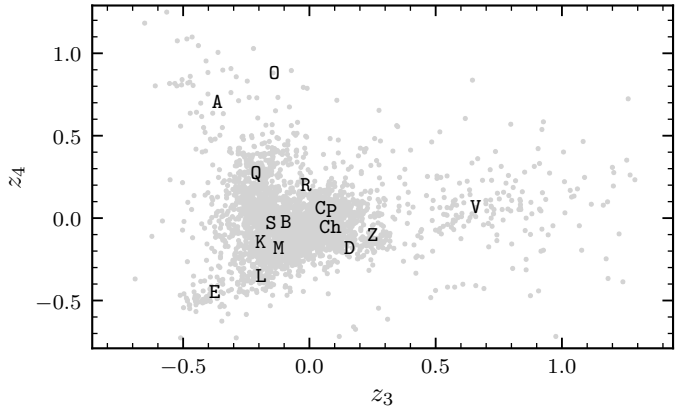}
  \caption{As in \cref{fig:latent_scores}, but giving the scores in the third and
    fourth latent components. For better readability, the mean score of class \class{S} has
    been shifted by -0.02 in $z_3$ and of classes \class{C} and \class{P} by 0.04 in
$z_4$.}
  \label{fig:latent_scores2}
\end{figure}

The latent scores of the asteroid observations are shown in
\cref{fig:latent_scores,fig:latent_scores2}. The input data depicts
a larger variance when projected along the first two components rather than
along the last two due to the initialisation of the latent components with
PPCA. It is clear that the featureless spectra will show little
variance when projected along the pyroxene- and olivine-like axes
$z_2$, $z_3$, and $z_4$.

\Cref{fig:latent_scores,fig:latent_scores2}
additionally indicate the mean latent scores of all asteroids assigned to a
given asteroid class, designated by the class letter and derived in the
following sections. As an example for the interpretation of latent scores, we
point out here that the degeneracy between  classes \class{E} and \class{S}
in the latent scores depicted in \cref{fig:latent_scores} is the result of the
large loading of the albedo in the second component, which offsets the generally
featureless \class{E}-types with respect to the feature-rich but darker
\class{S}-types. This degeneracy is resolved in other latent components, as can
be seen in \cref{fig:latent_scores2}.

\subsection{Clusters}
\label{sub:cluster}

\begin{figure*}[p]
  \centering
  \input{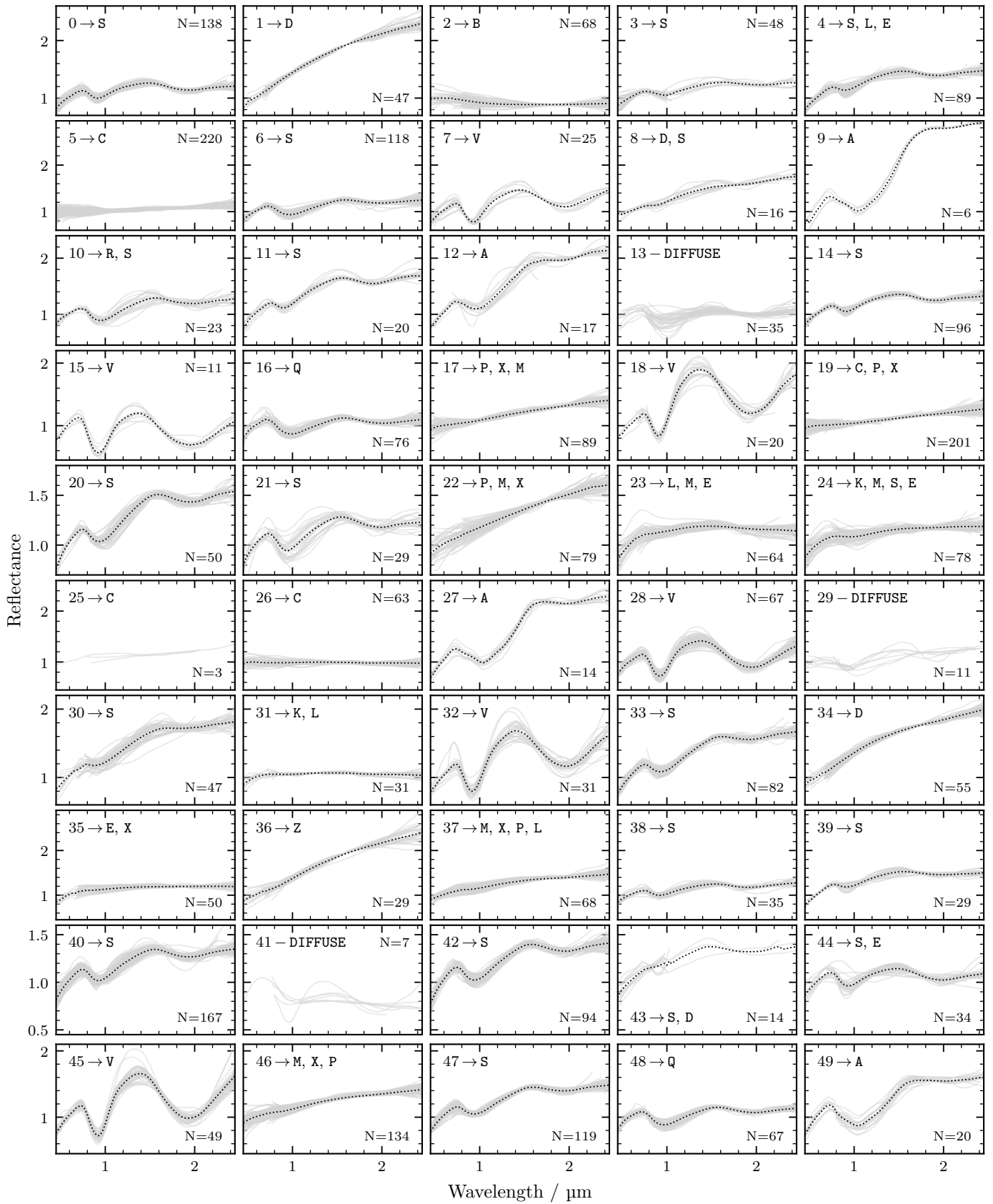}
  \caption{Overview of  asteroid spectra assigned to each cluster, including the number $N$ of
    spectra and the asteroid classes to which the cluster contributes,
    excluding classes with fewer than three contributed observations except for cluster
    25 which only has three members. The classes are sorted by the total
    number of observations the cluster contributed. The dotted line gives the
  mean value of the spectra per cluster except for diffuse clusters (defined in \cref{sub:cluster}) and
    cluster 25. The mean spectra are normalised to unity at
    \SI{0.55}{\micro\meter}. The y-axis limits change in each  row.}
  \label{fig:feature_overview_feature_spectra_which_cluster}
\end{figure*}

Concurrent with the dimensionality reduction, the input data is divided into
\num{\NCluster} Gaussian clusters during the model training.
The clusters are not constrained in their covariances, yielding a wide range of
cluster shapes and orientations in the latent space. Illustrating the distribution of the clusters in the
four-dimensional latent space is not practical due to their large number;
instead, we show the distributions of input spectra and albedos over the
clusters in \cref{fig:feature_overview_feature_spectra_which_cluster,fig:feature_overview_feature_albedo_which_cluster} respectively.

Most clusters occupy a narrow volume in the latent space and encompass Gaussian
populations in previously recognised classes such as \class{S} and \class{V}.
When building the asteroid classes from the clustering, we map the probability
of any sample to belong to either of these narrow clusters one-to-one to the
respective asteroid class. As an example, for all observations the probability
of belonging to cluster 0 is added to the probability $p_\texttt{S}$ of belonging
to the \class{S}-types (see
\cref{fig:feature_overview_feature_spectra_which_cluster}). Additional
\class{S}-like clusters such as cluster 6 further add to $p_\texttt{S}$; \num{\NCoreCluster} clusters are
mapped to a single asteroid class in this manner.

Other clusters either capture continuous trends between classes or the diffuse
background population. An example of the former is cluster 22, containing
spectra from both \class{M}-  and \class{P}-types, and of the latter cluster 13,
containing observations with varying spectral characteristics and albedos. For these
clusters, we implement decision trees to separate the observations into mostly two or
three distinct classes. These decision trees are described in
\cref{sec:discussion} on a per class basis at the end of each class
description. The probability of  belonging to either of these clusters is split and
added to the respective class probabilities following the decision trees. As an
example, cluster 22 is resolved via the albedo. If no albedo is present in the
observation, the cluster probability is added entirely to $p_{\texttt{X}}$,
otherwise it is split between $p_{\texttt{M}}$ and $p_{\texttt{P}}$
proportionally based on the albedo distribution of \class{M}-  and
\class{P}-types, derived in \cref{sub:resolving_the_xcomplex}.

For clusters 13, 29, and 41, we note that they capture objects with high
variability in their spectral and albedo features. These are either unique
objects, such as the only \class{O}-types \nuna{3628}{Boznemcova} and
\nuna{7472}{Kumakiri} in cluster 13, or spectra of questionable quality. We
resolve these clusters with decision trees based on GMMs into
different classes: cluster 13 into \class{C}, \class{O}, \class{Q}; cluster 41
into \class{B} and \class{V}; and cluster 29 into every class except for
\class{E}, \class{K}, \class{L}, \class{O}, \class{R}, \class{X}, and
\class{Z} (see \cref{tab:decision_tree}). Objects in either of the
three clusters are flagged in the classification output as \texttt{DIFFUSE} and
should undergo visual
scrutiny.

\subsection{Classes}
\label{sub:classes}

\subsubsection{Class continuity}
\label{subsub:class_continuity}

When deriving the mapping of the Gaussian clusters to the asteroid classes, we
strive to maximise the resemblance of the resulting taxonomy to the established
system by \citet{1984PhDT.........3T} and the Bus-DeMeo system. For
any change in the classes, we weigh the evidence in the data to
necessitate the change against the overall practicality of class
continuity, opting for the latter when in doubt. Furthermore,
we also take into account mineralogic and meteoritic interpretations established
in the literature using observables outside this feature space, allowing us to
derive classes which are more useful for communicating class properties within
the community. These influences from outside the data-driven approach are
stated in the description of the respective class in \cref{sec:discussion}.

The main drivers for the evolution of the class scheme are twofold. The first is  the
fundamental difference between the probabilistic clustering employed here and
the visual clustering used in previous schemes, affecting specifically classes
that reflect continuous trends in the asteroid population. The  second is  the
reintroduction of the albedo to the observables of the
taxonomy.

The fundamental division of asteroids into feature-poor and feature-rich
populations, the \class{C}- and \class{S}-complexes, is the baseline of our
scheme, as it has been since the first taxonomic efforts by
\citet{SurfacePropertChapma1975}. A small population of asteroids with faint
features occupies the space between these complexes in
\citet{AnExtensionOfDemeo2009}, separated into the classes \class{K}, \class{L},
\class{Xc}, \class{Xe}, \class{Xk}, and \class{T}. Thanks to the taxonomic
information provided by the albedo and targeted campaigns of these populations
\citep[\eg][]{TheCompositionOckert2010,TheCompositionNeeley2014}, this
population has grown considerably, to the point that we recognise it as a third
complex, which we dub the \class{M}-complex based on its most populous
class.

Taxonomic constants such as the \class{A}- and \class{V}-types represent no
challenge in identification. It is more difficult to prove the definition of the
\class{Q}-types, which represent a continuous trend towards smaller slopes
compared to the \class{S}-types  and as such does not separate clearly in the
latent space. In favour of class continuity, we still identify a population in
the \class{S}-complex as \class{Q}-types. Subclasses such as \class{Sa},
\class{Sq}, and \class{Sr} are not identified, however,  as we observe numerous
clusters with varying slopes and mineralogies in the \class{S}-complex. Labelling
 each cluster with a secondary letter would increase the entropy of the taxonomic
system, and would lead to more confusion than resolution. Furthermore, we note
an overall large variability between observations of single asteroids which often
exceeds the variability between these subclasses. Instead, we highlight the
different mineralogical interpretations of these clusters in \cref{sub:stypes}.

\begin{figure}[t]
  \centering
  \input{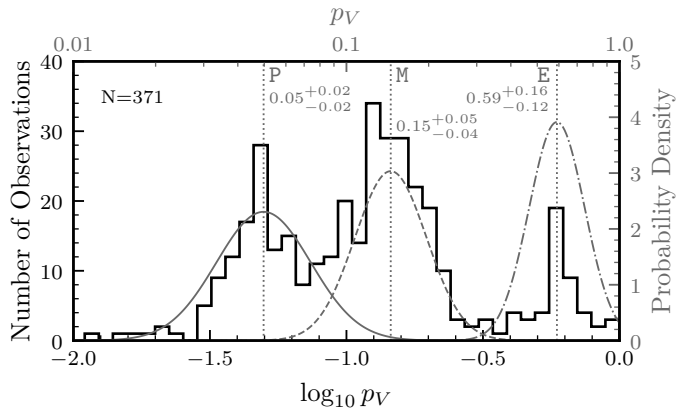}
  \caption{Distribution of  visual albedos in clusters associated with
    the \class{X}-complex. The spectral degeneracy of the
    \class{X}-complex is resolved by fitting a
    three-component Gaussian mixture model to its albedo distribution,
    consisting of $N$ observations and shown in the histogram. The fitted
    components are given by the solid, dashed, and dash-dotted grey lines in terms of the probability
    distribution. The vertical dotted lines give the mean values of
  components, labelled by the established class designations \class{P},
\class{M}, and \class{E} in order of ascending albedo. The numbers below the
class labels give the mean $p_V$ and the upper and lower \SI{1}{\sigma} limits
per class. We note that these values slightly change later as other class
members are added from clusters which are not assigned purely to the
\class{X}-complex. The final albedo distributions are given in
\cref{tab:class_descriptions}.} \label{fig:resolve_emp}
\end{figure}

\subsubsection{Resolving the X-complex}
\label{sub:resolving_the_xcomplex}

Solving the spectral degeneracy of the \class{X}-complex  in the Bus-DeMeo scheme
is the main motivation to reintroduce the albedo to the taxonomic system. We employ the system
established in \citet{1984PhDT.........3T}; asteroids in the
\class{X}-complex are differentiated based on their albedo values and are
labelled \class{P}, \class{M}, and \class{E} in ascending order of albedo, while
the letter \class{X} is retained for observations without albedo. However, instead of
applying strict limits,\footnote{\citet{1984PhDT.........3T} applied visual
albedo separations of $\sim$0.06 between \class{P} and \class{M} and $\sim$0.28
for \class{M} and \class{E}.} we model the joint albedo distribution of all
observations in clusters that  we consider to be \class{X}-like based on their
spectral appearance: clusters 17, 22, 35, 37, and 46. The employed model is a
GMM with three components. In \cref{fig:resolve_emp} we show the model fit to
the albedo distribution of the \class{X}-complex, as well as the derived mean
and standard deviations in $p_V$ for classes \class{E}, \class{M}, and
\class{P}. Any asteroid that falls into one of the these
clusters and has an albedo observation is assigned based on its probability in
this model to the respective class. The subclassification indicating the
presence of features in the spectra (\class{e} and \class{k}) is retained and
discussed in the following subsection.

\subsubsection{Feature flags}
\label{subsub:feature_flags}

The Bus-DeMeo system recognises four classes which are based on the presence of
distinct absorption features in addition to the overall shape of the spectra: (1) \class{Ch}, exhibiting
a feature around \SI{0.7}{\micro\meter} associated with possible surface hydration
\citep[\eg][]{TheChClassAsRivkin2015}; (2) \class{Xe}, showing a narrow feature
at \SI{0.5}{\micro\meter} \citep{2002Icar..158..146B}; (3) \class{Xk}, depicting a  faint broad feature
between 0.8\,-\,\SI{1.0}{\micro\meter} \citep{2002Icar..158..146B}; and (4) \class{Xn}, with a feature
around \SI{0.9}{\micro\meter} \citep{CompositionalDBinzel2019}. Example spectra carrying the \class{e}-,
\class{h}-, and \class{k}-features are shown in \cref{fig:feature_flags}.

The identification and flagging of these features by use of the secondary
letters in the class designation is carried over in this scheme with a slight
modification. First, we do not differentiate between the \class{k}- and the
\class{n}-feature. Both are  centred around \SI{0.9}{\micro\meter} and after
slope removal, we find no appreciable systematic difference between the features
in a sample of spectra previously classified as \class{Xk} or \class{Xn}. We do
not rule out that these features are imprinted by different surface
mineralogies; however, we chose the evidence in the data over class
continuity, we decided to drop the \class{n}-feature, and continued with only the
\class{k}-feature.

Second, we do not reserve unique classes for observations depicting the \class{e}- or
the \class{k}-feature. As discussed in \cref{sub:mcomplex}, both features are
prevalent in members of the \class{X}-complex showing a variety of spectral
slopes and albedos. We judge these two properties to be more important when
deriving classes than the presence of a single feature. Furthermore, we note
that \class{e} and \class{k} are not mutually exclusive; for example, \nuna{2035}{Stearns} depicts both features as shown in
\cref{fig:feature_flags}. We thus decided to flag the presence of these features
by appending the respective letter to the class designation without considering
the resulting combinations such as \class{Mk} or \class{Eek} as proper classes.

On the other hand, the \class{h}-feature is treated consistently with the
Bus-DeMeo system. It is exclusive to the members of the \class{C}-complex and
displays a much narrower, continuous distribution than the other two features,
as shown in \cref{sub:ccomplex}. Any sample depicting the \SI{0.7}{\micro\meter}
band is assigned to the \class{Ch}-class, regardless of the subclass in the
\class{C}-complex that the spectra falls in.

The features are identified in a semi-automated manner. For each feature we
 defined a wavelength interval
around the band centre in which the spectral continuum is removed and the
reflectance is fitted using a polynomial of fourth degree, following
\citet{AqueousAlteratFornas2014}. Both the interval and the expected band centre were
defined heuristically using a training sample of visually identified
features, and are given in \cref{tab:feature_flags}.
Using the polynomial fit, we estimate the band depth with respect to
the continuum, the band centre, and its signal-to-noise ratio. The last is
given by the ratio of the band depth to the reflectance uncertainty, which is
estimated using the residuals of the polynomial fit. The band is considered to
be present if the band centre is within three standard deviations of the
expected position derived from the training sample and the signal-to-noise ratio is
higher than one.

\begin{figure}[t]
  \centering
  \input{gfx/feature_flags.pgf}
  \caption{Example spectra carrying the \class{e}-, \class{h}-, or
    \class{k}-feature which are recognised in this taxonomic system. The mean
    band centres derived from all visually identified features in the spectral
    observations is indicated by the vertical dotted lines (\class{e}:
    \SI[round-mode=places, round-precision=2]{\BCe}{\micro\meter}, \class{h}:
    \SI[round-mode=places, round-precision=2]{\BCh}{\micro\meter}, \class{k}:
    \SI[round-mode=places, round-precision=2]{\BCk}{\micro\meter}). \nuna{2035}{Stearns} exhibits both
    the \class{e}- and the \class{k}-feature. Data from SMASS.\protect\footnotemark}
  \label{fig:feature_flags}
\end{figure}

\begin{table*}[t]
  \centering
  \caption{Distribution of observations and asteroids over taxonomic classes
    and orbital populations.
  The second column gives
    the number of observations assigned to each class, while the third and all
    following columns refer to the number of individual asteroids  assigned
    to the class. DM09 refers to \citet{AnExtensionOfDemeo2009}. The fractions in this column do
not add up to \SI{100}{\percent}, due to the missing \class{T}-class in this
scheme. The orbital classes use the following acronyms: NEA - near-Earth
asteroids; MC - Mars-crosser; H - Hungaria; IMB - inner main belt; MMB - middle
main belt; OMB - outer main belt; Cyb - Cybele; JT - Jovian trojans.
  }
  \label{tab:class_stats}
\begin{tabular}{lrrrrrrrrrrrrr}
        \toprule
        &\multicolumn{2}{c}{}&\multicolumn{2}{c}{Fraction}&\multicolumn{9}{c}{Orbital Class}\\
        Class & Samples & Asteroids & This Work & DM09 & NEA & MC & H & IMB & MMB & OMB & Cyb & Hilda & JT \\
        \midrule
\class{A} & \num{\NASamples} & \num{\NAFinal} & \num[round-mode=places,round-precision=1]{\NAFinalFraction}& \num[round-mode=places,round-precision=1]{\NATypesFractionDeMeo} & \num{2} & \num{3}&\num{2}&\num{7} &\num{10} &\num{8}  & - &-  &-\\
\class{B} & \num{\NBSamples} & \num{\NBFinal} & \num[round-mode=places,round-precision=1]{\NBFinalFraction}& \num[round-mode=places,round-precision=1]{\NBTypesFractionDeMeo} & \num{15} & \num{4}&\num{1}&\num{12} &\num{5} &\num{8}  & - &-  &-\\
\class{C} & \num{\NCSamples} & \num{\NCFinal} & \num[round-mode=places,round-precision=1]{\NCFinalFraction}& \num[round-mode=places,round-precision=1]{\NCTypesFractionDeMeo} & \num{69} & \num{8}&\num{2}&\num{89} &\num{72} &\num{79}  & \num{2} &\num{2}  &\num{5}\\
\class{Ch} & \num{\NChSamples} & \num{\NChFinal} & \num[round-mode=places,round-precision=1]{\NChFinalFraction}& \num[round-mode=places,round-precision=1]{\NChTypesFractionDeMeo} & \num{9} & \num{2}&-&\num{20} &\num{47} &\num{26}  & \num{2} &-  &\num{1}\\
\class{D} & \num{\NDSamples} & \num{\NDFinal} & \num[round-mode=places,round-precision=1]{\NDFinalFraction}& \num[round-mode=places,round-precision=1]{\NDTypesFractionDeMeo} & \num{6} & \num{1}&-&\num{1} &\num{4} &\num{5}  & \num{5} &\num{16}  &\num{44}\\
\class{E} & \num{\NESamples} & \num{\NEFinal} & \num[round-mode=places,round-precision=1]{\NEFinalFraction}& - & \num{7} & \num{4}&\num{27}&\num{4} &\num{3} &\num{1}  & - &-  &-\\
\class{K} & \num{\NKSamples} & \num{\NKFinal} & \num[round-mode=places,round-precision=1]{\NKFinalFraction}& \num[round-mode=places,round-precision=1]{\NKTypesFractionDeMeo} & \num{21} & \num{2}&-&\num{5} &\num{2} &\num{12}  & - &-  &-\\
\class{L} & \num{\NLSamples} & \num{\NLFinal} & \num[round-mode=places,round-precision=1]{\NLFinalFraction}& \num[round-mode=places,round-precision=1]{\NLTypesFractionDeMeo} & \num{20} & \num{4}&\num{3}&\num{4} &\num{22} &\num{3}  & - &-  &\num{2}\\
\class{M} & \num{\NMSamples} & \num{\NMFinal} & \num[round-mode=places,round-precision=1]{\NMFinalFraction}& - & \num{29} & \num{7}&\num{2}&\num{17} &\num{47} &\num{28}  & - &\num{2}  &\num{10}\\
\class{O} & \num{\NOSamples} & \num{\NOFinal} & \num[round-mode=places,round-precision=1]{\NOFinalFraction}& \num[round-mode=places,round-precision=1]{\NOTypesFractionDeMeo} & - & -&-&- &\num{1} &\num{1}  & - &-  &-\\
\class{P} & \num{\NPSamples} & \num{\NPFinal} & \num[round-mode=places,round-precision=1]{\NPFinalFraction}& - & \num{14} & \num{6}&\num{1}&\num{11} &\num{26} &\num{36}  & \num{12} &\num{12}  &\num{17}\\
\class{Q} & \num{\NQSamples} & \num{\NQFinal} & \num[round-mode=places,round-precision=1]{\NQFinalFraction}& \num[round-mode=places,round-precision=1]{\NQTypesFractionDeMeo} & \num{89} & \num{5}&-&\num{7} &\num{4} &\num{2}  & - &-  &-\\
\class{R} & \num{\NRSamples} & \num{\NRFinal} & \num[round-mode=places,round-precision=1]{\NRFinalFraction}& \num[round-mode=places,round-precision=1]{\NRTypesFractionDeMeo} & \num{7} & -&-&\num{2} &- &\num{1}  & - &-  &-\\
\class{S} & \num{\NSSamples} & \num{\NSFinal} & \num[round-mode=places,round-precision=1]{\NSFinalFraction}& \num[round-mode=places,round-precision=1]{\NSTypesFractionDeMeo} & \num{404} & \num{101}&\num{35}&\num{140} &\num{172} &\num{45}  & - &\num{1}  &-\\
\class{V} & \num{\NVSamples} & \num{\NVFinal} & \num[round-mode=places,round-precision=1]{\NVFinalFraction}& \num[round-mode=places,round-precision=1]{\NVTypesFractionDeMeo} & \num{28} & \num{2}&-&\num{104} &\num{4} &\num{4}  & - &-  &-\\
\class{X} & \num{\NXSamples} & \num{\NXFinal} & \num[round-mode=places,round-precision=1]{\NXFinalFraction}& \num[round-mode=places,round-precision=1]{\NXTypesFractionDeMeo} & \num{20} & \num{8}&\num{2}&\num{1} &- &\num{2}  & - &-  &-\\
\class{Z} & \num{\NZSamples} & \num{\NZFinal} & \num[round-mode=places,round-precision=1]{\NZFinalFraction}& - & \num{1} & -&\num{1}&\num{4} &\num{6} &\num{3}  & - &\num{1}  &\num{7}\\
\midrule \\$\Sigma$ & \num{2983} & \num{2125} &  100 &  98.9 & 741 & 157 & 76 & 428 & 425   & 264& 21 & 34 & 86\\
        \bottomrule
\end{tabular}
\end{table*}

The fitting procedure is run automatically to identify the \class{h}-feature in
spectra classified as members of the \class{C}-complex (\class{B}, \class{C},
\class{P}, and the degenerate class \class{X}; see  \cref{sub:ccomplex}) and
the \class{e}- and \class{k}-features for those belonging to the
\class{X}-complex (\class{E}, \class{M}, \class{P}, and \class{X}). In practice,
we find that relying on the automated band identification yields many false
positives given the low threshold of one in the signal-to-noise ratio and the general uncertainty of
the expected wavelength of the band centre. For example, \citet{ReflectanceSpeClouti2018} give band centres
between \SIrange{0.6}{0.75}{\micro\meter} for the \class{h}-feature. Hence, we
recommend a semi-automated approach where the bands are fitted automatically
and the observer visually confirms the quality of the fit and the
  presence or absence of the band. The fitting and confirmation are handled by the
classification tool presented in
\cref{sec:classification}. In the \num{\NSamples} spectra classified during the
clustering, \num{\NumberFeaturee} (\num{\NumberFeatureh}, \num{\NumberFeaturek})
carry the \class{e}-feature (\class{h}-feature, \class{k}-feature). For
\num{\NumberUnknownFeaturee} spectra (\num{\NumberUnknownFeatureh},
\num{\NumberUnknownFeaturek}), no conclusion could be made as the spectral
region is missing.

The \class{k}-feature is particularly challenging to observe as it falls in
the transition of visible and near-infrared spectra, which are acquired using
different instruments. Merging the spectral parts is  non-trivial and several
subjective decisions have to be made, as outlined in
\citet{SpectroscopyOfClark2009}. The unknown offsets between visible and
near-infrared can give rise to an artificial feature when joining the
observations. In the case of the \class{e}-feature,
  \citet{PhaseIiOfTheBusS2002} point out a systematic feature between
  \SI{0.515}{\micro\meter} and \SI{0.535}{\micro\meter} in the SMASS spectra,
  which are frequently used to complement acquired NIR-only spectra. Hence, we
note here that the \class{e}-feature should only be considered
present if its band centre is well below this wavelength range.

\subsubsection{Class per asteroid}
\label{subsub:asteroid_classes}

\footnotetext{\url{https://smass.mit.edu}}

A total of \num{\NMultipleSamples} of  \num{\NAsteroids} asteroids in the input data
have more than one sample in the input data. These observations may or may not have
been assigned to the same class, opening the possibility that asteroids have
different classes assigned. We resolve these ambiguities by computing the sum of
the class probabilities across all observations of the asteroid, weighted by the
fraction of observed data dimensions. Observations with albedo values received
an additional weight corresponding to 25 data dimensions, meaning that a
visible-only spectrum including albedo has approximately as much weight as a
VisNIR spectrum without albedo. If  one of the \class{e}-, \class{h}-, or
\class{k}-features is detected in any of the observations, the final class of the
asteroid carries the respective suffix letter.

In \cref{tab:class_stats}, we report the total number of observations per taxonomic
class, followed by the number of distinct asteroids in the class. The latter
number only includes asteroids which were assigned to the class after the
merging procedure outlined above in the case of multiple observations.

\section{Discussion}
\label{sec:discussion}

In the following, we discuss the main properties of
the \num{\NClasses} classes defined in this taxonomy in data and latent space,
structured into three complexes: \class{C}, \class{M}, and \class{S}.
We give our motivation for class scheme and point out where it aligns with or deviates
from the existing classifications,
in particular the taxonomy by \citet{1984PhDT.........3T} and the Bus-DeMeo
system \citep{2002Icar..158..146B,AnExtensionOfDemeo2009}, which are
the closest predecessors in terms of the observables.
We further outline the decision tree   used to derive the classes from the 50 clusters
that were fit to the input observations in the previous section. An overview of this decision tree is given
in \cref{tab:decision_tree}

A general overview of the class properties in data space is given in
\cref{fig:feature_overview_feature_spectra_which_classes,fig:feature_overview_feature_albedo_which_classes}.
\Cref{tab:class_stats} gives an overview
of the number of samples and asteroids per taxonomical and orbital class.
\Cref{tab:taxonomy_overview,tab:class_descriptions} at the end of this section
show an evolution of the taxonomic scheme and describe the classes defined in
this taxonomy, including an overview of the spectra of class prototype
asteroids, most of which are discussed in the text. The mean spectra and albedos
for each class (`class templates') are available in the CDS repository. The
\class{X}-class is not discussed separately as its members are covered by
classes \class{E}, \class{M}, and \class{P}.

\subsection{\class{C}-complex: \class{B}, \class{C}, \class{Ch}, \class{P}}
\label{sub:ccomplex}

The members of the \class{C}-complex are found throughout the main belt and
dominate the regions past the 3:1 mean-motion resonance in terms of number and
mass \citep{2014Natur.505..629D,DifferentOrigiVernaz2017}. Their spectral
appearance is generally feature-poor apart from the \class{h}-feature at
\SI{0.7}{\micro\meter} observed in about one-third of the population and
associated with phyllosilicates present on the surface
\citep{TheFractionOfRivkin2012}. Instead, the diversity of the complex
constituents is present in the slope and in the shape of the spectra, the former
ranging from blue over neutral to red and the latter from overall linear to a
concave appearance attributed to a carbonaceous surface composition including
magnetite
\citep{SurfacePropertChapma1975,1979aste.book..688G,ReflectanceSpeClouti1990}

Common meteorite linkages to the population of the \class{C}-complex involve
carbonaceous chondrites such as CI, CK, CM, and CO with different degrees of
thermal metamorphism or aqueous alteration
\citep{ThermalMetamorHiroi1996,2010JGRE..115.6005C,Cloutis2011CI,NearInfraredSDeLeo2012}.
However, the paucity of these meteorite groups among the falls even after
bias-correction is difficult to reconcile with the abundance of the complex
members in the main belt, leading \citet{Vernazza_2015} to suggest
\abbr{interplanetary dust particles}{IDPs} as analogues for the non-hydrated
asteroids. Using a radiative transfer model, the spectral
appearance of most \class{C}-complex asteroids is well matched using
constituents of chondritic-porous IDPs. The open question on the surface
composition is decisive for the behaviour of the asteroids under the influence of  spectral
weathering. Laboratory irradiation experiments
\citep{IonIrradiationLantz2017,SpaceWeatherinLantz2018} and statistical
approaches \citep{SpaceWeatherinThomas2021} both show opposite trends for different
initial surface compositions: while high-albedo material exhibits spectral
reddening and surface darkening, low-albedo assemblages become bluer in
slope and brighter.

Apart from the \class{C}-types, \citet{1984PhDT.........3T} defined three
smaller classes based on the albedo and UV distributions: \class{B}-types
are `bright-\class{C}' types with visual albedos around
\SI{10}{\percent}, while \class{F}- and \class{G}-types are characterised by their
behaviour in the UV wavelength region (the former flat, the latter showing
strong absorption behaviour). The Bus-DeMeo system retained  classes \class{B} and
\class{C} and extended the taxonomy by addition of the \class{Ch}-class for hydrated \class{C}-type asteroids, as well as the
classes \class{Cb}, \class{Cg}, and \class{Cgh}, which describe different
slope behaviours in different wavelength regions. Neither system counts the members of
\class{P}-class as members of  \class{C}, but rather as member of the
\class{X}-complex.

\begin{figure}[t]
  \centering
  \input{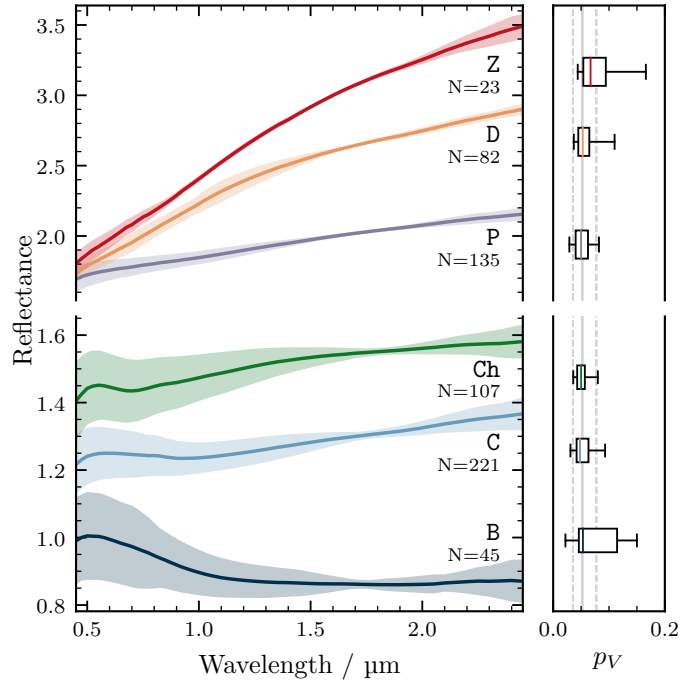}
  \caption{Mean (solid line) and standard deviation (shaded
    area) of the reflectance
    spectra for each class and endmember of the \class{C}-complex on the left
    hand side. The spectra are shifted along the y-axis for
    comparability. The reflectance scale changes between \class{B},
    \class{C}, \class{Ch} and \class{P}, \class{D}, \class{Z}. The number
    $N$ of individual asteroids assigned to each class is given below the class
    letter. On the right  side are given the median (solid
    line), the lower and upper quartiles (box), and the
    \num{5}th and \num{95}th percentiles of the distribution of visual albedos
    within the class. The vertical grey lines give the mean albedo (solid) and
    the upper and lower standard deviation (dashed) within the whole
    complex. These latter values are $\num[round-mode=places,
    round-precision=2]{\pVCCMean}^{+\num[round-mode=places,
    round-precision=2]{\pVCCStdUpper}}_{-\num[round-mode=places,
round-precision=2]{\pVCCStdLower}}$ for the \class{C}-complex.
}
\label{fig:class_spectra}
\end{figure}

\begin{figure*}[t]
  \centering
  \input{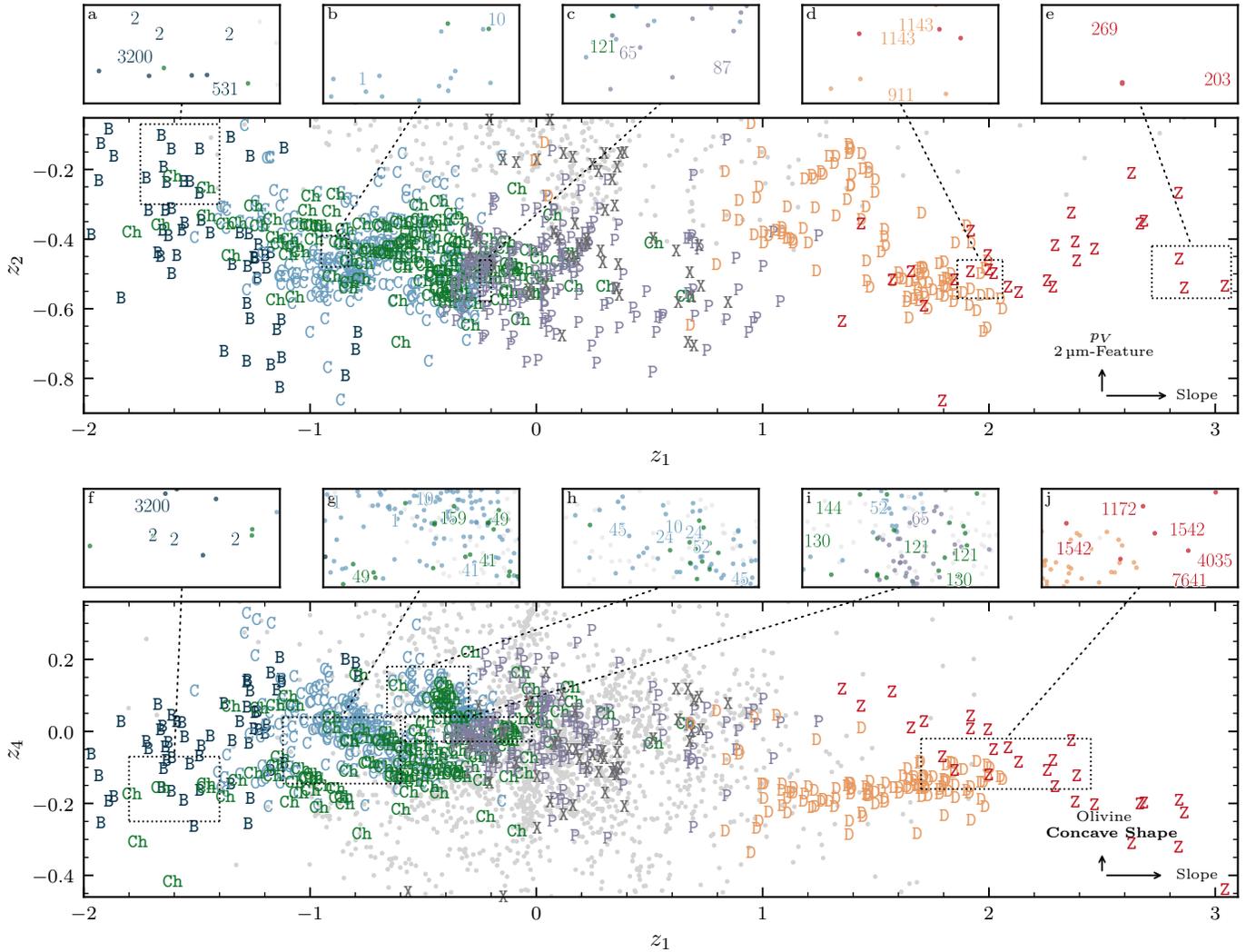}
  \vspace{-1.75em}
  \caption{Distribution of  \class{C}-complex and its endmember classes
    \class{D} and \class{Z} in the first latent component vs the second
    (top) and the fourth (bottom) latent components. The samples assigned to each class are
    given with the respective class letter. The latent scores of all samples
    outside these classes are shown as grey circles. Some outliers in $z_2$ and
    $z_4$ are not shown for readability. The five subpanels above each panel
  show regions of interest where a selection of asteroids are highlighted by
replacing the symbol with the respective asteroid's number. If more than one
spectrum of the asteroid is in the input data, its number may appear several
times.}
\label{fig:ccomplex}
\end{figure*}

In this taxonomy, we divide the \class{C}-complex into four classes: \class{B},
\class{C}, \class{Ch}, \class{P}. The \class{P}-class is here defined for the
first time in both albedo and spectral appearance, allowing us to move it from the
\class{X}-complex and firmly establish it as part of the \class{C}-complex. Any
object within the complex that exhibits the \class{h}-feature is
classified as a \class{Ch}-type, even if it falls in \class{B} or
\class{P}. The distribution of reflectance spectra and visual
albedos for each class is shown in \cref{fig:class_spectra}. The heterogeneous
yet continuous distribution of the \class{C}-complex members in latent space is
illustrated in \cref{fig:ccomplex}. As change in slope and a broad feature
around \SIrange{1}{1.3}{\micro\meter}  are the main differentiators, the
complex members split best in the $z_1$ and $z_4$ latent
dimensions.\footnote{We consider an absorption feature to be
concave, while other works such as \citet{SpaceWeatherinLantz2018} define it as
convex.} We note that the apparent diagonal gaps between the \class{C}- and
\class{Ch}-class members in the lower part of \cref{fig:ccomplex} are an
artefact of the spectral normalisation (see  \cref{subsub:spectra}) and are not
of a physical nature, as shown by the large number of asteroids which have samples
on either side of the gaps.

\subsubsection{\class{B}-types}
\label{subsub:btypes}

The \class{B}-class was first defined in \citet{1984PhDT.........3T} based on
their  average  albedo, which is higher in comparison to the other members of the
\class{C}-complex. With the disappearance of the UV wavelength region from
taxonomy, \class{F}-types are no longer distinguishable from
\class{B}-types, and the distribution of generally high albedos  of the latter
has become a broad distribution with a standard deviation of around
\SI{10}{\percent}, see \cref{fig:class_spectra}. This distribution is further
visible in the large variance in the $z_2$-scores of \class{B}-types
(\cref{fig:ccomplex}).
Instead, the bright \class{C} \citep{1984PhDT.........3T} are best identified
by another common interpretation of the class mnemonic, their blue slope
longwards of $\sim$\SI{0.7}{\micro\meter}, causing a readily apparent distinction
from other classes specifically in the $z_1$ latent score. Nevertheless, the
\class{B}-types do not separate entirely from the neighbouring \class{C}-types
and form a diffuse but continuous branch of the complex, as shown in
\cref{fig:ccomplex}.

The class variance in $z_1$-$z_2$ indicates that bluer \class{B}-types also
tend to be brighter. As shown in subpanel (a) in \cref{fig:ccomplex},
  the archetype \class{B}-type \nuna{2}{Pallas} and  near-Earth
asteroid \nuna{3200}{Phaethon} are among the bluest and brightest class
members. \nuna{531}{Zerlina} is further highlighted as a member of the Pallas collisional
family, for which \citet{DifferencesBetAliLa2016} note a
significantly higher average albedo compared to the remaining \class{B}-types.
In $z_4$ \class{B}-types have higher scores than the  other
\class{C}-complex members, with the  $z_1$ score due to the visible part of
the fourth latent component resembling the \class{B} spectral region (compare
\cref{fig:latent_loadings,fig:class_spectra}).

A total of \num{\NBFinal} asteroids
(\SI[round-mode=places,round-precision=1]{\NBFinalFraction}{\percent}) are
classified as \class{B}-types in this study.
The \class{B}-class is made up of a single cluster (2) and is not subject to any
decision tree.
We note that the Themis-like
\class{B}-types with a neutral-to-reddish slope in the NIR, as
described in \citet{2010JGRE..115.6005C} and \citet{NearInfraredSDeLeo2012}, are
\class{C}-types in this taxonomy, in agreement with their classification in the
Bus-DeMeo system (see subpanel (h) in \cref{fig:ccomplex}).

\subsubsection{\class{C}-types}
\label{subsub:ctypes}

The carbonaceous \class{C}-types present spectra with a neutral to small red
slope and are generally featureless except for a broad feature around
\SI{1.3}{\micro\meter,} which may give the spectrum an overall concave
shape. In the upper part of \cref{fig:ccomplex}, we observe a uniform distribution of the
\class{C}-types in $z_1$-$z_2$ with the class variance aligned with the $z_1$
axis; $z_2$ is not a suitable projection for the \class{C}-types as they are
featureless and present a narrow albedo distribution, as shown in
\cref{fig:class_spectra}. Instead, the concave feature shape is captured in
$z_4$, hence in the lower part of \cref{fig:ccomplex} we observe a more
structured clustering. The positive correlation of $z_1$ and $z_4$ scores among
the \class{C}-types indicates that the spectra on average get more concave as
they get redder. Nevertheless, the wide and continuous distribution around this general trend
prevents us from defining analogues to the classes \class{Cb}, \class{Cg}, and
\class{Cgh} in the Bus-DeMeo system as we aim to refrain from subjectively
partitioning the latent space.

Both \nuna{1}{Ceres} and \nuna{10}{Hygiea} are members of the \class{C}-class (see subpanel (b) in \cref{fig:ccomplex}). In subpanel
(h), we highlight \nuna{24}{Themis}, \nuna{45}{Eugenia}, and
\nuna{52}{Europa}. All these asteroids are well matched by the models composed
of IDP constituents as described in \citep{Vernazza_2015} and have on average
higher $z_4$ scores than the \class{Ch}-class members of comparable slope.

A total of \num{\NCFinal} asteroids
(\SI[round-mode=places,round-precision=1]{\NCFinalFraction}{\percent}) are
classified as \class{C}-types in this study. \class{C}-types are present in
three different clusters (5, 19, 26, where the first two are the two largest
of  the \num{\NCluster} clusters in the model). Cluster 19 contains
both prominent \class{C}-types such as \nuna{45}{Eugenia} and \nuna{52}{Europa}
as well as prominent \class{P}-types such as \nuna{65}{Cybele} and
\nuna{87}{Sylvia}, as shown in subpanel (c) in
\cref{fig:ccomplex}. The cluster resembles the \class{Cb}-class from the
Bus-DeMeo system. We split this cluster into two components
(\class{C} and \class{P}) using a GMM in
$z_1$-$z_4$. While we generally aimed to keep the number of
post-clustering decision trees to a minimum, we make the choice here to  follow the
mineralogical interpretation of the \class{C}-complex given in
\mbox{\citet{Vernazza_2015}} and \citet{2016A&A...586A..15M}, among others, and to increase class continuity for the objects in these clusters.

\subsubsection{\class{Ch}-types}
\label{subsub:chtypes}

Unlike for the other feature flags outlined in \cref{subsub:feature_flags}, we
reserve a unique class for the \SI{0.7}{\micro\meter} \class{h}-feature,
following the convention of the Bus-DeMeo system. We observe the continuous and
narrow distribution of samples carrying this feature similar to the other classes
in the \class{C}-complex. Furthermore, as above for the \class{C}-types, we
recognise the mineralogical and meteoritic interpretation of the
\class{C}-complex members in the literature
\citep[e.g.][]{Cloutis2011CI,Vernazza_2015,2016A&A...586A..15M}.

While degenerate with the distribution of \class{C}-types in $z_1$-$z_2$,
the \class{Ch}-types  generally have lower scores in $z_4$ than the
\class{C}-types, corresponding to linear rather than concave spectra.
In subpanels (g) and (i) of \cref{fig:ccomplex}, we highlight asteroids
 \nuna{41}{Daphne}, \nuna{49}{Pales}, \nuna{121}{Hermione}
\nuna{144}{Vibilia}, and \nuna{159}{Aemilia}, all of which are compatible with
CM chondrite spectra following the interpretation in
\citet{Vernazza_2015}. \nuna{130}{Elektra} is also linked to these
objects based on the density measurements
\citep{DensityOfAsteCarry2012,ExtremeAoObseYang2016,VolumesAndBulHanus2017}.

The \SI{0.7}{\micro\meter} \class{h}-feature has been observed in at least one
observation of \num{\NChFinal} asteroids
(\SI[round-mode=places,round-precision=1]{\NChFinalFraction}{\percent}). Members
of the \class{Ch}-class are found in clusters 2, 5, 17, 19, and 26. The
assignment requires the identification of the \SI{0.7}{\micro\meter}
\class{h}-feature. Within the \class{C}-complex only,
\SI[round-mode=places,round-precision=1]{\NFracChC}{\percent} of samples
present the \class{h}-feature. The actual number is likely higher as
\SI[round-mode=places,round-precision=1]{\RatioUnknownFeatureh}{\percent} of
samples in the \class{C}-complex are missing the visible wavelength
range, for example  a NIR-only spectrum of \nuna{41}{Daphne} indicated as
\class{C}-type in subpanel (g) of \cref{fig:ccomplex}.

\subsubsection{\class{P}-types}
\label{sub:ptypes}

The \class{P}-types have been absent from the taxonomic schemes since
\citet{2002Icar..158..146B}, and thus no definition of the
VisNIR behaviour exists. As part of the \class{X}-complex, the
`pseudo-\class{M}' types \citep{1982Sci...216.1405G}  are spectrally degenerate
to the \class{E}- and \class{M}-types in the visible wavelength
  range, specifically, the ECAS colours. In the NIR, \class{P}-types show a red
linear slope (see  \cref{fig:class_spectra}). We find that the spectral degeneracy
  between \class{P} and \class{M} continues in the NIR, while \class{E}-types
differentiate by showing overall neutral slopes. Classes \class{P} and \class{M} have to
be distinguished by visual albedo observations, which is about
$\SI[round-mode=places, round-precision=2]{5}{\percent}$ for \class{P}-types.

\begin{figure}[t]
  \centering
  \input{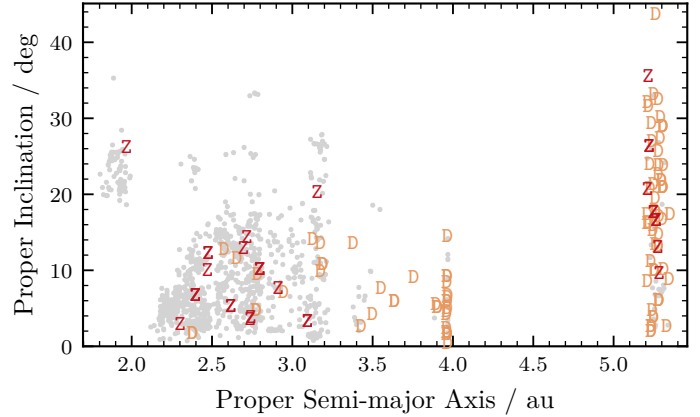}
  \vspace{-0.75em}
  \caption{Orbital distribution of
    \class{D}- and \class{Z}-types  given by the respective class letters. The
  grey dots show the orbital elements of all other asteroids in the input data.}
  \label{fig:orbital_parameters_classes_D_Z}
\end{figure}

As the \class{X}-complex is dissolved in this taxonomy, we assign the class to
the \class{C}-complex following the proximity to the other classes in
\cref{fig:ccomplex}. This assignment is also in line with the IDP
interpretation \citep{Vernazza_2015,2016A&A...586A..15M}. In $z_1$-$z_4$ space
we observe a high-density cluster of \class{P}-types immediately adjacent to
\class{C}-types. These samples are spectrally similar to the \class{Cb}-class
in the Bus-DeMeo system. Furthermore, there  is a more diffuse population of
\class{P}-types building a bridge between the \class{C}-complex and the
\class{D}-class.

The \class{P}-class is part  of the former \class{X}-complex and
of  the new \class{C}-complex. Observations assigned to the \class{P}-class are
thus inspected for all three features.
While \SI[round-mode=places,round-precision=1]{\NFracPh}{\percent} of samples in
the \class{P}-class present the \class{h}-feature, we note that no sample
carries the \class{k}-feature, which is most prominent in the \class{M}-class.
Three samples assigned to \class{P} show the \class{e}-feature, yet they belong
to asteroids which are later assigned to the
\class{M}-class: \nuna{4660}{Nereus} and \nuna{5645}{1990 SP}. The
\class{k}-feature may thus be a reliable differentiator between the spectrally
degenerate \class{M} and \class{P}. The distribution of these features is
 discussed further in \cref{sub:mcomplex}.

A total of \num{\NPFinal} asteroids
(\SI[round-mode=places,round-precision=1]{\NPFinalFraction}{\percent}) are
classified as \class{P}-types in this study. Class
\class{P} is built primarily from clusters 17, 19, and 22, where cluster 19
entails the continuous transition to class \class{C} and the first and third
 \class{M}-types. As mentioned above, we used the prototypes \nuna{65}{Cybele}
 and \nuna{87}{Sylvia} to differentiate between the classes, though assigning
 both to the \class{C}-types would have been justified as well given the cluster
 trend depicted in \cref{fig:ccomplex} (see subpanel (c)).

 \subsection{Endmembers: \class{D}, \class{Z}}
\label{sub:endmembers_d_z}

We refer to \class{D} and \class{Z} as endmembers, due to the visible gap
between their members and the \class{C}-complex in the latent space in
\cref{fig:ccomplex}; however, some of the \class{P}-types form a bridge population
between the two classes.

\subsubsection{\class{D}-types}
\label{subsub:dtypes}

The defining property of dark \class{D}-type asteroids is their featureless and
strongly red-sloped spectrum both in the visible and in the
NIR \citep{1984PhDT.........3T,AnExtensionOfDemeo2009}. They are
predominantly found beyond the outer main belt, especially among the Jupiter
trojan population, where they dominate the region in terms of mass
\citep{TheTaxonomicDDemeo2013,2014Natur.505..629D}.

\class{D}-types form a homogeneous population  in spectral  and in albedo
space, as shown in \cref{fig:class_spectra}. This homogeneity is mirrored in the
latent scores $z_1$ and $z_4$ as well (see \cref{fig:ccomplex}), where in
subpanel (d) we show the positions of \nuna{911}{Agamemnon} and
\nuna{1143}{Odysseus}. In the second latent score, the \class{D}-types appear to
split into a blue and a red population. We attribute this again to the
normalisation of the spectra, which can cause these spurious offsets. Comparing
the samples in the clusters showed no significant difference in the
observables, and   \nuna{2246}{Bowell} and \nuna{2674}{Pandarus} are
 present in the two clusters. Nevertheless, this serves as an example that the
normalisation algorithm we devised for the partial observations may require
further improvement. Furthermore, all clusters in latent space have to be
verified by comparing the members in the observed features.

A total of \num{\NDFinal} asteroids
(\SI[round-mode=places,round-precision=1]{\NDFinalFraction}{\percent}) are
classified as \class{D}-types in this study. \class{D}-types appear predominantly in two
clusters, the homogeneous main cluster 1 and a more diffuse cluster 34, which may
contain interlopers of classes \class{P} and \class{M}. Furthermore, there are two
small clusters containing both \class{D}- and \class{S}-types. Cluster 8 has 16 VisNIR
spectra of \class{D}-types and strongly-sloped \class{S}-types, which are
separated using a two-component GMM in $z_2$-$z_4$, where the feature-rich \class{S}-types
have higher scores in $z_2$. Cluster 43 contains 14 spectra, which are mainly
visible-only \class{S}-types but include five visible-only \class{D}-types, which we
separate in the same way as in cluster 8.

\subsubsection{\class{Z}-types}
\label{subsub:ztypes}

The clustering revealed a low-number diffuse cluster of featureless extremely
red objects, showing larger slopes than the \class{D}-types. \Cref{fig:ccomplex}
shows that in $z_1$ these objects form a continuum with the
\class{D}-types; however, the classes show different variances in the
$z_1$-$z_4$ space: unlike \class{D}-types, \class{Z}-types show a clear trend
towards a more convex shape with increasing slope.
In addition, the classes show distinct orbital distributions, as illustrated in
\cref{fig:orbital_parameters_classes_D_Z}. While \class{D}-types are mostly
situated among the Jupiter trojan  population and the Hildas, these extremely red
objects are largely scattered over the main belt.
Three members of this population, \nuna{3283}{Skorina}, \nuna{15112}{Arlenewolfe}, and \nuna{17906}{1999
 FG32}, have previously been recognised in SDSS observations
\citep[\eg][]{SdssBasedTaxoCarvan2010} and described in a follow-up study by
\citet{UnexpectedDTyDemeo2014}, who further identified \nuna{908}{Buda} as
a similar object.

The distinct distributions  in latent and in orbital space prompt us
to define a new class for this group of minor bodies. We propose the letter \class{Z},
which had previously been suggested by \citet{ExtraordinaryCMuelle1992} for the
extremely red Centaur \nuna{5145}{Pholus}.
The \num{\NZFinal} asteroids in the \class{Z}-class show overall low
albedos, though we note the presence of outliers in \cref{fig:class_spectra}.

The two reddest objects in this new class, \nuna{203}{Pompeja} and
\nuna{269}{Justitia}, have been proposed as implanted trans-Neptunian
objects by \citet{Hasegawa_2021}. The authors suggest that complex
organic material on the surface of these objects leads to the extremely red
appearance. The prevalence of \class{Z}-types in the inner and middle main belt
orbits of the objects could also indicate that  a surface process such as spectral
weathering is  responsible.

A total of \num{\NZFinal} asteroids
(\SI[round-mode=places,round-precision=1]{\NZFinalFraction}{\percent}) are
classified as \class{Z}-types in this study. They fall exclusively into cluster
36. Even so, we expect a certain number of \class{D}-type interlopers in this class
as we observe an overlap in the latent space (see  \cref{fig:ccomplex}) and
in subpanel (j), where we have highlighted the Trojan asteroids
\nuna{1172}{Aneas}, \nuna{1542}{Schalen}, \nuna{4035}{Thestor}, and
\nuna{7641}{Cteatus}, which spectrally match \class{D}-types.

\subsection{\class{M}-complex: \class{K}, \class{L}, \class{M}}
\label{sub:mcomplex}

The \class{M}-complex comprises classes that  fall  in terms
of spectra and  albedo between the \class{C}- and the \class{S}-complex. Compositionally, it
is the most diverse complex. For \class{C} and \class{S} the ensemble
properties can be regarded as carbonaceous, primitive for the former and
silicaceous, in part thermally metamorphosed for the latter
\citep{MetalSilicateClouti1990,Cloutis2011CI,MultipleAndFaVernaz2014}, while the likely
mineralogical properties of any \class{M}-complex member cannot be given
based solely on its complex membership. Meteorite analogues range from most
carbonaceous chondrite clans in the meteorite collection to stony-iron and
iron meteorites
\citep{AncientAsteroiSunshi2008,SpectroscopyOfClark2009,TheCompositionOckert2010,ARadarSurveyShepar2010,ESCHRIG2021114034}.
Indeed, the only unifying property of these objects appears to be the spectral
appearance with absent or generally faint features around \SI{0.9}{\micro\meter}
or \SI{1.9}{\micro\meter} and an albedo around \SI{15}{\percent} with the
exception of the endmember class \class{E}.

Devising the cluster-to-class decision tree proved challenging in this
complex. In combination with the faint features, we observe slight variations in
the slope in the NIR, and class degeneracies appear when the visible
information is missing. Furthermore, we cannot rely as much on previously
established terminology as this is a new complex in terms of taxonomic
systems, replacing the \class{X}-complex as a third complex in previous
taxonomic systems. Both the \class{K}- and the \class{L}-types are more recent
than the \citet{1984PhDT.........3T} taxonomy, which introduced the
\class{X}-complex \citep{1988Metic..23..256B,2002Icar..158..146B}. The Bus-DeMeo
system captures the diversity in the NIR in part in the  form of the
\class{X}- and \class{Xk} classes; however, no clear separation between the
\class{X}- and the \class{C}-complex is achieved due to the  lack of   albedo
information.

We split the complex into the three classes \class{K}, \class{L}, and \class{M},
shown in \cref{fig:class_spectra_mcomplex}.   Class \class{M}, in particular,
contains a wide distribution of spectral appearances and likely mineralogical
compositions. Nevertheless, we opt against a division of this class as no clear
separation presents itself in this study, and we advocate for
a division based on observables not included in this taxonomy.
The \class{T}-class, which was tentatively introduced by
\citet{1984PhDT.........3T} and carried over in the Bus-DeMeo taxonomy, is
dropped as prototypes \nuna{114}{Kassandra} and \nuna{308}{Polyxo} are well
described by  classes \class{P} and \class{M}.

\subsubsection{\class{K}-types}
\label{subsub:ktypes}

\begin{figure}[t]
  \centering
  \input{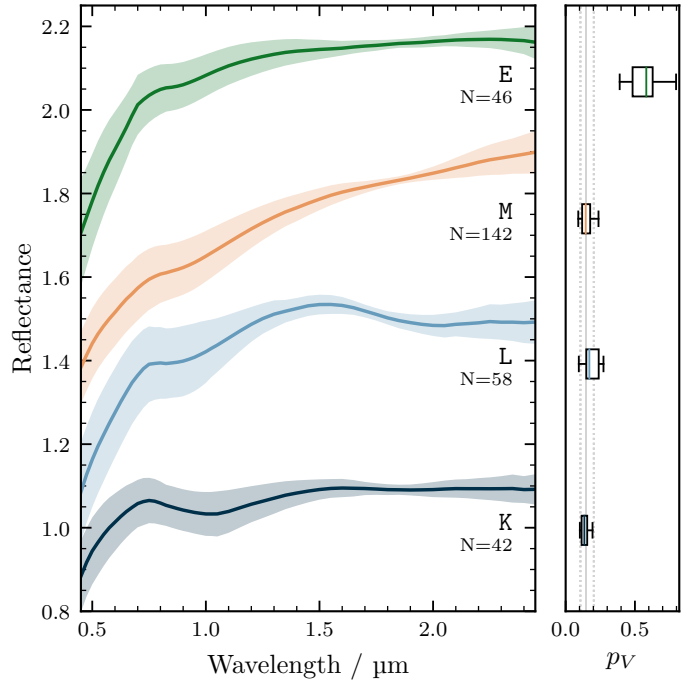}
  \vspace{-0.75em}
  \caption{As in \cref{fig:class_spectra}, but for  the data space properties of the
  \class{M}-complex. The \class{E}-class was excluded in the computation of the
albedo distribution of the complex, indicated by the dotted linestyle of the
upper and lower standard deviation. The albedo distribution of the
\class{M}-complex excluding the \class{E}-types is $\num[round-mode=places,
round-precision=2]{\pVCMMean}^{+\num[round-mode=places,
round-precision=2]{\pVCMStdUpper}}_{-\num[round-mode=places,
round-precision=2]{\pVCMStdLower}}$.}
  \label{fig:class_spectra_mcomplex}
\end{figure}

\begin{figure*}[t]
  \centering
  \input{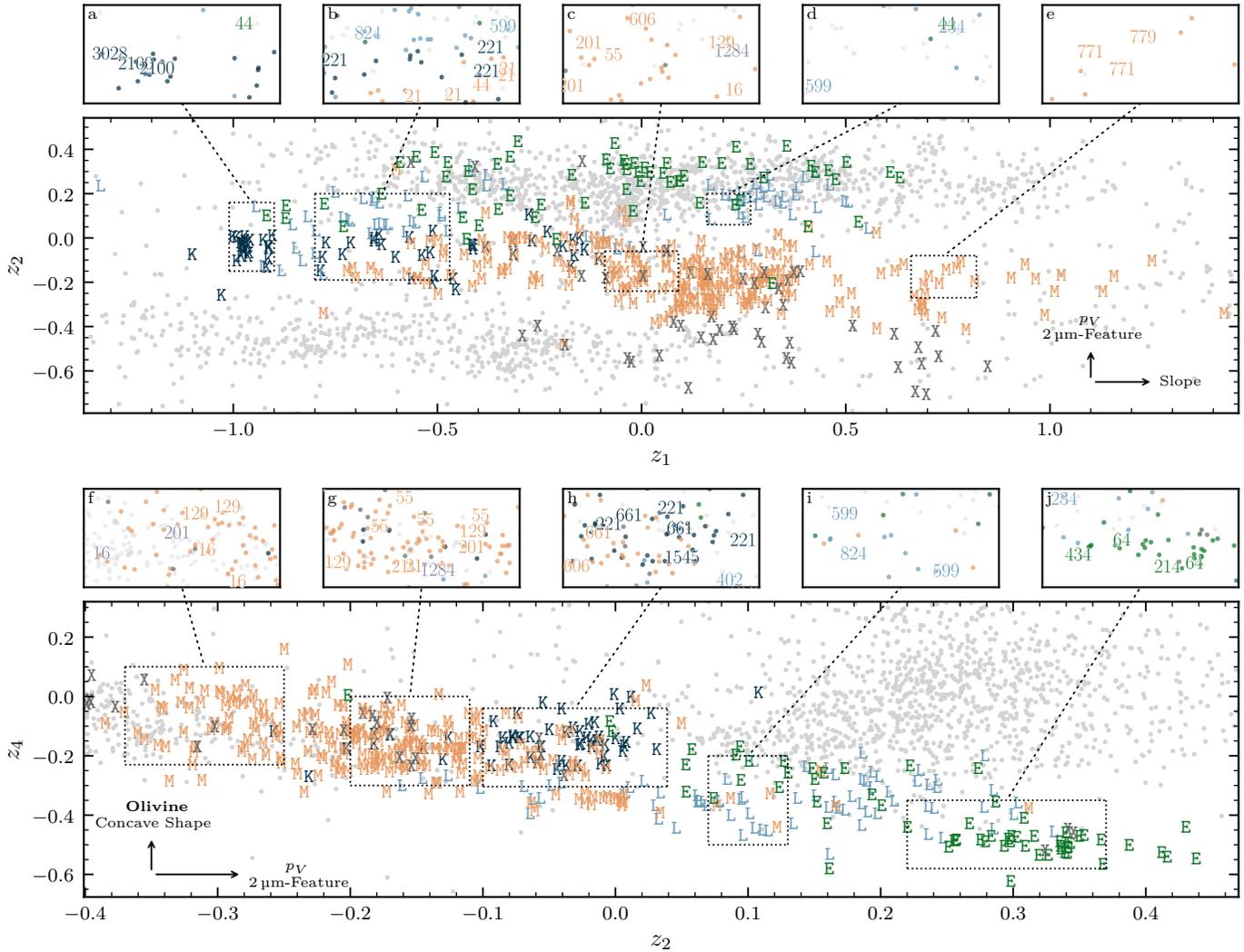}
  \vspace{-0.75em}
  \caption{As in \cref{fig:ccomplex}, but for the member classes of the
  \class{M}-complex and its endmember class, the \class{E}-types.}
  \label{fig:highlight_classes_klm}
\end{figure*}

Members of the \class{K}-class exhibit a red slope in the visible region with a
\SI{1}{\micro\meter} band associated with forsteritic olivine
\citep{MineralogicalAMothe2008} and a neutral slope in the NIR. They
have low $z_1$ and high $z_4$ scores in comparison with the complex
companion classes (see \cref{fig:class_spectra_mcomplex,fig:highlight_classes_klm}).
Most \class{K}-types have visual albedos in the range \SIrange{10}{15}{\percent}, a
narrow distribution which is comparable to the \class{M}-types and slightly
lower than the \class{L}-types.

Dynamically, most main belt \class{K}-types are associated with the
Eos family and depict on average a deeper \SI{1}{\micro\meter} band than
\class{K}-types outside the family based on the $z_4$ score
\citep{SpectroscopyOfClark2009}, compare for example  \nuna{402}{Chloe} and
\nuna{1545}{Thernoe} to \nuna{221}{Eos} and \nuna{661}{Cloelia} in subpanel
(h) in \cref{fig:highlight_classes_klm}.

The class-averaged slope is neutral to slightly red in the NIR. However, some
members, including the class archetype \nuna{221}{Eos} and
\nuna{3028}{Zhangguoxi}, have a blue NIR slope, indicated by their low
$z_1$ scores in subpanels (a) and (b) of
  \cref{fig:highlight_classes_klm}. As the NIR spectrum is featureless above
  $\sim$\SI{1}{\micro\meter}, this leads to a spectral degeneracy with the
  \class{B}-types, and the brighter part of the \class{B}-population
  requires the visible wavelength range information to be separated from the
  \class{K}-class. In subpanel (a) of \cref{fig:highlight_classes_klm},
  we see that \nuna{2100}{Ra-Shalom} is classified as a \class{K}-type, based on
  two NIR spectra. The only VisNIR sample of \nuna{2100}{Ra-Shalom} in this
  study is classified as a \class{B}-type. We note that \nuna{2100}{Ra-Shalom}
  is classified both as \class{B} and as \class{K} in the literature, based
on its VisNIR spectrum (\class{B}:
\citealt{NearInfraredSDeLeo2012,CompositionalDBinzel2019} and  \class{K}:
\citealt{MultiWavelengtShepar2008}).
 The same degeneracy has been reported for
  \class{B}- and \class{K}-types  in NIR spectra
  \citep{SpectroscopyOfClark2009} and in the colour-space of the VISTA MOVIS
  survey \citep{TaxonomicClassPopesc2018}.

The distribution in $z_1$-$z_2$ shows a considerable overlap between M and K, with a
slight gap between the populations around $z_1=0.3$. We considered
whether the  redder \class{K}-types may be \class{Mk} instead; however,
among them are
Eos family members such as \nuna{579}{Sidonia} and \nuna{653}{Berenike}, and thus  we consider this slope variability to indicate \class{K}-types. The overlap is
further resolved in $z_3$-$z_4$, where the \class{K}-class forms a denser
population than the sparsely distributed \class{Mk}-types (not shown).

A total of \num{\NKFinal} asteroids
(\SI[round-mode=places,round-precision=1]{\NKFinalFraction}{\percent}) are
classified as \class{K}-types in this study. \class{K}-types are found in two
clusters, neither of which they populate entirely on their own. Cluster 24 is
shared with \class{M}-types with neutral NIR slopes, while cluster 31
contains NIR-only observations of \class{B}-types as well as
\class{L}-types. We resolve   cluster 24 into \class{K}- and
\class{M}-types using a two-component GMM fit to the cluster
distribution in $z_2$-$z_3$, where \class{K}-types separate due to the large
\SI{1}{\micro\meter} band. Cluster 31 is only split into \class{K} and \class{L}
members as the degeneracy with NIR observations of \class{B} cannot
be resolved with the observables in this taxonomy. The cluster members are assigned
based on their probability of belonging to cluster 23 (\class{L}) or 24 (\class{K})
in $z_2$-$z_3$.

\subsubsection{\class{L}-types}
\label{subsub:ltypes}

\class{L}-type asteroids are associated with large abundances of spinel-bearing
calcium--aluminium-rich inclusions due to a wide absorption feature around
\SI{2}{\micro\meter} \citep{AncientAsteroiSunshi2008}. This composition would
imply that the \class{L}-type parent bodies were among the first planetesimals
to form in the accretion disk,
making them of high interest for formation scenario studies
\mbox{\citep{NewPolarimetriDevoge2018}}. However, in addition to
the \SI{2}{\micro\meter} feature, \class{L}-types are spectrally heterogeneous
in slope and shape of the visible and \SI{1}{\micro\meter} region and  in
their albedo distribution, shown in \mbox{\cref{fig:class_spectra_mcomplex}}. The
diversity of \class{L}-types makes it difficult to reliably identify them in a
taxonomy based on spectral features and opens up degeneracies with a handful of
neighbouring classes, such as \class{K}, \class{M}, and \class{S}.

We find that many previously classified \class{L}-types cluster in
dimensions $z_2$-$z_4$, where they branch off of the \class{M}-complex below the
\class{S}-complex together with the \class{E}-types (see
\cref{fig:highlight_classes_klm}). The second latent component matches the
spinel-associated \SI{2}{\micro\meter} band best, giving \class{L}-types higher
$z_2$ scores compared to the other classes in the complex, while compared to the
\class{S}-types the \SI{0.9}{\micro\meter} contribution to the $z_4$ score is
missing.

In \cref{fig:highlight_classes_klm} we see that the \class{L}-types identified
in $z_2$-$z_4$ exhibit a bimodality in terms of their slope in $z_1$, further
shown in the spectral domain in
\cref{fig:feature_overview_feature_spectra_which_classes}. This dichotomy is not
caused by the normalisation of the spectra. Instead, we find that previously
classified \class{L}-types with intermediate slope such as \nuna{606}{Brangane}
are classified either as \class{M} or \class{S} as they lack the
\SI{2.0}{\micro\meter} feature (see subpanels (c) and (h) in \cref{fig:highlight_classes_klm}). We regard the slope variability of the
\class{L}-types classified here as intrinsic to the class, supported by
\nuna{599}{Luisa}, which has both a blue and a red spectrum (see subpanels
(b), (d), and (i) in \cref{fig:highlight_classes_klm}).

Of particular interest among the \class{L}-types are  the subgroup members
referred to as Barbarians after \nuna{234}{Barbara}, which show anomalously high
inversion angles in their negative polarisation branch
\citep{ASuccessfulSeCellin2014,NewPolarimetriDevoge2018}. We find that this
group of asteroids has a large variance in latent space. In $z_1$-$z_2$,
Barbarians such as \nuna{234}{Barbara}, \nuna{824}{Anastasia},
\nuna{599}{Luisa}, and \nuna{606}{Brangane} and \nuna{1284}{Latvia} (which  are
classified as \class{M}) are found in both the \class{M}- and
\class{S}-complexes and at the transition region (see subpanels (b)--(d) and (g)--(j)) in \cref{fig:highlight_classes_klm}). We also do not
find a reliable clustering in the remaining latent scores. The spectral
\class{L}-types do not include all Barbarians, among which we observe a diversity that is too large
 to derive a unique class in this taxonomy. Of the \num{\NBarbarians}
Barbarians from \citet{NewPolarimetriDevoge2018}, \num{\NLBarbarians} are
\class{L}-types and \num{\NMBarbarians} are \class{M}-types. An extension of the
taxonomy observables with polarimetric observations is required to reliably
identify Barbarians.

A total of \num{\NLFinal} asteroids
(\SI[round-mode=places,round-precision=1]{\NLFinalFraction}{\percent}) are
classified as \class{L}-types in this study. \class{L}-types occur predominantly
in clusters 4 and 23. As for the \class{K}-class, these two clusters are populated by
members from other classes as well. For cluster 4 we split the  \class{L}- and
\class{S}-types based on a two-component GMM in $z_3$-$z_4$
trained on the distribution of the members of cluster 23 and cluster 40 in this
space. For cluster 23, we split the  \class{L}- and \class{M}-types based on a
two-component GMM in $z_1$-$z_4$. A small fraction of \class{L}-types
are also in cluster 37, which consists largely of \class{M}-types. The \class{L}-types are recovered
using a two-component GMM in $z_2$-$z_4$.

\subsubsection{\class{M}-types}
\label{subsub:mtypes}

The \class{M}-class is one of  the oldest asteroid designations \citep{1976AJ.....81..262Z}. Originally introduced to describe asteroids representing presumably
metallic cores of disrupted planetesimals \citep{1979aste.book..688G,1989aste.conf..921B},
dedicated observational efforts have revealed a variety of objects based on their densities \citep{DensityOfAsteCarry2012,VltSphereImagVernaz2021}, hydration
\citep{3MmSpectrophoRivkin1995,TheNatureOfMRivkin2000}, radar albedos \citep{ARadarSurveyShepar2010,ARadarSurveyShepar2015}, and silicate spectral features
\citep{SpectroscopyOfClark2004,TheCompositionOckert2010,TheCompositionNeeley2014,SpectroscopicSFornas2010}.

In spectral space, \class{M}-types asteroids are red with either linear or
convex shapes, as shown in \cref{fig:class_spectra_mcomplex}. The convex trend
may even result in an overall blue slope longwards of \SI{1.5}{\micro\meter}, as
is the case for \nuna{21}{Lutetia} in four out of five observations in this
study. \class{M}-types in the lower $z_1$ region around \nuna{21}{Lutetia},
highlighted in subpanels (b) and  (g) in \cref{fig:highlight_classes_klm}, closely
resemble the \class{Xc}-class in the Bus-DeMeo system. At the other end of the
class in $z_1$, asteroids like \nuna{771}{Libera} and \nuna{779}{Nina} are
examples of red, linear slopes in the NIR, shown in subpanel (e) in
\cref{fig:highlight_classes_klm}. \class{M}-types have an albedo distribution of
\SIrange{10}{20}{\percent}. We note that \nuna{55}{Pandora} has an albedo of 0.34,
and one of its samples is classified as \class{E}, visible in the upper part of
\cref{fig:highlight_classes_klm}, around $(z_1, z_2) = (0.3, -0.2)$.

\begin{figure}[t]
  \centering
  \input{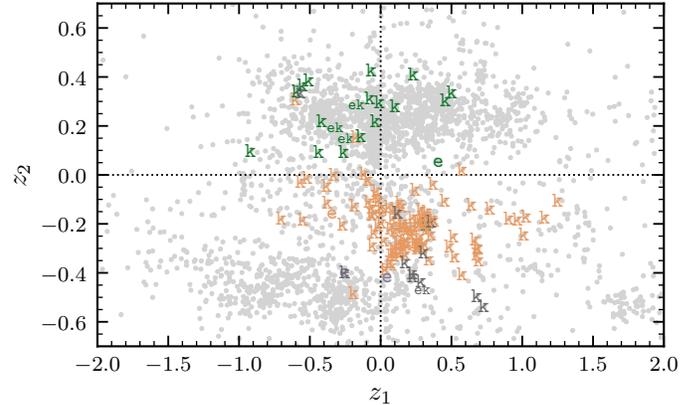}
  \vspace{-0.75em}
  \caption{Distribution of observations which carry the \class{e}-
    and \class{k}-feature in the first two latent scores,
      colour-coded by the class they are assigned to:
  green - \class{E}, orange - \class{M}, purple - \class{P}, grey -
\class{X}. A smaller font size is used if the observation carries both \class{e}
and \class{k}.}
  \label{fig:feature_distribution}
\end{figure}

Silicate features at \SI{0.9}{\micro\meter} or \SI{1.9}{\micro\meter} are
likely more common than an entirely featureless spectrum among \class{M}-types,
with \SI[round-mode=places,round-precision=1]{\RatioMk}{\percent} of
\class{M}-type samples exhibiting the
\class{k}-feature. Of the samples, \SI[round-mode=places,round-precision=1]{\RatioMknan}{\percent}
 lack the
corresponding wavelength region observed. In \cref{fig:feature_distribution}, we
display the first two latent scores of samples with the \class{e}- and
\class{k}-feature. The latter feature is ubiquitous among \class{M}-types,
and a concentration in latent space around \nuna{16}{Psyche} is
visible. \nuna{55}{Pandora}, \nuna{129}{Antigone}, and
  \nuna{201}{Penelope} further show the \class{k}-feature in one or several
  samples, and are highlighted in subpanels (c), (f), and (g) in \Cref{fig:highlight_classes_klm}. The bands are linked to
different pyroxenes \citep{NearIrSpectraHarder2005}, and the presence of the
\SI{1.9}{\micro\meter} band is accompanied by the \SI{0.9}{\micro\meter}
band, but not vice versa \citep{ARadarSurveyShepar2015}.

The distribution of \class{M}-types in latent space and the results acquired in
the studies cited above suggest that there are at least two populations of
\class{M}-types, the chondritic population, of which \nuna{21}{Lutetia} may be the
archetype, and  the metallic population, of which \nuna{16}{Psyche} is the  prototype
\citep{Asteroid21LVernaz2011,AdaptiveOpticsViikin2017}. We see this as
a reasonable division of the \class{M}-class to further dissolve the compositional
degeneracy of the \class{X}-complex. However, this division cannot be done based
on spectra alone. To not increase the entropy of the taxonomy in a false
direction, we refrain here from dividing the \class{M}-class.

A total of \num{\NMFinal} asteroids
(\SI[round-mode=places,round-precision=1]{\NMFinalFraction}{\percent}) are
classified as \class{M}-types in this study.
The main clusters containing \class{M}-types are clusters 22, 37, and 46. Smaller
contributors are clusters 17 and 35. All these clusters make up the
\class{X}-complex in this taxonomy, and the spectra are split into \class{E},
\class{M}, and \class{P} as described in \cref{sub:resolving_the_xcomplex}.
Additional members of the \class{M}-class are found in clusters 23 and 24, which
are spectrally close to \class{L}- and \class{K}-types.

\subsection{Endmembers: \class{E}-types}
\label{sub:endmembers:_e_x}

\class{E}-type asteroids are linked to the enstatite achondrites
\citep{RelationshipOfGaffey1992}. Their standout feature is a visual albedo generally above
\SI{50}{\percent}, see \cref{fig:class_spectra_mcomplex}.
This unique property makes them easy to recognise in the reduced latent space,
where they exhibit large absolute values in $z_2$ and $z_3$, with the former
shown in \cref{fig:highlight_classes_klm}.

Spectrally, \class{E}-types have a steep visible slope before flattening out in
the NIR.
In the case where the albedo observation is missing, \class{E}-types are degenerate
with all classes of the \class{M}-complex. As an example, we observe
samples of \nuna{44}{Nysa} located in subpanels (a) and (d), around the
\class{K}- and the \class{L}-types, correctly identified as an \class{E}-type, due
to the albedo observation. However, the third sample in subpanel (b)
lacks an associated albedo value and is classified as an \class{M}-type. As for
\class{L} and \class{M}, we find a large intrinsic variability of the samples
of individual asteroids in the \class{E}-class.

Most \class{E}-types in \citet{1984PhDT.........3T} are classified as \class{Xe}
in the Bus-DeMeo system due to the presence of the \class{e}-feature at
\SI{0.5}{\micro\meter}. In \cref{fig:feature_distribution}, we see that the
\class{e}-feature is overall sparse compared to the
\class{k}-feature. Thirteen samples in the \class{M}-complex exhibit the
feature, while \SI[round-mode=places,round-precision=1]{\RatioMCenan}{\percent}
of samples lack the corresponding wavelength region observed.
Of the \num{\NMCe} samples, \num{\NEe} are classified as
\class{E}-type. Considering the relative sizes of the \class{M}- and
\class{E}-class, the latter are hence more likely to exhibit the feature. We do
not observe a clustering of the \class{e}-feature.

The bias towards \class{E} over \class{M} for \class{e}-feature presence may be
of observational nature. As an abundance of metal on the surface of
\class{M}-types may lead to a drop-off of the spectral reflectance in the
UV, the \SI{0.5}{\micro\meter} feature might not be observed as the reflectance
does not increase again towards smaller wavelengths. The band is associated
with the sulfide mineral oldhamite present in aubrites
\citep{1979LPSC...10.1073W} or to titanium-bearing pyroxene
\citep{TitaniumBearinShesto2010}. The prototype for this feature is
\nuna{64}{Angelina}, while the \class{E}-class archetype \nuna{434}{Hungaria}
does not present it. The \class{k}-feature is present in
\SI[round-mode=places,round-precision=1]{\RatioEk}{\percent} of \class{E}-type
samples, while \SI[round-mode=places,round-precision=1]{\RatioEknan}{\percent}
of samples lack the corresponding wavelength region observed.

\nuna{214}{Aschera} highlights the benefit of resurrecting the visual albedo.
Since its classification as \class{E}-type in \citet{1989aste.conf..298T}, it
has been classified as \class{X}, \class{B}, \class{Cgh}, and \class{C} in
different works \citep{SosTheVisiblLazzar2004,AnExtensionOfDemeo2009,NearInfraredSDeLeo2012}. With a visual
albedo above \SI{50}{\percent}, \nuna{214}{Aschera} is here classified as
\class{Ek}-type and concludes its spin through the proverbial `alphabet
soup'. Observations of \nuna{64}{Angelina}, \nuna{214}{Aschera}, and
\nuna{434}{Hungaria} are highlighted in subpanel (j) of
\cref{fig:highlight_classes_klm}.

A total of \num{\NEFinal} asteroids
(\SI[round-mode=places,round-precision=1]{\NEFinalFraction}{\percent}) are
classified as \class{E}-types in this study. They are predominantly located in
cluster 35, though other clusters of the \class{M}-complex may also contain
single samples of \class{E}-types. These are identified and assigned to the \class{E}-class in a late branch of the
decision tree using the albedo distributions of \class{E}, \class{M}, and
\class{P} given in \cref{fig:resolve_emp}. \class{E}-types also  appear in
cluster 44 among the  \class{S}-types, where they are identified based on a
two-component GMM fitted to the albedo distribution of the cluster.

\subsection{\class{S}-complex: \class{S}, \class{Q}}
\label{sub:the_scomplex}

The \class{S}-complex is by far the largest complex in terms of individual asteroids,
in this work and  in previous taxonomies. This can be attributed
to observational biases such as the numeric dominance of the \class{S}-types in
the inner main belt and near-Earth space
\citep{TheTaxonomicDDemeo2013,2014Natur.505..629D,CompositionalDBinzel2019} and
the high average albedo of more than \SI{20}{\percent}.

The abundance of \class{S}-types makes their homogeneity
both in spectra and albedos as shown in
\cref{fig:class_spectra_scomplex} even more remarkable. While trends in the
slope and the silicate features at \SI{0.9}{\micro\meter},
\SI{1.0}{\micro\meter}, and \SI{1.9}{\micro\meter} are observable, these are
primarily continuous trends and well explained by variations in the mineral
composition, in particular olivine and pyroxene, as well as trends of
thermal alteration in ordinary chondrites \citep{MultipleAndFaVernaz2014,InvestigatingSEschri2022},
). \class{S}-types are one of two classes of
asteroids that  have an established meteorite analogue;  they were linked to
ordinary chondrites by the JAXA Hayabusa mission
\citep{ItokawaDustPaNakamu2011}. This linkage in combination with the
wealth of data on ordinary chondrites and \class{S}-types gives a solid
understanding of the spectral weathering processes occurring on the surfaces of
the minor bodies \citep{ElasticCollisiBrunet2005,SpaceWeatherinThomas2012,2108.00870v1}, which, unlike  the \class{C}-complex members, shows a universal
trend of surface darkening and spectral reddening with the surface age.

We divide the \class{S}-complex into two classes: \class{S} and
\class{Q}. Including the endmember classes \class{A}, \class{R}, and \class{V},
we establish all classes defined in the \citet{1984PhDT.........3T} system while
extending it with the \class{O}-class. Compared to the Bus-DeMeo system, we
reduce the taxonomy by subclasses of the \class{S}-class, as we explain
in the following.

\subsubsection{\class{S}-types}
\label{sub:stypes}

While class \class{C} has been split into subclasses since early taxonomic efforts
\citep{1982Sci...216.1405G,1984PhDT.........3T}, the \class{S}-class was not
divided until \citet{2002Icar..158..146B} as the silicaceous surfaces are
particularly subject to changes in slope and band structure induced by
phase-angle effects \citep{2012Icar..220...36S} and space weathering
\citep{SpectralAlteraStrazz2005}.

\begin{figure}[t]
  \centering
  \input{gfx/class_spectra_scomplex.pgf}
  \vspace{-0.75em}
  \caption{As in  \cref{fig:class_spectra}, but for the data space properties of the
  \class{S}-complex. The albedo distribution of the \class{S}-complex is $\num[round-mode=places,
round-precision=2]{\pVCSMean}^{+\num[round-mode=places,
round-precision=2]{\pVCSStdUpper}}_{-\num[round-mode=places,
round-precision=2]{\pVCSStdLower}}$.}
  \label{fig:class_spectra_scomplex}
\end{figure}

\begin{figure*}[t]
  \centering
  \input{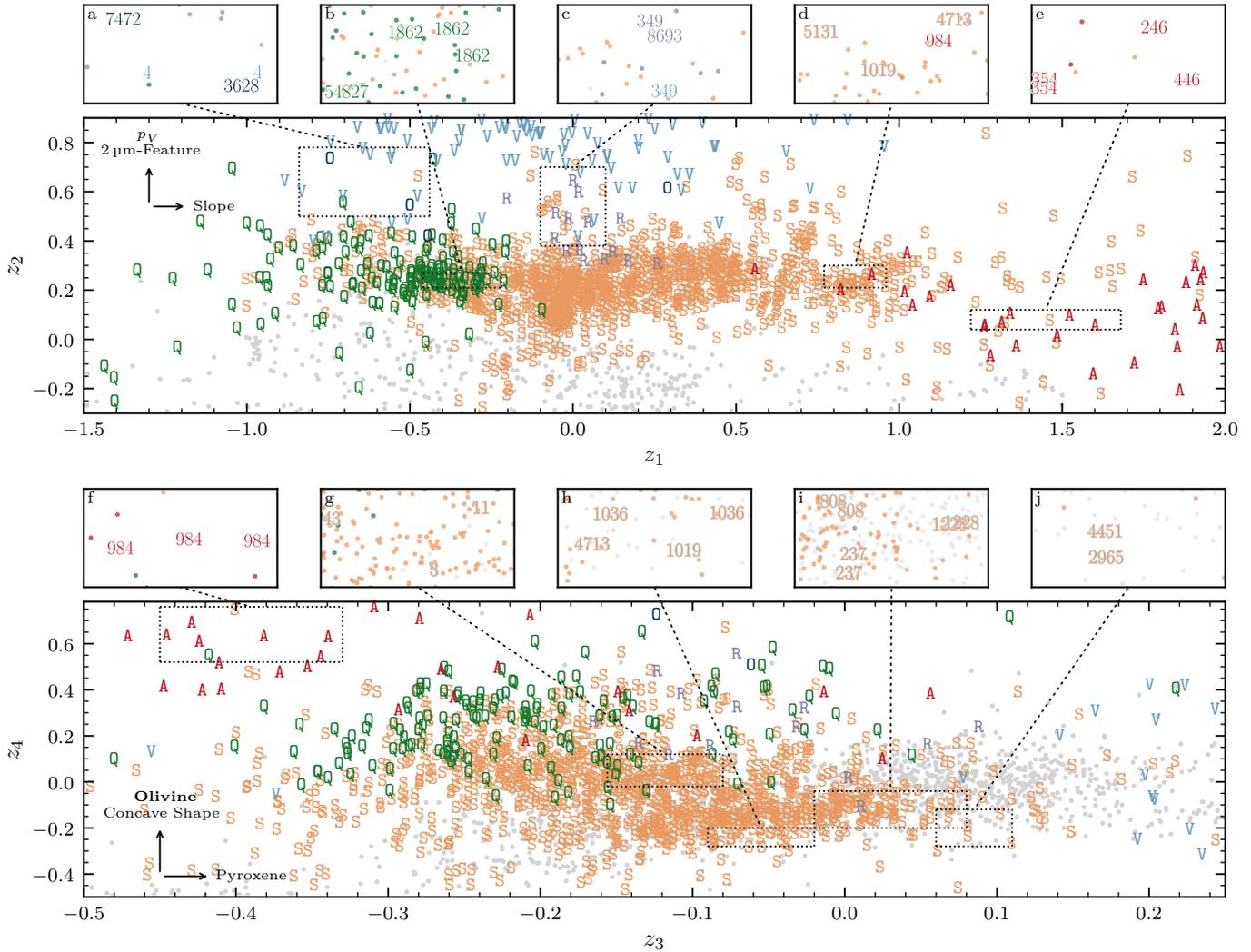}
  \vspace{-0.75em}
  \caption{As in \cref{fig:ccomplex}, but for the member classes of the
  \class{S}-complex. For increased resolution of the \class{S}-class, the
\class{A}- and \class{V}-class are only shown partially.}
  \label{fig:highlight_classes_s}
\end{figure*}

The Bus-DeMeo system accounts for these effects by subtracting the spectral
slope before classification; however, as outlined in previous sections, the
partial observations prevent us from applying  this taxonomy. Instead,
we rely on the interpretation of the latent components to serve as vectors in
the compositional analysis of the \class{S}-types.

The second and
third latent components both resemble pyroxene as this mineral dominates the
\class{S}-class, in addition to the large contribution in terms of variance
provided by the \class{V}-types. The first component resembles the slope, hence
we can approximate the vector of space weathering within the \class{S}-complex
with it \citep[\eg][]{ModelingAsteroBrunet2006}. \class{S}-types denoted with
the \class{w}-suffix for weathered in the Bus-DeMeo system exhibit higher
$z_1$ scores than their class siblings with fresh surfaces. The degeneracy
between a weathered \class{S}-type and an olivine-rich \class{S}-type
(\class{Sa} in the Bus-DeMeo system), which is redder by mineralogy rather than
by surface alteration, is resolved in the third and fourth latent component,
which separates the pyroxene-olivine composition of objects.

As a practical example, in subpanel (d) in
\cref{fig:highlight_classes_s} we show the Bus-DeMeo \class{Sa}-types
\nuna{984}{Gretia} and \nuna{5131}{1990 BG} and the \class{Sw}-types
\nuna{1019}{Strackea} and \nuna{4713}{Steel}. The subpanel (h) shows that both
\class{Sw}-types have below average olivine components, indicating that the
red surface is indeed due to weathering;
also shown in this subpanel is the
\class{S}-type \nuna{1036}{Ganymed}. \nuna{984}{Gretia} is classified as
\class{A}-type in this study due to its high  $z_4$ score (see
subpanel (f) in \cref{fig:highlight_classes_s}).

The Bus-DeMeo system further recognises \class{Sq}-, \class{Sr}-, and
\class{Sv}-types in addition to  the regular \class{S}-types. The prototypes given in
\citet{AnExtensionOfDemeo2009} for these subclasses are highlighted respectively in subpanels (g) (\nuna{3}{Juno}, (11) \textit{Parthenope}, \nuna{43}{Ariadne}), (i)
(\nuna{237}{Coelestina}, \nuna{808}{Merxia}, (1228) \textit{Scabiosa}), and (j) (\nuna{2965}{Surikov}, \nuna{4451}{Grieve}) in
\cref{fig:highlight_classes_s}. The continuous distribution between the main
\class{S}-complex and the subclasses confirms our decision to not subdivide the
\class{S}-class.

A total of \num{\NSFinal} asteroids
(\SI[round-mode=places,round-precision=1]{\NSFinalFraction}{\percent}) are
classified as \class{S}-types in this study. 
The class is made up of several clusters: 0, 3, 6, 11, 14, 20, 21, 30, 33, 38,
39, 40, 42, and 47. Clusters 4, 8, 10, 43, and 44 contain members from other classes,
which we divide via GMMs, as described in the respective
class descriptions (\class{L}, \class{D}, \class{R}, \class{D}, and \class{E},
in order of the clusters).

\subsubsection{\class{Q}-types}
\label{sub:qtypes}

\class{Q}-type asteroids are mostly found in the near-Earth asteroid population
and resemble spectrally the ordinary chondrites in the meteorite collection
\citep{ObservedSpectrBinzel2004}. Compared to \class{S}-types, \class{Q}-types have a wider
\SI{1}{\micro\meter} band and a neutral to blue slope over the whole spectral
range (see \cref{fig:class_spectra_scomplex}). The albedo distribution is
more extended towards higher albedo values, with values of
\SIrange{20}{35}{\percent}, in agreement with space weathering models
  predicting darkening of silicaceous asteroids with increasing surface age
\citep{ModelingAsteroBrunet2006}.

In latent space, \class{Q}-types occupy the blue end of the \class{S}-complex in
$z_1$. They are also distinguished from the less weathered \class{S}-types in the
$z_2$-$z_3$ space based on their high $z_3$ scores due to the wide
\SI{1}{\micro\meter} band. The archetype \nuna{1862}{Apollo} and class member
\nuna{54827}{Kurpfalz} are highlighted in subpanel (b) in
\cref{fig:highlight_classes_s}.

A total of \num{\NQFinal} asteroids
(\SI[round-mode=places,round-precision=1]{\NQFinalFraction}{\percent}) are
classified as \class{Q}-types in this study,
\SI[round-mode=places, round-precision=1]{\RatioQNea}{\percent} of which are
near-Earth asteroids, which is considerably higher than the average
  of \SI[round-mode=places, round-precision=1]{\RatioNea}{\percent} over all
asteroids in the input data.
They populate clusters 16 and 48, as well as the diffuse cluster 13, further
outlined in \cref{sub:rtypes}.
We considered merging the \class{Q}-class into the \class{S}-class as it
represents the overall continuity in the \class{S}-complex. However, as for the
\class{Z}-class,  the orbital distribution of the \class{Q}-types
convinced us to keep this class.

\subsection{Endmembers: \class{A}, \class{O}, \class{R}, \class{V}}
\label{sub:endmembers:_v_a}

The endmembers of the \class{S}-complex are the well-established classes
\class{A} and \class{V} and  the two classes that  were initially built around
single objects, \class{O} and \class{R}. Their distribution in the third and
fourth latent scores is given in \cref{fig:endmembersaorv}.

\subsubsection{\class{A}-types}
\label{subsub:atypes}

\class{A}-type asteroids are differentiated asteroids linked to brachinite
achondrites
\citep{TheMeteoriteACruiks1984,2002aste.book..653B,OlivineDominatDemeo2019}
 and are easily recognised in spectral space by their
strong red slope and deep olivine imprint at \SI{1}{\micro\meter} (see
\cref{fig:class_spectra_scomplex}). The albedo is within the complex average of
about \SIrange{20}{30}{\percent}.

In latent space, the red colour of \class{A}-types leads to a high score in
$z_1$, forming a diffuse branch off the \class{S}-type population. We highlight
the prototypes \nuna{246}{Asporina}, \nuna{354}{Eleonora}, and
\nuna{446}{Aeternitas} in subpanel (e) in \cref{fig:highlight_classes_s}.
Further characteristic of \class{A}-types is a high $z_4$ score due to the
high olivine content (see \cref{fig:endmembersaorv}).
Of all the  classified asteroids, \class{A}-types have  the highest $z_1$ and $z_4$ scores.
We note that all three spectra of Mars-Crosser \nuna{1951}{Lick} are
exceptionally red, even among \class{A}-types \citep{TestingSpaceWBrunet2007}.

A total of \num{\NAFinal} asteroids
(\SI[round-mode=places,round-precision=1]{\NAFinalFraction}{\percent}) are
classified as \class{A}-types in this study.
They fall into clusters 9, 12, 27, and 49.

\subsubsection{\class{O}-types}
\label{sub:otypes}

The class \class{O} was introduced in 1993 for supposedly ordinary-chondritic
(3628) \textit{Boznemcova} \citep{DiscoveryOfABinzel1993}. Its noteworthy
characteristic is the wide round \SI{1}{\micro\meter} feature as shown in
\cref{fig:class_spectra_scomplex}, placing it
between the known \class{A}-, \class{Q}-, and \class{V}-types.
The albedo is close to the \class{S}-complex average at
\SI{25}{\percent}.

None of the previously classified \class{O}-types, except for  archetype
(3628) \textit{Boznemcova} remains as an \class{O}, and a comparison of these objects
in spectral space showed little resemblance.
While we assign with \nuna{7472}{Kumakiri} a second asteroid to the class, we
find in this work that (3628)\,\textit{Boznemcova} remains without a true
spectral sibling.
\nuna{7472}{Kumakiri} was previously classified as
\class{V} \citep{AvastSurvey0Solont2012}; however, its spectral resemblance to
\nuna{3628}{Boznemcova} has been pointed out by
\citet{2011LPI....42.2483B}.

The unique appearance of the \class{O}-types can be seen by their position in
the latent space shown in \cref{fig:highlight_classes_s,fig:endmembersaorv}. The
depth and shape of the \SI{1}{\micro\meter} band in combination with the lack of
overall slope place the \class{O}-types \nuna{3628}{Boznemcova} and
\nuna{7472}{Kumakiri}  between the classes \class{Q} and \class{V} in
$z_1$-$z_2$ (see subpanel (a) in \cref{fig:highlight_classes_s}),
while in $z_3$-$z_4$, they are closest to \class{A}-types.

Two asteroids
(\SI[round-mode=places,round-precision=1]{\NOFinalFraction}{\percent}) are
classified as \class{O}-types in this study.
We debated whether keeping the \class{O}-class in the taxonomy is compatible
with the overall approach of data-driven clustering. In the end, the unique
feature and position of \nuna{3628}{Boznemcova} convinced us, although  an
argument against single-object classes can be made. The
\class{O}-class was difficult to carve out from the clusters using the given
method. It is derived from a three-component mixture model of the already
diffuse cluster 13, which is split into \class{C}, \class{O}, and \class{Q}.
Any assignment of the \class{O}-class by the classification tool
should undergo visual scrutiny and direct comparison to the spectrum of
\nuna{3628}{Boznemcova}.

\subsubsection{\class{R}-types}
\label{sub:rtypes}

\begin{figure}
  \centering
  \input{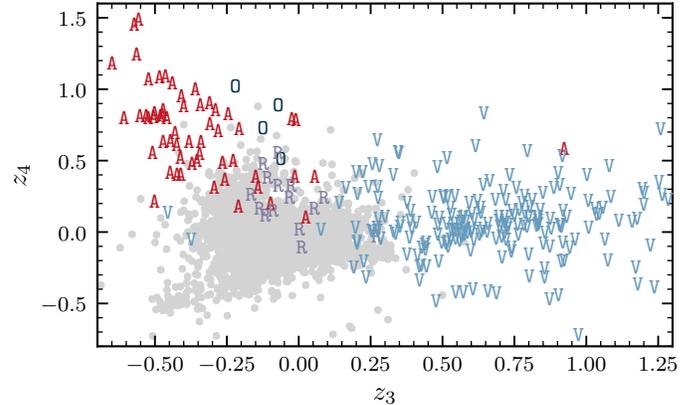}
  \vspace{-0.75em}
  \caption{Distribution of the \class{S}-complex endmember classes \class{A},
    \class{O}, \class{R}, and \class{V} in the last two components of the
  latent space.}
  \label{fig:endmembersaorv}
\end{figure}

The \class{R}-types are the second niche class of this taxonomy, built around
\nuna{349}{Dembowska}. The unique nature of (349) \textit{Dembowska} is recognised
jointly with that of \nuna{4}{Vesta} in early works of taxonomy
\citep{SurfacePropertChapma1975,1976AJ.....81..262Z} and the \class{R}-class was
introduced in \citet{1978Icar...35..313B}. However, the \class{A}-class, which
was split off the \class{R}-class in \citet{TheRAsteroidsVeeder1983}, has since been  absorbed into
most \class{R}-types.
The continuity between \class{A} and \class{R} is visible in \cref{fig:endmembersaorv}.

\class{R}-types show \SI{1}{\micro\meter} and \SI{2}{\micro\meter} features
which are deeper than those in \class{S}-types. The width of the
\SI{1}{\micro\meter} is between the \class{V}- and the \class{Q}-types. They have
albedos at the upper end of the \class{S}-complex distribution, around
\SI{28}{\percent} (see \cref{fig:class_spectra_scomplex}). The spectral
appearance is associated with low-iron ordinary chondrites
\citep{1976AJ.....81..262Z}. We note that of the four samples of
\nuna{349}{Dembowska} two are classified as \class{R} and another two as
\class{V} (see subpanel (c) in \cref{fig:highlight_classes_s},
where we  also give the position of \class{R}-class member \nuna{8693}{Matsuki}).

A total of \num{\NRFinal} asteroids
(\SI[round-mode=places,round-precision=1]{\NRFinalFraction}{\percent}) are
classified as \class{R}-types in this study. The class is derived from cluster
10 in a two-component GMM fit in $z_1$-$z_2$, where objects
with lower $z_2$ scores are assigned to the \class{S}-class.

\begin{table}[h]
  \centering
  \caption{Evolution of taxonomic scheme from
  \citet{1984PhDT.........3T} to Bus-DeMeo  to this
work.
  Arrows are used to indicate the overall evolution of each
class. The \class{T}-class is not present in this taxonomy and the feature
characteristic of the \class{Xn} has been grouped into the \class{k}-feature. The evolution of the
\class{X}-complex between the taxonomies is unclear as the visual albedo is not
taken into account in the Bus-DeMeo system. No analogues for \class{K}
and \class{L} were defined in \citet{1984PhDT.........3T}.
}
  \label{tab:taxonomy_overview}
  \begin{tabular}{lcccr}
    Tholen    &               & Bus-DeMeo   &               & This Work  \\ \toprule
    \class{B} & $\rightarrow$ & \class{B}   & $\rightarrow$ &  \class{B} \\
    \class{F} & $\nearrow$    &             &               &            \\
    \class{G} & $\rightarrow$ & \class{Cg}  & $\searrow$    &            \\
              & $\rightarrow$ & \class{Cgh} & $\searrow$    &            \\
    \class{C} & $\rightarrow$ & \class{C}   & $\rightarrow$ & \class{C}  \\
              & $\rightarrow$ & \class{Ch}  & $\rightarrow$ & \class{Ch} \\
              & $\rightarrow$ & \class{Cb}  & $\nearrow$    &            \\
    \class{D} & $\rightarrow$ & \class{D}   & $\rightarrow$ &  \class{D} \\
              &               &             & $\rightarrow$ &  \class{Z} \\
              &               &             &               &            \\
    \class{P} & \dots         & \class{Xc}  & \dots         &  \class{P} \\
    \class{M} & \dots         & \class{Xk}  & \dots         &  \class{M} \\
    \class{X} & \dots         & \class{X}   & \dots         &  \class{X} \\
    \class{E} & \dots         & \class{Xe}  & \dots         &  \class{E} \\
              & \dots         & \class{Xn}  & $\vert$       &            \\
    \class{T} & $\rightarrow$ & \class{T}   & $\vert$       &            \\
              &               & \class{K}   & $\rightarrow$ &  \class{K} \\
              &               & \class{L}   & $\rightarrow$ &  \class{L} \\
              &               &             &               &            \\
    \class{Q} & $\rightarrow$ & \class{Q}   & $\rightarrow$ &  \class{Q} \\
              &               & \class{Sq}  &               &            \\
              & $\nearrow$    & \class{Sr}  & $\searrow$    &            \\
    \class{S} & $\rightarrow$ & \class{S}   & $\rightarrow$ &  \class{S} \\
              & $\searrow$    & \class{Sa}  & $\nearrow$    &            \\
              &               & \class{Sv}  &               &            \\
    \class{O} & $\rightarrow$ & \class{O}   & $\rightarrow$ &  \class{O} \\
    \class{R} & $\rightarrow$ & \class{R}   & $\rightarrow$ &  \class{R} \\
    \class{A} & $\rightarrow$ & \class{A}   & $\rightarrow$ &  \class{A} \\
    \class{V} & $\rightarrow$ & \class{V}   & $\rightarrow$ &  \class{V} \\

  \end{tabular}
\end{table}

\subsubsection{\class{V}-types}
\label{subsub:vtypes}

\nuna{4}{Vesta} was the first asteroid to be observed spectrophotometrically
\citep{AsteroidVestaMccord1970} and the \class{V}-types have been an established
and easily-recognizable class in all asteroid taxonomies since
\citet{1984PhDT.........3T}. They are the  second class, in addition to  \class{S}, with an
established meteoritic analogue, the HED meteorites
\citep[\eg][]{QuantifiedMineKelley2003}. The class makes no exception here; its
members are differentiated easily in both $z_2$ and $z_3$ due to the large
contribution of pyroxene to the spectral appearance giving rise to the
characteristic deep \SI{1}{\micro\meter} and \SI{2}{\micro\meter} features
(see
\cref{fig:class_spectra_scomplex,fig:highlight_classes_s,fig:endmembersaorv}). The class
archetype \nuna{4}{Vesta} is highlighted in subpanel (a) in
  \cref{fig:highlight_classes_s}.

The large class variance in $z_2$ and $z_3$ represents
 high variability in terms of band depth and position in the
\SI{0.9}{\micro\meter} and \SI{2.0}{\micro\meter} features. However, we do not
identify a subpopulation based on the band parameters, as was suggested by
\citet{ChipsOffOfAsBinzel1993}.

A total of \num{\NVFinal} asteroids
(\SI[round-mode=places,round-precision=1]{\NVFinalFraction}{\percent}) are
classified as \class{V}-types in this study. \class{V}-types populate clusters
7, 15, 18, 28, 32, and 45. \class{V}-types with a blue slope in the
NIR further share the diffuse cluster 41 with the \class{B}-types.

\begin{table*}[p]
  \centering
  \caption{Description of taxonomic classes defined in this work.
  Listed are the spectral appearance, visual albedo distribution giving
  the mean value,  the lower and upper standard deviation, and the spectral prototypes of the \num{\NClasses} classes defined in this taxonomy excluding the \class{X}-types.
  }
  \label{tab:class_descriptions}
\setlength{\tabcolsep}{4pt}\renewcommand{\arraystretch}{1.5}\begin{tabular}{llrr}
        \toprule
        Class & Spectrum & Albedo & Prototypes \\
        \midrule
\texttt{A} & \makecell[l]{Broad and deep absorption feature at \SI{1}{\micro\meter},\\ strong red slope in the near-infrared.} & \num[round-mode=places, round-precision=2]{0.2530308194681455}\,$^{+\num[round-mode=places, round-precision=2]{0.08883308398979983}}_{-\num[round-mode=places, round-precision=2]{0.06574987242134128}}$ &\raisebox{-0.5\height}{\input{gfx/spectrum_name_Asporina.pgf}
}\raisebox{-0.5\height}{\input{gfx/spectrum_name_Nenetta.pgf}
}\raisebox{-0.5\height}{\input{gfx/spectrum_name_Eleonora.pgf}
} \\
\texttt{B} & \makecell[l]{Neutral to blue slope in the visible, blue slope\\ in the near-infrared.} & \num[round-mode=places, round-precision=2]{0.0625027179928539}\,$^{+\num[round-mode=places, round-precision=2]{0.0528134736922265}}_{-\num[round-mode=places, round-precision=2]{0.028625517407156298}}$ &\raisebox{-0.5\height}{\input{gfx/spectrum_name_Pallas.pgf}
}\raisebox{-0.5\height}{\input{gfx/spectrum_name_Zerlina.pgf}
}\raisebox{-0.5\height}{\input{gfx/spectrum_name_Phaethon.pgf}
} \\
\texttt{C} & \makecell[l]{Red visible slope with a possible broad feature\\ around \SI{1}{\micro\meter} and a red near-infrared slope. The\\spectrum might have an overall concave shape.} & \num[round-mode=places, round-precision=2]{0.0512018968619827}\,$^{+\num[round-mode=places, round-precision=2]{0.02096461908700361}}_{-\num[round-mode=places, round-precision=2]{0.014874325719178398}}$ &\raisebox{-0.5\height}{\input{gfx/spectrum_name_Ceres.pgf}
}\raisebox{-0.5\height}{\input{gfx/spectrum_name_Hygiea.pgf}
}\raisebox{-0.5\height}{\input{gfx/spectrum_name_Themis.pgf}
} \\
\texttt{Ch} & \makecell[l]{Absorption feature at \SI{0.7}{\micro\meter}. The near-infrared\\slope is red while the overall shape might\\be convex.} & \num[round-mode=places, round-precision=2]{0.0506377511131056}\,$^{+\num[round-mode=places, round-precision=2]{0.016716445295485896}}_{-\num[round-mode=places, round-precision=2]{0.012567638566031304}}$ &\raisebox{-0.5\height}{\input{gfx/spectrum_name_Egeria.pgf}
}\raisebox{-0.5\height}{\input{gfx/spectrum_name_Fortuna.pgf}
}\raisebox{-0.5\height}{\input{gfx/spectrum_name_Daphne.pgf}
} \\
\texttt{D} & \makecell[l]{Featureless with steep red slope with a possible\\ convex shape longwards of \SI{1.5}{\micro\meter}.} & \num[round-mode=places, round-precision=2]{0.0574815454283042}\,$^{+\num[round-mode=places, round-precision=2]{0.025816571411693696}}_{-\num[round-mode=places, round-precision=2]{0.017815245754652702}}$ &\raisebox{-0.5\height}{\input{gfx/spectrum_name_Achilles.pgf}
}\raisebox{-0.5\height}{\input{gfx/spectrum_name_Agamemnon.pgf}
}\raisebox{-0.5\height}{\input{gfx/spectrum_name_Odysseus.pgf}
} \\
\texttt{E} & \makecell[l]{Strong red slope in the visible with a feature\\around \SI{0.9}{\micro\meter} of varying depth and\\a neutral near-infrared continuation.} & \num[round-mode=places, round-precision=2]{0.5657434492084065}\,$^{+\num[round-mode=places, round-precision=2]{0.1515270518234968}}_{-\num[round-mode=places, round-precision=2]{0.11951618925311569}}$ &\raisebox{-0.5\height}{\input{gfx/spectrum_name_Angelina.pgf}
}\raisebox{-0.5\height}{\input{gfx/spectrum_name_Aschera.pgf}
}\raisebox{-0.5\height}{\input{gfx/spectrum_name_Hungaria.pgf}
} \\
\texttt{K} & \makecell[l]{Strong red slope in the visible with a broad\\feature around \SI{1}{\micro\meter} followed by a blue to\\neutral near-infrared slope.} & \num[round-mode=places, round-precision=2]{0.1309762836062341}\,$^{+\num[round-mode=places, round-precision=2]{0.0361749711788191}}_{-\num[round-mode=places, round-precision=2]{0.028345963006124186}}$ &\raisebox{-0.5\height}{\input{gfx/spectrum_name_Eos.pgf}
}\raisebox{-0.5\height}{\input{gfx/spectrum_name_Sidonia.pgf}
}\raisebox{-0.5\height}{\input{gfx/spectrum_name_Berenike.pgf}
} \\
\texttt{L} & \makecell[l]{Variable appearance apart from a red visible slope.\\A small feature around \SI{1}{\micro\meter} and a possible\\one at \SI{2}{\micro\meter}. The near-infrared slope is blue or red.} & \num[round-mode=places, round-precision=2]{0.1821748530012528}\,$^{+\num[round-mode=places, round-precision=2]{0.06640544693859021}}_{-\num[round-mode=places, round-precision=2]{0.0486659744857004}}$ &\raisebox{-0.5\height}{\input{gfx/spectrum_name_Barbara.pgf}
}\raisebox{-0.5\height}{\input{gfx/spectrum_name_Vienna.pgf}
}\raisebox{-0.5\height}{\input{gfx/spectrum_name_Luisa.pgf}
} \\
\texttt{M} & \makecell[l]{Linear red slope with possible faint features around\\ \SI{0.9}{\micro\meter} and \SI{1.9}{\micro\meter}. Might show convex shape in the\\near-infrared.} & \num[round-mode=places, round-precision=2]{0.1432294737097776}\,$^{+\num[round-mode=places, round-precision=2]{0.051123678407832995}}_{-\num[round-mode=places, round-precision=2]{0.037675836345740094}}$ &\raisebox{-0.5\height}{\input{gfx/spectrum_name_Psyche.pgf}
}\raisebox{-0.5\height}{\input{gfx/spectrum_name_Kalliope.pgf}
}\raisebox{-0.5\height}{\input{gfx/spectrum_name_Kleopatra.pgf}
} \\
\texttt{O} & \makecell[l]{Broad, bowl-shaped \SI{1}{\micro\meter} absorption feature\\ and a weaker feature at \SI{2}{\micro\meter}.} & \num[round-mode=places, round-precision=2]{0.256907104617992}\,$^{+\num[round-mode=places, round-precision=2]{0.021721084982650296}}_{-\num[round-mode=places, round-precision=2]{0.020027769121467087}}$ &\raisebox{-0.5\height}{\input{gfx/spectrum_name_Boznemcova.pgf}
}\raisebox{-0.5\height}{\input{gfx/spectrum_name_Kumakiri.pgf}
}\hspace{7em} \\
\texttt{P} & \makecell[l]{Linear red slope and generally featureless.\\ Less red than \texttt{D}-types.} & \num[round-mode=places, round-precision=2]{0.0489506319670027}\,$^{+\num[round-mode=places, round-precision=2]{0.020304376851979992}}_{-\num[round-mode=places, round-precision=2]{0.014351482954806503}}$ &\raisebox{-0.5\height}{\input{gfx/spectrum_name_Cybele.pgf}
}\raisebox{-0.5\height}{\input{gfx/spectrum_name_Sylvia.pgf}
}\raisebox{-0.5\height}{\input{gfx/spectrum_name_Hilda.pgf}
} \\
\texttt{Q} & \makecell[l]{Broad absorption at \SI{1}{\micro\meter} and a shallow feature\\ at \SI{2}{\micro\meter}. An overall blue slope in the near-infrared.} & \num[round-mode=places, round-precision=2]{0.2388921360759776}\,$^{+\num[round-mode=places, round-precision=2]{0.11906405803133363}}_{-\num[round-mode=places, round-precision=2]{0.07946074860895519}}$ &\raisebox{-0.5\height}{\input{gfx/spectrum_name_Apollo.pgf}
}\raisebox{-0.5\height}{\input{gfx/spectrum_name_Daedalus.pgf}
}\raisebox{-0.5\height}{\input{gfx/spectrum_name_Heracles.pgf}
} \\
\texttt{R} & \makecell[l]{Strong feature at \SI{1}{\micro\meter} and a feature at \SI{2}{\micro\meter}.\\ The latter feature is shallower than in \texttt{V}-types.} & \num[round-mode=places, round-precision=2]{0.3018764177419406}\,$^{+\num[round-mode=places, round-precision=2]{0.046094285573039406}}_{-\num[round-mode=places, round-precision=2]{0.039988360153893676}}$ &\raisebox{-0.5\height}{\input{gfx/spectrum_name_Dembowska.pgf}
}\raisebox{-0.5\height}{\input{gfx/spectrum_name_Abehiroshi.pgf}
}\raisebox{-0.5\height}{\input{gfx/spectrum_name_1998_WM.pgf}
} \\
\texttt{S} & \makecell[l]{Moderate features around \SI{1}{\micro\meter} and \SI{2}{\micro\meter}\\ and a neutral to red near-infrared slope.} & \num[round-mode=places, round-precision=2]{0.2391920577832534}\,$^{+\num[round-mode=places, round-precision=2]{0.10311018339483588}}_{-\num[round-mode=places, round-precision=2]{0.07205076092910528}}$ &\raisebox{-0.5\height}{\input{gfx/spectrum_name_Juno.pgf}
}\raisebox{-0.5\height}{\input{gfx/spectrum_name_Astraea.pgf}
}\raisebox{-0.5\height}{\input{gfx/spectrum_name_Irene.pgf}
} \\
\texttt{V} & \makecell[l]{Deep absorption features at \SI{1}{\micro\meter} and \SI{2}{\micro\meter}.\\ The former is much narrower than the latter.} & \num[round-mode=places, round-precision=2]{0.2903473463555044}\,$^{+\num[round-mode=places, round-precision=2]{0.11249287003876252}}_{-\num[round-mode=places, round-precision=2]{0.081079308793992}}$ &\raisebox{-0.5\height}{\input{gfx/spectrum_name_Vesta.pgf}
}\raisebox{-0.5\height}{\input{gfx/spectrum_name_Kollaa.pgf}
}\raisebox{-0.5\height}{\input{gfx/spectrum_name_Kamo.pgf}
} \\
\texttt{Z} & \makecell[l]{Extremely red slope, redder than the \texttt{D}-types.\\ Featureless but may exhibit concave shape\\ in the near-infrared.} & \num[round-mode=places, round-precision=2]{0.0725179104620768}\,$^{+\num[round-mode=places, round-precision=2]{0.0437546817196032}}_{-\num[round-mode=places, round-precision=2]{0.027289303796383398}}$ &\raisebox{-0.5\height}{\input{gfx/spectrum_name_Pompeja.pgf}
}\raisebox{-0.5\height}{\input{gfx/spectrum_name_Justitia.pgf}
}\raisebox{-0.5\height}{\input{gfx/spectrum_name_Buda.pgf}
} \\
        \bottomrule
\end{tabular}
\end{table*}

\section{Classification}
\label{sec:classification}

In this section we introduce the classification tool described in this work. We
demonstrate the probabilistic classification results using asteroid observations
with different wavelength regions covered. We further
investigate degeneracies in the classification space. Finally, we compare the
results obtained in this taxonomy to the previous systems.

\subsection{Classification tool: classy}
\label{sub:classification_tool_classy}

To facilitate the classification of asteroid observations within the framework
of this taxonomy, we provide the \emph{CLAssification of a Solar System bodY}
(\texttt{classy}\footnote{\url{https://github.com/maxmahlke/classy}})  tool written in \texttt{Python}. It is able to interactively smooth
the input spectral observations prior to resampling them to the
required wavelength grid, to automatically apply the necessary
pre-processing steps outlined in \cref{sec:method} to both spectra and albedo,
to identify features in the spectra as outlined in \cref{subsub:feature_flags} (either
fully automated or guided by the user),
to execute the cluster-to-class decision tree,
and to return the probabilistic classifications for each observation.

The \texttt{classy} tool provides a command-line interface written in \texttt{Python} and
is available for Windows, MacOS, and Linux. The software is actively maintained
and developed by the authors.

\subsection{Class degeneracies}
\label{sub:class_degeneracies}

The probabilistic nature of the classifications in this taxonomy allow  the degeneracies between classes to
be quantified in certain wavelength regions and in
albedo. One example is given in \cref{subsub:btypes}, where we point out the
degeneracy of \class{B} and \class{K} in the case of a NIR-only
observation.

\begin{figure}[t]
  \centering
  \input{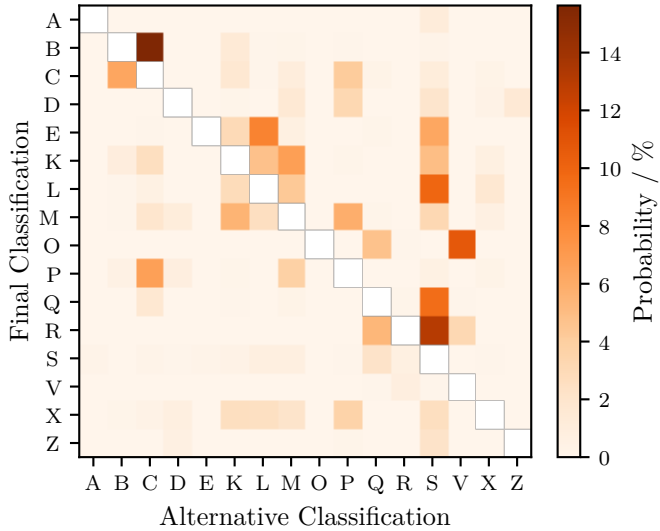}
  \caption{Confusion matrix between the classes defined in this
    taxonomy in the visible-near-infrared and albedo input space. For each
    class in the taxonomic scheme, we give the average probability of its
    samples to be classified as any other class based on the
  complete dataset. The \class{Ch}-class is missing as it relies on
  the detection of the \SI{0.7}{\micro\meter} \class{h}-feature and does not
  have an associated class probability. For better readability, the main matrix
  diagonal corresponding to the equal-class cases is left empty. These values
  are generally above \SI{80}{\percent} and lowest for \class{K},
\class{L}, \class{M}, and \class{R}.}
  \label{fig:degeneracy_matrix}
\end{figure}

We can quantify class degeneracies for three datasets in this work, with the aim of
reflecting the most commonly available observation ranges of asteroid spectra: the
\num{\NSamples} spectra used to devise the clustering, the
\num{\NSamplesVisOnly} visible-only spectra shown in grey in
\cref{fig:input_data} with
\SI[round-mode=places,round-precision=1]{\NAlbedoVisOnly}{\percent} albedos
observed, and the  \num{2813} spectra from the clustering sample which have NIR
information. For the last we remove all observations of wavelengths below
\SI{0.8}{\micro\meter} and the albedo information present in the samples. We
 refer to these samples as the complete, the visible-only, and the
NIR-only datasets;  however, this wording is not entirely accurate as more
than \SI{50}{\percent} of the samples in the complete sample are NIR-only spectra
and the visible-only sample contains more than \SI{80}{\percent} of albedo
observations.

\subsubsection{Complete sample}
\label{subsub:complete_sample}

To estimate the class degeneracy in the complete dataset we compute the average probability of belonging to any
other
class for all
samples assigned to a given class. This comparison is given in \cref{fig:degeneracy_matrix}. The
\class{Ch}-class is missing as it relies on the detection of the
\class{h}-feature, and as such does not have an associated class probability. Larger matrix element values indicate a higher degeneracy between
the classes. A large sum per matrix row indicates that the class assignment is
overall less certain.

\begin{figure}[t]
  \centering
  \input{gfx/degeneracy_matrix_visonly.pgf}
  \caption{As in \cref{fig:degeneracy_matrix}, but using the dataset of
    \num{\NSamplesVisOnly} visible-only
  spectra with
  \SI[round-mode=places,round-precision=1]{\NAlbedoVisOnly}{\percent} albedo
  observations. The  colourbar scale is different   to that in
  \cref{fig:degeneracy_matrix}.
The main matrix diagonal values are between \SIrange{63}{91}{\percent} and
lowest for \class{K}, \class{L}, \class{M}, and \class{O}.}
\label{fig:degeneracy_matrix_visonly}
\end{figure}

\begin{figure}[t]
  \centering
  \input{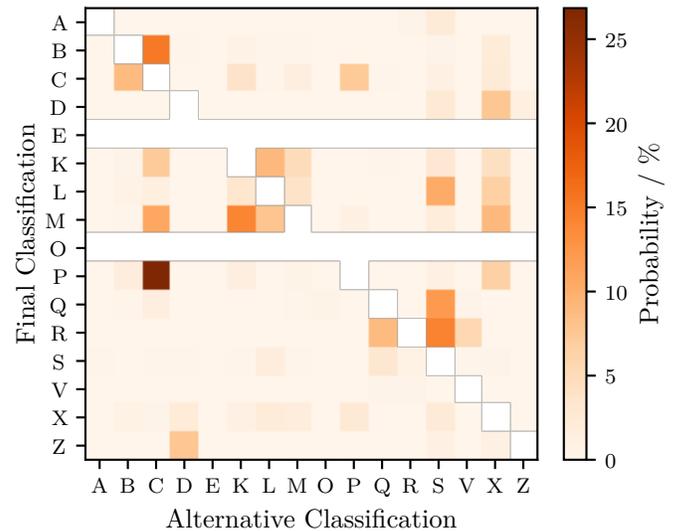}
  \caption{As in \cref{fig:degeneracy_matrix}, but using the dataset of
    \num{2813} NIR-only
  spectra without albedo information. The  colourbar scale is different   to that in
\cref{fig:degeneracy_matrix}. No observation in this sample is classified as \class{E} or \class{O}.
The main matrix diagonal values are between \SIrange{55}{99}{\percent} and
lowest for \class{M} and \class{P}.}
  \label{fig:degeneracy_matrix_nironly}
\end{figure}

\begin{figure}[t]
  \centering
  \input{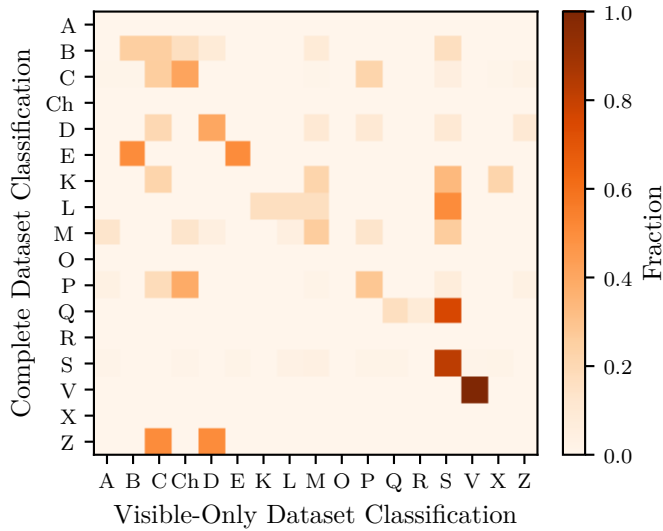}
  \caption{Comparison of the classifications of \num{\NSamplesBothSamples}
  samples of \num{\NAsteroidsBothSamples} individual asteroids resulting from visible-only
  spectra with \SI[round-mode=places,round-precision=1]{\NAlbedoVisOnly}{\percent}
  observed albedos to the classifications of the same asteroids resulting from
  the complete sample classifications.   The sample size iis different in each row: the intersection of asteroids present in both datasets gives
  \num{\NEVisOnly} samples classified as \class{E} using complete samples as
  well as \num{\NZVisOnly} samples classified as \class{Z}, while there are
  \num{\NSVisOnly} samples entering the calculation in the row of
\class{S}-types. No \class{A}-, \class{O}-, or \class{X}-types are present in
both samples.}
  \label{fig:complete_vs_vis}
\end{figure}

\begin{figure}[t]
  \centering
  \input{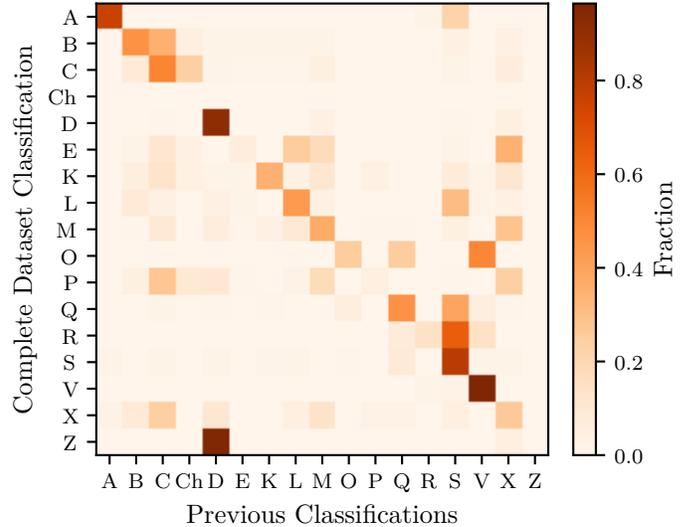}
  \caption{Comparison of the classifications of \num{\NSamplesPrev} samples of
    \num{\NAsteroidPrev} individual asteroids
  classified in the complete dataset to the classifications in the
literature. The literature classifications were mapped into this taxonomy scheme
following \cref{tab:taxonomy_overview}. The number of samples differs between the rows.}
  \label{fig:complete_vs_previous}
\end{figure}

\Cref{fig:degeneracy_matrix} shows the intuitive result that endmember classes
such as \class{A}, \class{V}, and \class{Z} are assigned with a large
probability. The largest degeneracies in pairs of classes are between \class{B}-
and \class{C}-types and \class{R}- and \class{S}-types. Neither result is
surprising as they  overlap in latent space, and even in visual
inspection these classes can be difficult to tell apart.
The largest uncertainty overall for a single class (given by the sum per row in
\cref{fig:degeneracy_matrix}) is around \SI{20}{\%} for \class{K}, \class{L},
and \class{M} and also for the \class{R}-types. For the first three, we
already pointed out in \cref{sub:mcomplex} the similarity in data space between
these classes, hence this result is again expected.

\subsubsection{Visible-only sample}
\label{subsub:vis_only}

The estimation of the class degeneracy is repeated for the visible-only dataset
after classifying the samples therein using the \texttt{classy}
tool. \Cref{fig:degeneracy_matrix_visonly} shows a  result similar to that  for the
complete dataset, except that the overall values of uncertainty increase. Instead
of \SIrange{80}{99}{\percent} certainty in the class assignment, we  obtain
values between \SIrange{63}{91}{\percent}.
Except for  this overall change in scale, we
do not observe significant differences between the results for the visible-only
 and the complete datasets.
\class{K}, \class{L}, and \class{M} are among the least-certain classes,
while \class{O}-types have the largest uncertainty due to the missing
\SI{1}{\micro\meter} band. For classes from the \class{M}- and \class{S}-complex, we see an overall
increasing probability to be classified as \class{S}-type.

\subsubsection{NIR-only sample}
\label{subsub:nironly_sample}

The class degeneracy is next calculated for the NIR-only spectra that are part of the input observations
used to train the MCFA model. We remove the albedo information present in
\SI[round-mode=places,round-precision=1]{78.5}{\percent} of the samples prior to
classifying them. The confusion matrix is shown in
\cref{fig:degeneracy_matrix_nironly}. The overall scale of the uncertainty in
the class assignment is between the results for the complete and the
visible-only dataset, with the maximum average degeneracy
between two classes just over \SI{25}{\percent} between \class{P} and
\class{C}, likely due to both the missing albedo information and the truncation
of the broad \SI{1.3}{\micro\meter} feature in the \class{C}-types. We note that
no sample is classified as \class{E}-type, due to the missing albedo information,
and no sample is classified as \class{O}-type, as both \nuna{3628}{Boznemcova}
and \nuna{7472}{Kumakiri} are classified as \class{Q} without the
visible-wavelength information. The
bowl-shaped \SI{1}{\micro\meter} band of the \class{O}-types extends below the
\SI{0.8}{\micro\meter} limit we apply to this dataset, hence this
misclassification is acceptable.
The expected degeneracy between \class{B} and \class{K} in NIR-only data is not visible in \cref{fig:degeneracy_matrix_nironly}
due to the presence of the \SI{1}{\micro\meter} information. While visible-only
spectra lead to uncertainty among the \class{M}- and \class{S}-complexes, in
particular with respect to the \class{S}-types,
  this calculation shows that NIR-only
spectra lead to greater confusion between the \class{C}- and the \class{M}-complexes.

We conclude that the class degeneracies in the complete, visible-only, and NIR-only
samples follow an intuitive behaviour: the largest classes in terms of number of
samples (\class{S}, \class{C}, and \class{M}) become more probable with
decreasing observational data.
This is in line with the established
classification guideline that, when in
doubt, assignment to small classes should only be done on the basis of convincing
observational evidence.

\subsubsection{Complete versus visible-only sample}
\label{subsub:showdown}

Another way to investigate class degeneracies is the comparison of
classifications resulting from samples with different wavelength regions
observed. There are \num{\NAsteroidsBothSamples} asteroids   present in both the
complete and the visible-only datasets with a total of
\num{\NSamplesBothSamples} observations. For these asteroids, we compare the resulting
classifications based on the samples in both datasets, shown in
\cref{fig:complete_vs_vis}. Each row gives for each class in the taxonomy the
fraction of asteroids classified as any class based on the visible-only dataset.
We note that the figure does not account for the different samples sizes: there are
\num{\NEVisOnly} samples classified as \class{E} in the intersection of the
dataset and \num{\NSVisOnly} classified as \class{S}. No samples classified as
\class{A}, \class{O}, and \class{X} are present in both samples.

\Cref{fig:complete_vs_vis} shows that \class{Ch}, \class{S}, and \class{V} are
the most reliable when classified using visible-only data. \class{Ch} benefits
from the binary classification which takes place once the \class{h}-feature is
observed. The members of the \class{M}-complex show increasing degeneracy with
the \class{S}-class with decreasing near-infrared coverage. The least-expected
degeneracies are \class{Z} and \class{C}, as well as \class{E} and \class{B},
however, they  are all based on a single sample.

Both results in \cref{fig:degeneracy_matrix_visonly,fig:complete_vs_vis}
show that visible-only spectra in combination with the albedo place a strong
constraint on the taxonomic class, as is well-established from the previous
taxonomies which relied on the visible wavelength ranges exclusively. This
highlights the strengths of the new method employed here: NIR-spectra are not strictly
necessary to derive a classification as incomplete
observations can be classified and the albedo as an accessible observable is
accounted for.

We do not repeat this comparison for the complete and the NIR-only samples as
the latter make up a significant fraction of the former, hence the agreement
between the samples would be overestimated.

\subsection{Comparison to previous taxonomies}
\label{sub:comparison_to_previous_taxonomies}

Class continuity was one of the aspects which we considered when designing the
scheme of classes in this taxonomy. We quantify this goal as above using a
confusion matrix, except that we
compare the classes assigned based on the complete dataset to the most-probable
previous classification of the asteroid in the literature, retrieved for \num{\NSamplesPrev}
samples of \num{\NAsteroidPrev} individual asteroids from the SsODNet database. We
convert the previous classifications done mostly in the Bus-DeMeo scheme to this
scheme using the mapping given in \cref{tab:taxonomy_overview}.

\Cref{fig:complete_vs_previous} shows an overall good agreement of the classes
assigned in this work with the ones from the literature. Notable exceptions are
the \class{O}-type, which has no legacy members apart from
\nuna{3628}{Boznemcova} as pointed out in \cref{sub:otypes}, and the new
\class{Z}-class, which hosts almost exclusively previous
\class{D}-types. Furthermore, \class{L} loses members to \class{S} as
well as \class{O} to \class{V}.

\section{Conclusion}
\label{sec:conclusion}

The taxonomic scheme for minor body classification has been in development for close
to 50 years. During this time, numerous efforts to categorise the observational
properties of asteroids have been driven forwards through dedicated observational
campaigns and instrumental advancement. We focused on the methodology and
statistical foundation, allowing us to increase the sample size by an order of
magnitude compared to the previous taxonomy by \citet{AnExtensionOfDemeo2009}
and to reintroduce the albedo into the classifying observables as done in
\citet{1984PhDT.........3T}.

The dimensionality reduction and clustering applied to \num{\NSamples} spectra of
\num{\NAsteroids} asteroids
revealed three main complexes: the well established \class{C}-
and \class{S}-complexes and a restructured \class{M}-complex. While the
\class{S}-complex is well understood in terms of mineralogy and meteoritic
analogue material, both the \class{C}-complex and \class{M}-complex show a large
degree of variability of so far unknown origin. We derive \num{\NClasses} classes from
the three complexes, where the data-driven clustering is guided by the previous
taxonomies and the goal of class continuity.

A classification tool named \texttt{classy} is available online and allows the
user to classify asteroid observations covering the spectral
VisNIR region and the visual albedo either completely or
partially. The resulting array of class probabilities for each sample serves to
estimate classification uncertainty and possible taxonomic trends.

We  established a methodology for asteroid taxonomy which is well suited for
the current and future datasets of asteroid observations. The ongoing MITHNEOS
survey, the upcoming
Gaia Data Release 3 \citep[including visible spectra,][]{AsteroidSpectrDelbo2012},
and the planned NEO Surveyor mission \citep{SurveySimulatiMainze2015}
 and SPHEREx survey
\citep{SimulatedSpherIvezic2022}
will provide or continue to provide spectral and albedo observations of asteroids
in different wavelengths, which are able to be classified within the framework
of this taxonomy.

The dimensionality reduction and clustering are able to resolve more features
and find more meaningful clusters when fed with more data. It may be
worthwhile exploring how the model properties described in \cref{sec:results}
change when fed with significantly more data. Nevertheless, during this work, we
found that the latent space properties show little change whether we
train with 500, 1000, or all samples in the dataset. Instead, we anticipate that
a future taxonomy-revision will benefit more from an increased feature set. In
particular the UV information offered by the Gaia data may solve
degeneracies in the \class{C}- and \class{M}-complex. A further improvement should
be the addition of polarimetric data, provided the amount of observations is
comparable to the availability of the other features. The \class{M}-complex
could benefit, and we consider that most work is left to be done in this
complex. Extension of the spectral space into the \SI{3}{\micro\meter} region
is promising as well.

\section*{Acknowledgements}%
\label{sec:acknowledgements}%
The authors thank the numerous members of the community who shared their
spectral observations of minor bodies to support this work and the referee
Pierre Vernazza for helpful comments during the review process. Rémi Flamary
provided valuable support in the exploration of machine learning methods during
the initial stage of the project.
Malheureusement, la nature de l'astéroïde B612 reste un mystère.

This research has made use of IMCCE's SsODNet/Quaero VO tool.

This research has made use of IMCCE's Miriade VO tool.

This research has made use of the SVO Filter Profile Service
supported from the Spanish MINECO through grant AYA2017-84089.

All (or part) of the data utilised in this publication were obtained and made
available by the MITHNEOS MIT-Hawaii Near-Earth Object Spectroscopic Survey. The
IRTF is operated by the University of Hawaii under contract 80HQTR19D0030 with
the National Aeronautics and Space Administration. The MIT component of this
work is supported by NASA grant 80NSSC18K0849.

\begin{refcontext}[sorting=nyt]
\printbibliography
\end{refcontext}

\onecolumn

\renewcommand\thefigure{\thesection.\arabic{figure}}
\setcounter{figure}{0}
\renewcommand\thetable{\thesection.\arabic{table}}
\setcounter{table}{0}

\begin{appendix}

\section{Distribution of albedos in cluster}%
\label{app:distribution_albedos_cluster}%

\begin{figure}[h] \centering
  \input{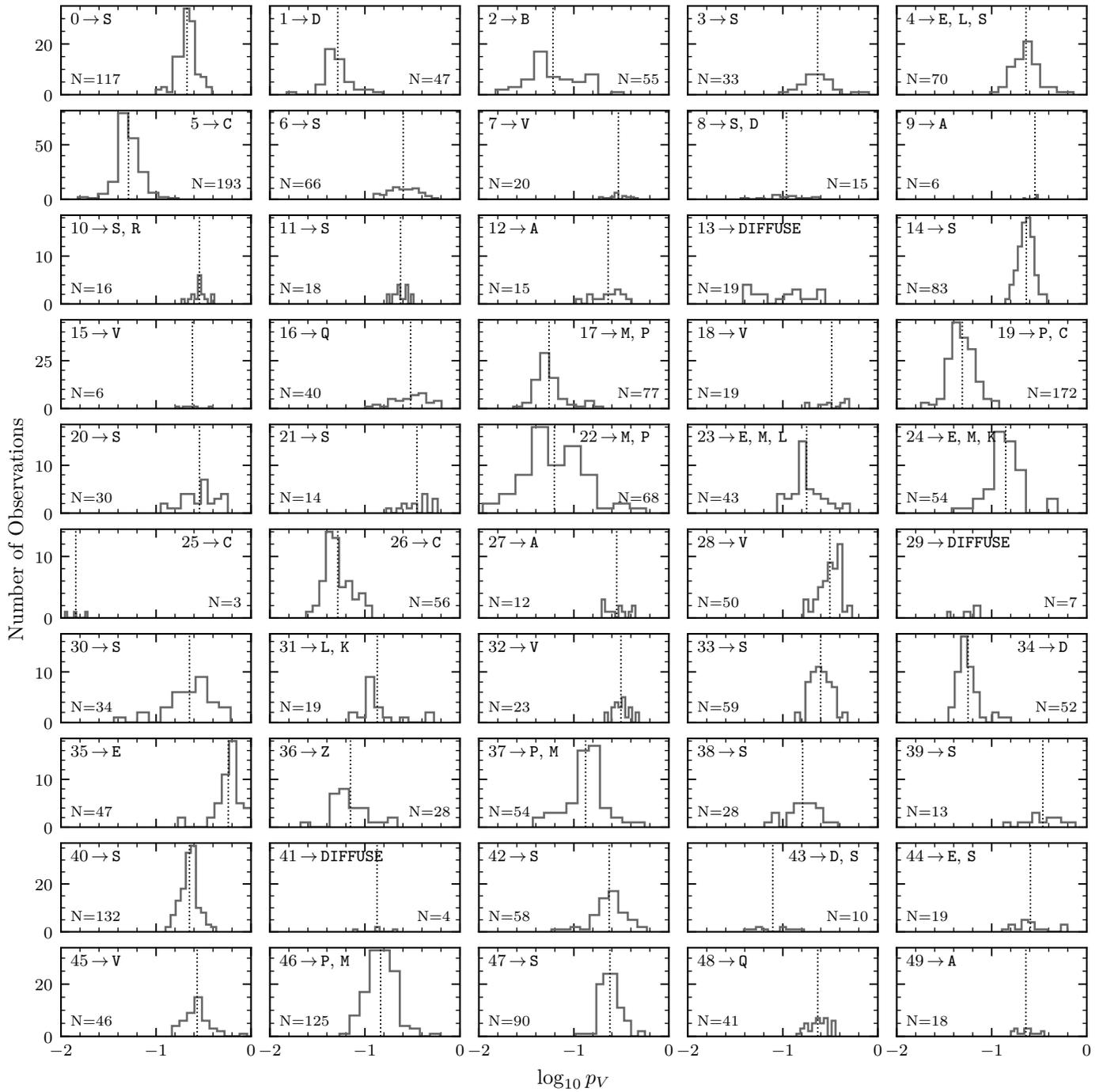}
  \caption{Overview of the albedo distribution per cluster, including the number
    $N$ of albedos and the asteroid classes to which the cluster contributes ,
    excluding classes with fewer than three contributed observations except for
    cluster
      25 which has only three observations. The classes are sorted by the total
         number of observations the cluster contributed. The dotted line gives the
       mean value of the albedos per cluster except for diffuse clusters and
     cluster 25. The y-axis limit is different in each
    row.} \label{fig:feature_overview_feature_albedo_which_cluster}
 \end{figure} \clearpage

\section{Feature centres and windows}
\label{sec:feature_centers_and_windows}

\begin{table}[h]
  \centering
  \caption{Listed  are the mean band centres and the mean upper and lower band limits
  determined using the visually identified features in the input data.
These values are applied when using the automatic feature detection
  with the \texttt{classy} tool.
}
  \label{tab:feature_flags}
\begin{tabular}{lrrr}
        \toprule
        Feature & Centre / \textmu m & Lower Limit / \textmu m & Upper Limit / \textmu m \\
        \midrule
\texttt{e} & \num[round-mode=places,round-precision=2]{0.49725}\,$\pm$\,\num[round-mode=places,round-precision=2]{0.0055} & \num[round-mode=places,round-precision=3]{0.45} & \num[round-mode=places,round-precision=3]{0.539} \\
\texttt{h} & \num[round-mode=places,round-precision=2]{0.69335}\,$\pm$\,\num[round-mode=places,round-precision=2]{0.011} & \num[round-mode=places,round-precision=3]{0.549} & \num[round-mode=places,round-precision=3]{0.834} \\
\texttt{k} & \num[round-mode=places,round-precision=2]{0.90596}\,$\pm$\,\num[round-mode=places,round-precision=2]{0.017} & \num[round-mode=places,round-precision=3]{0.758} & \num[round-mode=places,round-precision=3]{1.06} \\
        \bottomrule
\end{tabular}
\end{table}

\section{Distribution of spectra and albedos in classes}%
\label{app:distribution_spectra_albedos_classes}%

\begin{figure}[h]
  \centering
  \input{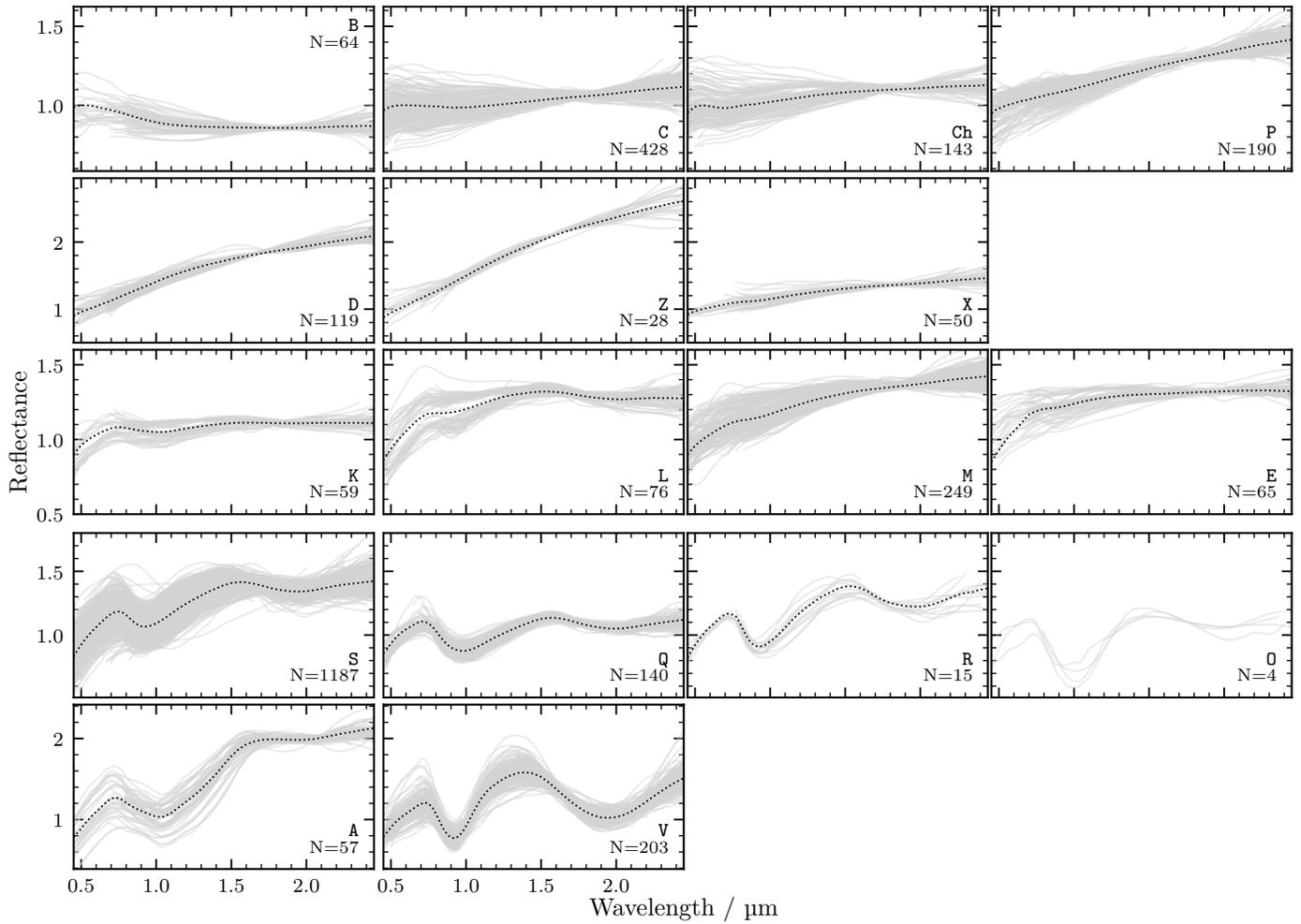}
  \caption{Distribution of spectral observations over the \num{\NClasses}
    classes assigned in this taxonomy. The number $N$ of spectral observations assigned
    to the class is given under the respective letter. Spectra contributed by diffuse
  clusters are excluded.}
  \label{fig:feature_overview_feature_spectra_which_classes}
\end{figure}

\begin{figure}[t]
  \centering
  \input{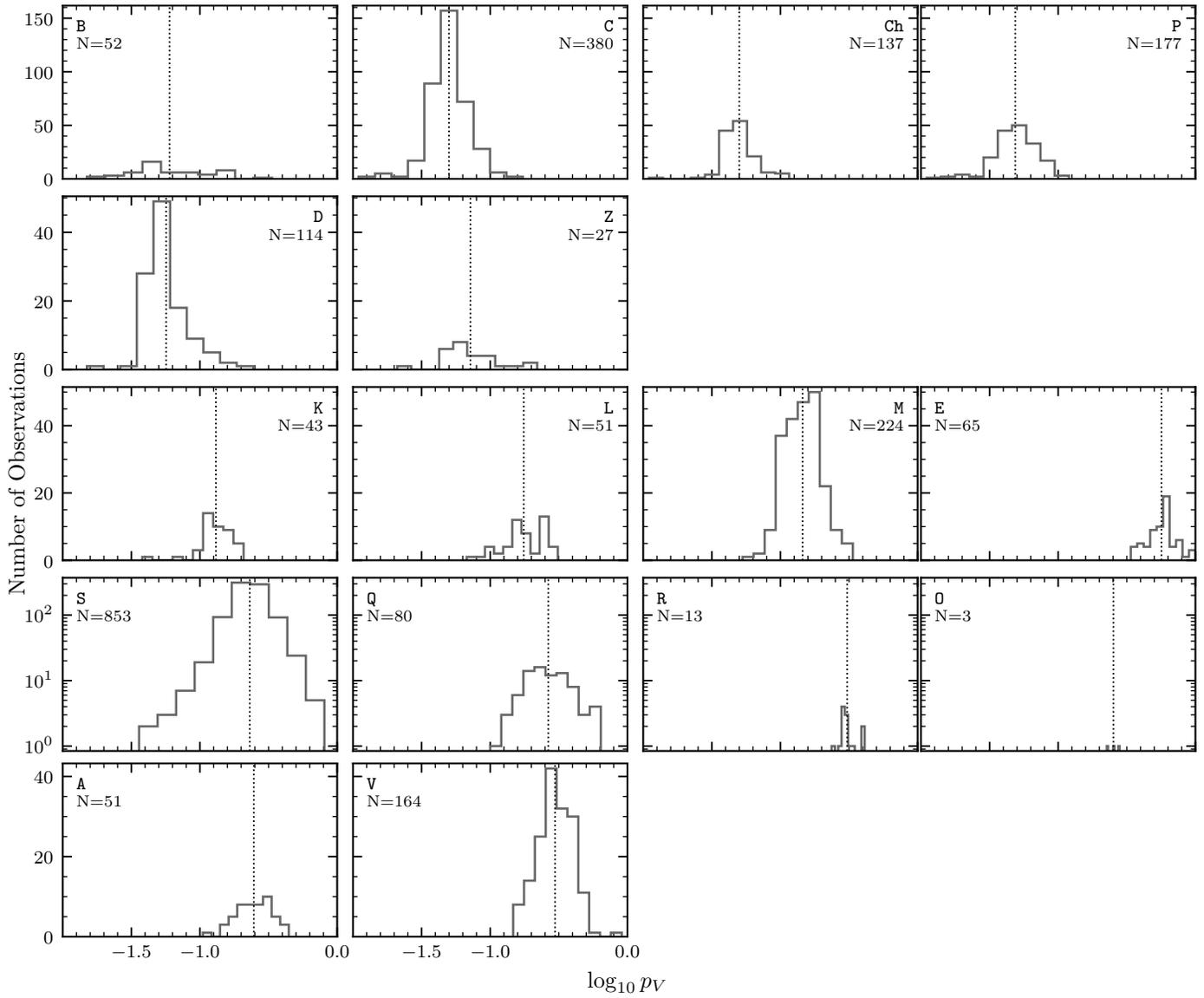}
  \caption{Distribution of albedo observations over the \num{\NClasses}
    classes assigned in this taxonomy excluding the \class{X}-class. The number
    $N$ of albedo observations assigned to the class is given under the respective
    letter. Albedos contributed by diffuse clusters are excluded.}
  \label{fig:feature_overview_feature_albedo_which_classes}
\end{figure}
\clearpage

\section{Cluster-to-class decision tree}
\label{sec:summaries_of_preprocessing_clustering_and_classification}

\begin{table}[h]
  \centering
  \caption{Cluster-to-class decision tree.
Overview of the computation of the asteroid-class
  probability for each observation based on its cluster probabilities. The
      upper part of the table contains clusters whose members are mapped to a
      single asteroid class. The lower part contains clusters where the
      resulting asteroid class probabilities
depend on the criterion given in the middle column. GMM($z_x$, $z_y$) means
that the cluster probability is split based on a Gaussian mixture model with
$N$ components fit to all cluster members in $z_x$ and $z_y$, where $N$ is
equal to the number of possible outcome classes (\ie each mixture component
represents one candidate class). $P_X(y)$ refers to the probability of belonging
to the class or cluster $X$ given the value of $y$. The last line
gives the definition of the \class{Ch}-class, which is the last step of the classification.
  }
  \label{tab:decision_tree}
\begin{tabular}{lcl}
\toprule
Cluster & & Class \\ \midrule
0, 3, 6, 11, & &  \\
\hspace{0.5em}14, 20, 21, 30, & \textrightarrow & \class{S} \\
\hspace{0.5em}33, 38, 39, 40, & &  \\
\hspace{0.5em}42, 47 & &  \\
1, 34 & \textrightarrow & \class{D} \\
2 & \textrightarrow & \class{B}\\
5, 25, 26 & \textrightarrow & \class{C}\\
7, 15, 18, & \textrightarrow & \class{V} \\
\hspace{0.5em} 28, 32, 45 & & \\
9, 12, 27, 49 & \textrightarrow & \class{A} \\
16, 48 & \textrightarrow & \class{Q} \\
36 & \textrightarrow & \class{Z} \\
\midrule
4 & $P_{23}(z_3, z_4) / P_{40}(z_3, z_4)$ & \class{L}, \class{S} \\
8, 43 & GMM($z_2$, $z_4$) & \class{D}, \class{S} \\
10 &    GMM($z_1$, $z_2$) & \class{R}, \class{S}  \\
13 &    GMM($z_2$, $z_4$) & \class{C}, \class{O}, \class{Q}\\
17, 22, 35, &  & \class{E}, \class{M}, \class{P}, \class{X}\\
\hspace{0.5em}37, 46 & $P_\class{E}(p_V) / P_\class{M}(p_V) / P_\class{P}(p_V)$ & \\
19 &    GMM($z_1$, $z_4$) & \class{C}, \class{P} \\
23 &    GMM($z_1$, $z_4$) & \class{L}, \class{M} \\
24 &    GMM($z_2$, $z_3$) & \class{K}, \class{M} \\
 & & \class{A}, \class{B}, \class{C}, \\
29 & GMM($z_1$, $z_2$) & \hspace{0.5em}\class{D}, \class{M}, \class{P}, \\
 & & \hspace{0.5em}\class{S}, \class{Q}, \class{V} \\
31 & GMM($z_3$, $z_4$) & \class{K}, \class{L}  \\
37 & GMM($z_2$, $z_4$) & \class{L}, \class{M} \\
41 &    GMM($z_1$, $z_2$) & \class{B}, \class{V} \\
44 & $P_\class{E}(p_V) / P_\class{M}(p_V)$  & \class{E}, \class{S}  \\
\midrule

\multicolumn{2}{l}{Class is \class{B}, \class{C}, \class{P}, or \class{X} and \class{h}-feature is present} & \class{Ch} \\
\bottomrule
\end{tabular}

\end{table}
\clearpage

\section{References of spectra and visual albedos}
\label{sec:references_of_spectra_and_visual_albedos_spectra_albedo}

\begin{table}[h]
  \centering
  \caption{Spectroscopic data references}
  \label{tab:ref_spectra}
\begin{tabular}{p{1\textwidth}}
\toprule \\
\cite{0725IMIWong2017,2006PDSS...45.....V,ModelingOfAstPopesc2012,
2007PDSS...51.....L,
2009PDSS..107.....W,
TheInnerRegioAlvare2006,
2009PDSS..108.....M,
2010PDSS..127.....R,
2010PhDT.......134F,
2011PDSS..144.....F,
2016PDSS..242.....R,
2017PDSS..277.....R,
2021pdss.data....6G,
2021pdss.data....8P,
ACommonOriginMoskov2019,
ANearInfraredYang2011,
ANewInvestigaLandsm2015,
ASpectroscopicMoskov2010,
ASpectroscopicPerna2018,
AnRTypeAsterMarchi2004,
AnalysisOfNeaVernaz2005,
AncientAsteroiSunshi2008,
AqueousAlteratFornas2014,
SpexAMediumRayner2003,
BasaltOrNotHarder2018,
BasalticMateriIeva2018,
CompositionalHVernaz2016,
ConstraintsOnEmery2003,
ETypeAsteroidNedelc2007,
HighAlbedoCCKasuga2013,
HungariaAsteroLucas2017,
VisibleSpectroDevoge2019,
IrtfSpectraFoOstrow2011,
MineralogicalCDeSan2011,
IrtfObservatioGietze2012,
MineralogicalCDuffar2004,
MineralogicalCReddy2011,
MineralogicalIJasmim2013,
MoreChipsOffHarder2014,
MultiWavelengtShepar2008,
MultipleAndFaVernaz2014,
NearEarthAstePopesc2019,
NearInfraRedBirlan2006,
NearInfraredSArredo2021,
NearInfraredSEmery2011,
TheMXAsteroHarder2011,
NearInfraredSFieber2011,
NearInfraredSFieber2012,
NearInfraredSFieber2014,
NearInfraredSFieber2015,
NearInfraredSKasuga2015,
NearIrSpectroBirlan2004,
NewObservationDeLeo2011,
PrimassVisitsDePra2018,
TheOlivineDomBoriso2017,
1215141999UBoriso2018,
TheVisibleAndLicand2018,
NewPolarimetriDevoge2018,
NewVTypeAsteMarchi2005,
ObservationsCDeLeo2010,
ObservationsOfOckert2008,
ObservationsOfPolish2014,
OlivineDominatSanche2014,
OlivineDominatSunshi2007,
PhysicalAndDyYang2020,
PhysicalCharacVernaz2006,
PhysicalProperHasega2018,
PortraitOfThePinill2016,
PyroxeneMineraBurbin2009,
SimilarOriginMarsse2014,
TheDebiasedCoMarsse2022,
SmallDTypeAsBarucc2018,
SpectralAndMiDeSan2011,
SpectralCharacMiglio2017,
SpectralProperBinzel2009,
SpectralProperBirlan2007,
SpectralProperBirlan2011,
SpectralProperLucas2019,
SpectralProperPopesc2011,
SpectralProperPopesc2014,
SpectralVariabFornas2016,
SpectroscopicILazzar2005,
SpectroscopicOBendjo2004,
SpectroscopicSYang2007,
SpectroscopyAnBirlan2014,
SpectroscopyOfClark2004,
SpectroscopyOfClark2009,
SpectroscopyOfMiglio2018,
SpinRatesOfVOszkie2020,
SurfaceComposiReddy2018,
SurfaceComposiSanche2013,
TheCompositionNeeley2014,
TheCompositionOckert2010,
TheMariaAsterFieber2011,
TheNatureOfBHasega2021,
TwoNewVTypeDuffar2009,
VestoidsPartHarder2015,
VisibleAndNeaLazzar2004,
VisibleSpectroFornas2007,
VisibleWavelenKuroda2014,
InfraredSpectrRivkin2004,
ObservedSpectrBinzel2004,
DynamicalAndCBinzel2004,
SpectralObservBinzel2004,
2001MPSA..36S..20B,
SpectralProperBinzel2001,
2000PhDT.......355B,
1999PhDT........50B,
2002Icar..158..146B,
PhaseIiOfTheBusS2002,
SmallMainBeltXuSh1995,
1994PhDT.........1X,
SpectralCharacMatlov2020} \\
\bottomrule
\end{tabular}
\end{table}

\begin{table}[h]
  \centering
  \caption{Data references for albedos, diameters, and absolute magnitudes}
  \label{tab:ref_albedos}
\begin{tabular}{p{1\textwidth}}
\toprule \\
\cite{1994IAUS..160..477B,
TheShapeOfGaThomas1994,
GalileoPhotomeHelfen1994,
TheShapeOfIdThomas1996,
GalileoPhotomeHelfen1996,
MathildeSizeThomas1999,
NearPhotometryClark1999,
TheShapeOfErThomas2000,
NearAtErosIVeverk2000,
RadarObservatiBenner2002,
KeckObservatioDelbo2003,
65CybeleInThMuller2004,
TheRubblePileFujiwa2006,
ARadarSurveyMagri2007,
MultiWavelengtShepar2008,
RadarObservatiShepar2008,
TriaxialEllipsDrummo2008,
ThermalInertiaDelbo2009,
ExploreneosITrilli2010,
10.1088/0004-6256/140/4/933,
ETypeAsteroidKeller2010,
ExploreneosIiMuelle2011,
Masiero_2011,
IWiseINeoGrav2011,
2011ApJ...741...90M,
DeterminationOMatter2011,
AsteroidCataloUsui2011,
ImagesOfAsterSierks2011,
IWiseIIGrav2012,
IWiseINeoGrav2012,
PreliminaryAnaMasier2012,
PhysicalParameMainze2012,
AbsoluteMagnitPravec2012,
Asteroid2867Jorda2012,
MultipleAsteroMarchi2012,
DawnAtVestaRussel2012,
PhysicalProperAliLa2013,
AThermophysicaRoziti2013,
EvidenceOfAMMatter2013,
GlobalPhotometLiJi2013,
TheGingerShapHuang2013,
ThermalInfrareMuller2014,
PhysicalCharacRoziti2014,
ThePopulationMainze2014,
InitialPerformMainze2014,
Masiero_2014,
PhysicalAndDyBerthi2014,
TheKilometerSRyan2015,
2015ApJ...814..117N,
PhysicalModeliBecker2015,
ThermophysicalHanus2015,
CharacterizingKoren2015,
SizeAndAlbedoLicand2016,
DifferencesBetAliLa2016,
NearEarthAsteHanus2016,
NeowiseReactivNugent2016,
Neosurvey1InTrilli2016,
SurfaceAlbedoLiLiJian2016,
ReflectanceOfDongF2016,
DawnArrivesAtRussel2016,
VolumesAndBulHanus2017,
SizesAndAlbedAliLa2017,
AdaptiveOpticsViikin2017,
NeowiseReactivMasier2017,
SurfaceThermopYuLi2017,
TheAkariIrcAAliLa2018,
3200PhaethonHanus2018,
TheTriaxialElDrummo2018,
ThermophysicalMasier2019,
2019pdss.data....3H,
PhysicalProperMasier2020,
AsteroidDiametMasier2020,
VltSphereImagVernaz2021,
VisNirDiskInTatsum2018,
ThermophysicalJiang2021,
AThermophysicaChavez2021,
AsteroidDiametMasier2021,
ThermalPropertHung2022} \\
\bottomrule
\end{tabular}
\end{table}

\end{appendix}
\end{document}